%% file: ms.tex
\documentclass[useAMS,usenatbib]{mn2e}
\usepackage{multirow} 
\usepackage{array}
\usepackage{graphicx}
\usepackage[applemac]{inputenc}
\usepackage[T1]{fontenc} 
 \usepackage{aecompl}

\def\ignore#1\endignore{}
\newcolumntype{L}{@{}>{\ignore}l<{\endignore}} %% blind out some colums

\usepackage{epsf} 
\usepackage{pdfpages}
\usepackage{lscape}	
\usepackage{multicol} 
\usepackage{float}

\usepackage{etex}
\reserveinserts{18}
\usepackage{morefloats}

\usepackage{color}
\usepackage{longtable}
\usepackage{topcapt}
\usepackage{caption}
\usepackage{array}

\usepackage{fancyheadings}
\usepackage{array}
\usepackage{hyperref}
\usepackage{footnote}

\usepackage{amssymb}
\usepackage{amsmath}
 
\everymath{\displaystyle\everymath{}}
\newcommand{\hi}{\mbox{H{\sc i}}}
\newcommand{\hii}{\mbox{H{\textsc{ii }}}}
\newcommand{\hd} {H$\mathrm{_2}$}
\newcommand{\ha} {H$\alpha$}

\newcommand{\kms}{km~s$^{-1}$}

\newcommand{\msol}{\rm M$_\odot$}

\definecolor{Zach}{rgb}{1,0,1}

\newcommand {\hI} {H\textsc{i}}
\newcommand {\hII} {H\textsc{ii}}

\title{Kinematics and Mass Modeling of M33:   \ha\ Observations}

\author[Kam et al.]{Kam, Z. S.$^{1,2,3}$\thanks{E-mail:kam@astro.umontreal.ca}, Carignan, C.$^{1,2,3}$, Chemin, L.$^{4}$, Amram, P.$^{5}$ \& Epinat, B.$^{5}$ \\
$^1$Laboratoire d'Astrophysique Exp\'erimentale, D\'epartement de physique,\\
Universit\'e de Montr\'eal, C.P. 6128, Succ. centre-ville, Montr\'eal, Qu\'ebec, Canada, H3C 3J7\\
$^2$Observatoire d'Astrophysique de l'Universit\'e de Ouagadougou, BP 7021, Ouagadougou 03, Burkina Faso\\
 $^3$Department of Astronomy, University of Cape Town, Private Bag X3, Rondebosch 7701, South Africa\\
 $^4$Universit\'e de  Bordeaux \& CNRS, LAB, UMR 5804, F-33270, Floirac, France\\
 $^5$ Universit\'e Aix Marseille , CNRS, LAM (Laboratoire d'Astrophysique de Marseille), 13388, Marseille, France}

\begin{document}
\pagerange{\pageref{firstpage}--\pageref{lastpage}} \pubyear{2015}
\maketitle
\label{firstpage}

\begin{abstract} 
As part of a long-term project to revisit the kinematics and dynamics of the large disc galaxies
 of the Local Group, we present the first deep, wide-field ($\sim42' \times 56'$) 3D-spectroscopic 
 survey of the ionized gas disc  of Messier 33.  Fabry-Perot interferometry has been used to map
  its \ha\ distribution and kinematics at unprecedented angular resolution ($\lesssim$3\arcsec) 
  and resolving power ($\sim12600$), with the 1.6~m telescope at the Observatoire du Mont M\'egantic.  
  The ionized gas distribution follows a complex, large-scale spiral structure, unsurprisingly coincident
   with the already-known spiral structures of the neutral and molecular gas discs. The kinematical 
   analysis of the velocity field shows that the rotation center of the \ha\ disc is  distant from 
   the photometric center by $\sim170$~pc (sky projected distance) and that the kinematical major-axis position angle and  disc inclination are in excellent agreement with photometric values. The \ha\ rotation curve agrees very well with the \hi\ rotation curves for $0 < R < 6.5$ kpc,  but the \ha\  velocities are $10-20$ \kms\ higher for $R >$ 6.5 kpc. The reason for this discrepancy is not well understood. The velocity dispersion profile is relatively flat around $16$~\kms, which is at the low end of velocity dispersions of nearby star-forming galactic discs.  A strong relation is also found between the \ha\ velocity dispersion and the \ha\ intensity. Mass models were obtained using the \ha\ rotation curve but, as expected, the dark matter halo's parameters are not very well constrained since the optical rotation curve only extends out to 8~kpc.
\end{abstract}

\begin{keywords}
techniques: interferometric, integral field, Fabry-Perot -- galaxies: individual: M33 -- galaxies: kinematics and dynamics -- galaxies: ISM: Bubbles, HII regions -- cosmology: dark matter
\end{keywords}
\section{ Introduction}

Late-type spiral and dwarf irregular (dIrr) galaxies present an extended \hI\ emission. The \hI\ emission  usually extends to larger radii than the optical emission. This makes of  \hI\ a  powerful tool for probing the dark matter (DM) dominated regions and for  characterizing the parameters of the flat part of the rotation curves (RCs). However, it is the rising inner part of the RCs that constrains the parameters of the mass models  \citep[see e.g.][]{Blaisou1999}. 
 Two-dimensional \ha\ observations  are ideally suited  to give  more accurate velocities in the inner parts of galaxies having a better spatial resolution than the \hI\  \citep{Amram1992, Amram1994,  Swaters2000}. The ideal approach is then to  combine high spatial resolution \ha\ observations in the inner parts  and sensitive  lower spatial  resolution \hI\  observations in the outer parts to derive the most accurate mass models for the  luminous disk and the  dark halo. The objective of this project is to perform a new mass distribution model of Messier 33 (M33), combining high-sensitivity \ha\ and HI interferometric data. In particular, this article  presents the first \ha\ survey devoted to the large-scale distribution and kinematics of the M33 ionized gas disk.  
  
M33 is the second most luminous spiral (SA(s)cd) galaxy in our neighborhood after M31. With an absolute magnitude of M$_V$$\sim$-19.3, it presents  several arms \citep{sandage1980}. Although two main arms are well known, M33 has not a clearly defined  grand-design pattern. A system of seven spiral arms  is pointed out  by \cite{Ivanov1985}.  In the optical bands,  it could be considered as a flocculent spiral galaxy, but  UV and IR observations show more prominent arms.  
The M33 profile presents a  small bulge-like component in the IR bands. Its bulge and nucleus  are the subject of numerous studies (e.g.: \citealt{McLean1996, Lauer1998, Gordon1999, Corbelli2007a}).   The optical parameters of M33 are given in Table~\ref{tab:oparam}.
\begin{table}
\centering

\caption{Optical parameters of M 33.}
\begin{tabular}{@{}lllllllllllll@{}}
\hline
\hline
Parameter & Value & Source \\ \hline
Morphological type	& SA(s)cd & RC3	 \\ 
R.A. (2000) 				& 01$\rm^h$ 33$\rm^m$ 33.1$\rm^s$&RC3 \\
Dec. (2000)				& +30$^o$ 39$^{'}$ 18${''}$&RC3 \\
Optical inclination, {\it i}		& 52°$\pm$3°& WW \\
Optical major axis, PA  		& 22.5°$\pm$1°& WW \\ 	
Apparent magnitude, m$_V$ 	& 5.28& RC3\\ 
Absolute magnitude, M$_V$ 	&-19.34& \\
(J-K) &0.89 				&JTH\\
Optical radius, R$_{25}$ (\arcmin) 	& 35.4 $\pm$ 1.0&RC3 \\
Systemic velocity, V$_{sys}$		 &-179 $\pm3$ \kms& RC3\\
\hline
\end{tabular}
\label{tab:oparam}

References:WW: \cite{warner1973}, RC3: \cite{deVaucouleurs1991},  JTH: \cite{Jarrett2003}
\end{table}

In the literature, one can find distances to M33 varying from 700 to 940~kpc. Table~\ref{tab:distances} gives recent distance measurements for M33. The adopted distance for this study is D = 0.84~Mpc (scale 1' = 244~pc). It is the  mean distance estimated by using resolved sources techniques  such as Cepheids,  Planetary  Nebulae Luminosity Function (PNLF) and  Tip of the Red Giant Branch (TRGB). The optical disk of M33 has a scale length of $\sim$9.2\arcmin\ (2.25 kpc) in the V band \citep{Kent1987,Guidoni1981} and its optical radius is  R$_{25}$ = 35.4\arcmin\  (RC3). A cutoff in the radial surface brightness  profile appears at $\sim$ 36\arcmin\  ($\sim 9$ kpc) in the I band \citep{Ferguson2007}.   
\begin{table}
\centering
\caption{Distances for M 33.}
\begin{tabular}{@{}llclllllllllll@{}}
\hline
Method	&Distance Modulus& Distance & Source \\
(indicator)		& m-M [mag]		& [Mpc]    & \\
\hline		
    TRGB  & 24.54 $\pm$ 0.06   & 0.81   		& (1)  \\
    TRGB  & 24.75 $\pm$ 0.10   & 0.89  		& (2)   \\
    TRGB  &24.64 $\pm$ 0.15&0.85			& (3)  \\
    PNLF  & 24.62 $\pm$ 0.25   & 0.84  		& (4)  \\
    Cepheids & 24.64 $\pm$ 0.06   & 0.85 	 	& (5)  \\
    Cepheids & 24.70 $\pm$ 0.13   & 0.87	  	& (6)  \\
    Cepheids    & 24.52 $\pm$ 0.14   & 0.80  	& (7)  \\
    Cepheids & 24.55 $\pm$ 0.28   & 0.81	  	& (8) \\
    Cepheids & 24.58 $\pm$ 0.10   & 0.84   	& (9)   \\ 
    Cepheids & 24.62 $\pm$ 0.01   & 0.84	  	& (10)   \\  
     \multicolumn{2}{l}{Mean adopted distance} & 0.84 $\pm 0.04$&\\ 
   & & 1\arcsec\ $\equiv 4$ pc &\\ 
\hline
\end{tabular}
\label{tab:distances}

References: (1) \cite{McConnachie2005}, (2) \cite{Stonkute2008}, (3) \cite{Galleti2004}, (4) \cite{Magrini2000}, 
(5) \cite{Saha2006}, (6) \cite{Paturel2002}, (7) \cite{McConnachie2004}, (8) \cite{An2007}, (9) \cite{Freedman2001}, (10) \cite{Gieren2013}
\end{table}

Mapping the environment of the Local Group galaxies, as is done by the “Pan-Andromeda Archaeological Survey” (PAndAS) \citep{McConnachie2009}, confirms that there was many mergers and interactions between them. The discovery of  dwarf  galaxies around the Milky Way and M31 and the tidal streams between M31 and M33 (PAnDAS)  confirm our ideas about galaxy formation.  The particular structure (star streams) seen in M33 could be associated with this history of mergers and interactions. In fact, many of the structures presented in \cite{McConnachie2009, McConnachie2010, Cockcroft2013} and \cite{ Wolfe2013}  around M31 and M33 can only be explained by these interactions.  
Deep \ha\  observations of M33  reveal the presence of low density \hII\ regions outside the optical disk \citep{Hoopes2001}. This suggests recent star formation activity, possibly due to recent interactions. 
 
 Studying  the kinematics of such a galaxy will provide a better understanding of the contribution of dark matter and of the best functional form describing its distribution. 
 Still today, the cusp--core problem  remains as one compares observations to predictions, especially for dwarf systems and this, despite the numerous studies on the DM distribution in galaxies in the last 30 years.  While, on large scales, N-body cosmological  simulations reproduce well the observations, it is more problematic at galaxy scales. The NFW profile, derived from those simulations, predicts a cuspy distribution in the central parts of the DM halos (e.g.  \citealt{Fukushige1997, Moore1999, Navarro1997, Navarro2010, Ishiyama2013}), while observations especially of dwarf systems show more a core distribution \citep{oh2011}. Those results can be explained by the gravitational potential related to the gas in those simulations, since  the gas, which is important in the inner parts, is not accurately introduced in those simulations.  Phenomena such as stellar feedbacks, galactic winds or massive clumps are not often well handled and reproduced \citep{Goerdt2010, Inoue2011, Ogiya2011,Pontzen2012,Teyssier2013}.
 
 Several studies  of the kinematics of M33  exist  (e.g. \citealt{Corbelli2000, Corbelli2003, Putman2009, Corbelli2014}), mainly based on \hI. %  .
 It appears from those studies that M33 is dark matter dominated (the dark matter mass is $>$ 10 times larger than the mass of the stellar disk) and that its \hi\ disk is strongly warped.  The gas in the disk of M33 is estimated  $\sim$1.4-3$\times$$10^9$~\msol\ \citep{Corbelli2003,Putman2009}. In the mass model derived  by \cite{Corbelli2003}, the stellar mass is  estimated at $\sim3-6\times$$10^9$~\msol\ and the dark halo mass at $\sim$5$\times$$10^ {10} $~\msol.
Most of the M33 kinematical studies use  \hI\ or CO (\hd), with a probable beam smearing impact on the innermost parts of the RC due to the lower spatial resolution of the \hI\ data. Among  the few \ha\ (optical) studies, the results presented by \cite{Carranza1968}  show small velocity dispersions from 5~\kms\  to 9~\kms\ in the disk. However, they argue that these values could even be lower if taking into account the instrumental corrections.  
The mass-to-light ratio is an important parameter in the determination of the dark matter halo shape. In M33,  \cite{Boulesteix1970} show very low values of M/L in the central ($<$ 50\arcmin) part of the galaxy and increasing in the outer parts of the disk.   The kinematics of the inner regions was studied using the warm gas (\ha) \citep{Boulesteix1970, Boulesteix1974, Zaritsky1989} and the cold molecular gas (\hd\  via CO) \citep{Wilson1989, Gratier2010, Gratier2012, Kramer2013, Druard2014}. 

The precise determination of the rising parts of RCs with high resolution data and a better estimate of the M/L ratio for the stellar disk could define more accurately the shape of its dark matter halo.  High resolution optical observations of a nearby galaxy such as M33 is complex in view of the large size of the galaxy compared to the small field of view of  high resolution integral field spectroscopic instruments.  For M33,  there is a lack of high resolution optical  data available for the kinematical study. An exception is the study by \cite{Corbelli2007a} on the central 5\arcmin\ using the gas and stellar kinematics obtained by long-slit spectroscopy.

Mass models are sensitive to the rising part of the RCs (see, e.g \citealt{Amram1996}, \citealt{Swaters1999}, \citealt{Blaisou1999}). High resolution \ha\ RCs  give a more accurate determination of the kinematical parameters for the derivation of the RCs and subsequently a more accurate determination of the mass model parameters .  The  high resolution Fabry-Perot (FP) integral field observations at   \ha\ with a resolution $\ga~10000$ give  optimal kinematical data for the optical disk.   The \ha\ RC can be used to test the shape of the DM halo allowing us to compare the derived DM distribution with CDM predictions. 

To address these problems, this study presents Fabry-Perot (FP) mapping of M33 obtained at the Observatoire du mont Mégantic (OMM).
\cite{Relano2013} have studied the Spectral Energy Distribution (SED) of the \hii\ regions of M33 and the star formation rate (SFR)  and star formation efficiency (SFE) have been investigated by \cite{Gardan2007} and \cite{Kramer2011}.   More than 1000 \hII\ regions can be resolved by the  \ha\ observations; \cite{Courtes1987} gave a catalogue of 748 \hII\ regions. Observing those regions  allows us to derive the ionised gas (optical) kinematics of M33.
 Determination of the M33 kinematics with a spatial resolution $\lesssim$3\arcsec\ ($\sim12$~pc) using the  \ha\ velocity field and the derivation of an accurate RC in the inner parts are the main goals of this paper. The 3D data will be used to derive  mass models  with a dark halo component  (ISO and NFW).
   
%==================================	

Section~\ref{sec:obsred} describes the data acquisition and the instrumentation used while section~\ref{sec:reduction} discusses the data reduction techniques. Section~\ref{sec:result} details the kinematical analysis and section~\ref{sec:modmass} gives the details of the mass models analysis. A discussion of the M33 velocity dispersion, a comparison with other studies and the results from the mass modeling can be found in Section~\ref{sec:discussion}. Finally, a summary and the general conclusions are given in section~\ref{sec:conclusion} .

\section {Fabry-Perot observations}
\label{sec:obsred}
\input{observationsV2}

\input{reductionV2}

\section {\ha\ distribution and kinematics}
\label{sec:result}
\input{analyseV2}
\section {Mass modeling}
\label{sec:modmass}
\input{massmodV2}

\section {Discussion}
\label{sec:discussion}
\input{discussionV2}

\section {Summary and conclusions}
\label{sec:conclusion}

New \ha\ Fabry-Perot mapping of the nearby galaxy M33 has been presented. The data were obtained on the 1.6~m telescope of the Observatoire du Mont Mégantic (OMM) using a high resolution FP etalon (p = 765)  and a very sensitive photon counting EMCCD camera (QE$\sim90\%$). The ten fields observed cover a field of $\sim42' \times 56' $ (10 kpc$\times$13.5 kpc) with a spatial resolution $\sim$3\arcsec. The \hII\ regions inside this area are  well  defined spatially and spectrally. This study provides for the whole field, as well as for each \hII\ regions and nebulae (e.g: NGC 604, NGC 595, NGC 588, IC 137, IC 136) velocity and dispersion maps. The data was flux calibrated using \cite{Relano2013} data.

Looking at the velocity dispersion $\sigma$ profile as a function of radius, we found that it is essentially flat at an average value of $16 \pm 2$ \kms.  From $R \sim 7.5$ kpc, the velocity dispersions increase to $\sim 20-25$ \kms. This radius is the location of the beginning of the warp of the \hi\ disk where the twist of the  position angle starts. The mean velocity dispersion of M33 was found to be small when compared to a sample of nearby star-forming galactic discs (GHASP sample) which have $\sigma$ varying from 15 to 35 \kms.  Finally, a velocity dispersion versus intensity  ($\sigma-I$) diagram shows clearly a strong relation between the \ha\ velocity dispersion and the \ha\ intensity. 

The main aim of this study was to derive a high spatial resolution \ha\ RC for M33 in order to try to constrain 
the best functional form representing the dark matter distribution between the cored ISO models and the cosmologically motivated cuspy NFW models. The rotation center found is very close to the optical center. The other kinematical parameters found are  $\rm V_{sys}$ =  $-178\pm3$ \kms, PA = 202\degr$\pm 4$\degr and {\it i} = $52\degr \pm 2\degr$. The RC was derived separately for the approaching, receding and for both sides. The rotation curve was computed using a constant 5" step to perform the mass modeling.
Besides the intrinsic error in each ring, we added an error to represent the asymmetry of both sides since this RC will be compared to axisymmetric mass models. After a steep rise in the inner 1 kpc, the RC  rises slowly  to its maximum value of $\sim$123$\pm$3 \kms\ at the last point (33\arcmin).

Comparing our adopted RC to \hi\  rotation curves in the literature, we found that our \ha\ data agree very well with the \hi\ from the center out to $\sim$ 6.5 kpc. The only exception is the first inner point of \cite{Corbelli1997}, which may a bit suffer from beam smearing due to the large Arecibo beam. With our high resolution data we bring more data points in the inner part of the RC, which is very useful for the mass modeling. On the other hand, for R $>$ 6.5 kpc, this comparison suggests that our \ha\ data may overestimate the velocities in the outer parts. %Finally, the origin of the higher CO velocities from 5 to 15 kpc is difficult to explain. 

For the mass model analysis, the Spitzer/IRAC $3.6\mu m$ profile was used to represent the stellar disk mass contribution. The disk parameters found are a scale length $R_d=1.82\pm0.02$~kpc and an extrapolated surface brightness $\mu{_0}=18.01\pm0.03$ $\rm mag.arcsec^{-2}$. A bulge--disk decomposition was also done even if it will not alter significantly the results of the mass models with a bulge to disk ratio of only 0.04. Besides best-fit models, models with fixed M/L  based on IR colors and population synthesis models, following the method used by \cite{Oh2008} and 
\cite{deBlock2008} are also explored.

For the mass models, we decided to favor the disk-only models because of the extra uncertainties introduced by adding a bulge which contributes very little (B/D = 0.04) to the luminous mass.
In this case, the ISO models give a better fit despite the fact that they seem to underestimate the velocities in the inner 2 kpc (see the residual curves at the bottom of Figure~\ref{fig:m33dm_all}). This may point out to the fact that the determination of the M/L values using the IR colors and population synthesis models slightly underestimates the M/L ratio of the disk or that an inner mass component  has been omitted in the modeling.
 
An ideal RC using our high spatial resolution \ha\ RC in the inner 6.5 kpc and a high sensitivity but low spatial resolution \hI\ RC in the outer parts should allow to constrain much better the parameters of the mass models. This should be done in a subsequent study presenting new deep \hi\ observations.

\section*{Acknowledgment} 

SZK's work was supported by a grant of the CIOSPB of Burkina Faso, CC's Discovery grant of the Natural Sciences and Engineering Research Council of Canada and CC's South African Research Chairs Initiative (SARChI) grant of the Department of Science and Technology (DST), the SKA SA and the National Research Foundation (NRF). 
L.C. acknowledges a financial support from CNES.
We would like to thank Tom Jarrett for providing the WISE I data, the staff of the OMM for their support and Yacouba Djabo for observing with us at OMM.  A part of the montage package of IPAC have been used in the process of our reduction.
The optical image in blue band comes from the Digitized Sky Surveys (DSS images). The  IR archives images are from the Spitzer  and  WISE Space Telescopes.

%\bibliography{MyBiblio}
\bibliography{ham33_mnras}
 
%==================FIGURES 
%==================TABLES
\appendix{\large Appendices}

\section{ \label{appendTable} Data for the rotation curve and the \ha\ dispersion profile: online}
\input{M33RCadoptedV2}

\label{lastpage}
\end{document}

%% file: observationsV2.tex
\subsection {Telescope and instrument configuration}
The observations took place at the 1.6-m telescope of the Observatoire du mont Mégantic (OMM, Québec),  in September 2012. 
  A scanning Fabry-Perot (FP) etalon interferometer has been used during the observations  with the device IXON888,  a commercial Andor EM-CCD camera of 1024$\times$1024 pixels . The details of this  camera, based on e2v chips, are given in Table~\ref{tab:cameras}.

 \begin{table}

\caption{Photon counting cameras at OMM.}
    \begin{tabular}{lll}
    \hline
Cameras 		&   IXON888 $^{a} $\\ 
 \hline
 Pixels size 	& 0.84\arcsec $\times$ 0.84\arcsec 	\\ 

Active Pixels 	& 1024$\times$1024		\\ 
Quantum Efficiency (QE)  	&$\ga 90\%$				 \\ 
Cooling 		&  -85 °C										\\	
 Max. Readout Speed &10 MHz								\\	
 RON 		& $<1~e^{-}$ with EM gain						\\
 Detectors 	& CCD201-20$^b$	 					\\ 
 CIC$^{c}$  levels & $5-8\times 10^{-3}$$^{d}$		\\ 
  \hline  
  
  \end{tabular}%
 \label{tab:cameras}  
 
 1\arcsec\ $\equiv 4$ pc, 0.8\arcsec\ $\equiv 3.3$ pc at a distance of 0.84 Mpc\\
 $^{a} $ http://www.andor.com \\
 $^{b}$ http://www.e2v.com/products-and-services\\
 $^{c}$ Clock-induced charges per pixel per frame\\
 $^{d} $\cite{Daigle2009b} \\
RON : Read-Out Noise  given per pixel per frame
 
\end{table}%
The IXON888 camera provides a large field-of-view (FOV) and was set to Electron Multiplying (EM) mode, 14 bits read-out resolution and its detector was cooled to $-85$~K during the observations.  The  camera clocks, gains and read-out speeds were adjusted in order to reduce the noise.  
 
The order-sorter filter is a narrow band interference filter,  centered at $\rm \lambda_{c}$ = 6557\AA\ 
(nearly at the systemic velocity $\sim -$180 \kms) with a FWHM of 30\AA.  Its maximum transmission is $\sim$80\%.  The interference order of the FP interferometer is $p = 765$ at  \ha. 
The FP has a Free Spectral Range (FSR) of 8.16\AA\  ($\sim$373 \kms), which has been scanned  
through 48 channels, corresponding to a spectral  sampling of $\sim$ 0.17\AA\ (7.8 \kms).  
The finesse $f$ of the Fabry-Perot etalon provides the spectral resolution $\Delta \lambda = FSR/f$. 
Our observations were done with a mean finesse of $f=16.5$, as determined from datacubes of a Neon calibration lamp.
 The FWHM spectral resolution is thus $\Delta \lambda_{max} \sim$0.53\AA\ (resolving power of $\sim 12620$ at \ha). This corresponds to a FWHM instrumental broadening of 23 \kms\ (dispersion of $\sim 10$ \kms) at the scanning wavelength of the observation of 6558.8 \AA.

\subsection {\label{sec:data}Data acquisition}

\begin{figure*} 
\centering
\includegraphics[width=5.2in]{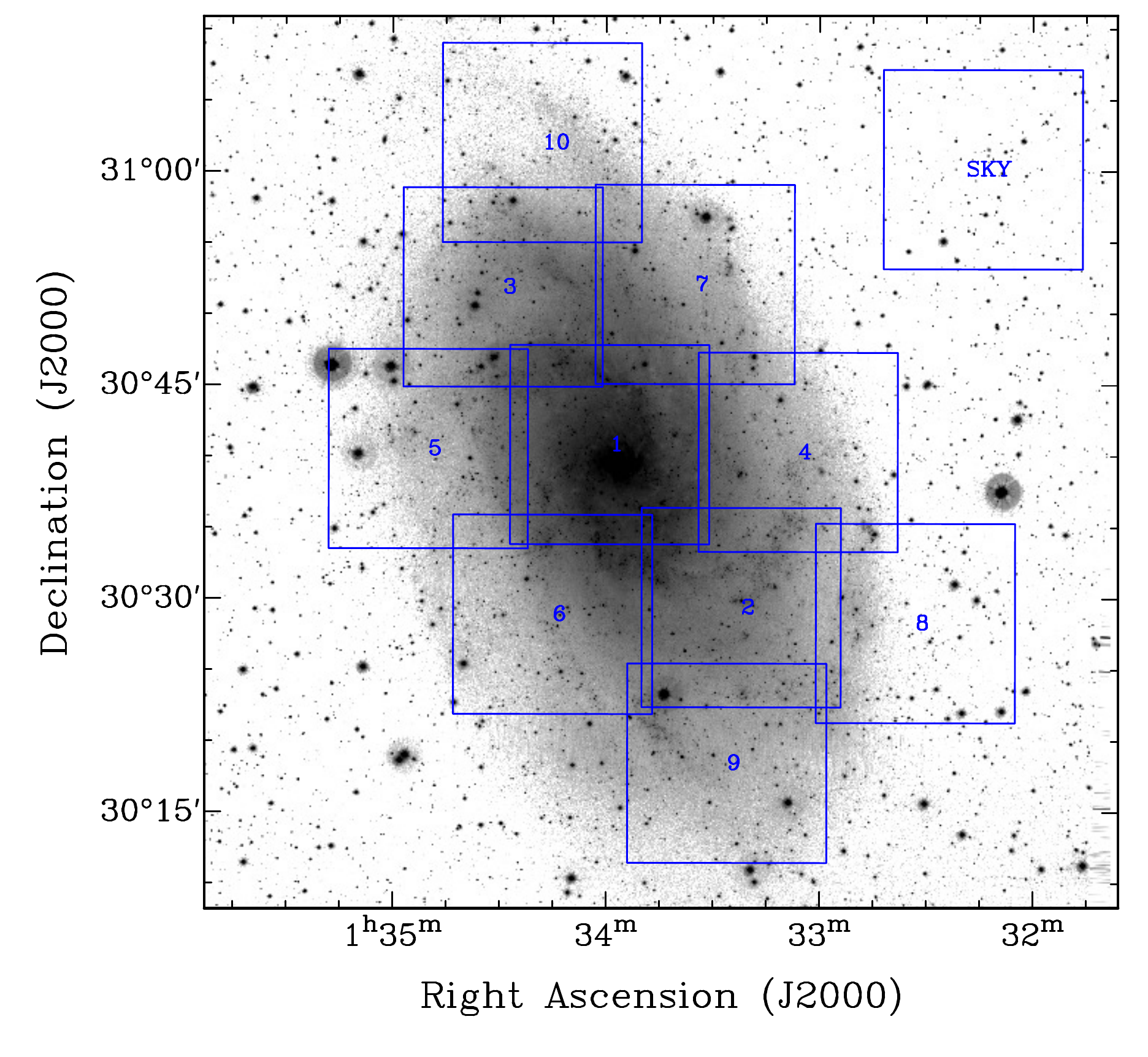}
 \caption[Andor FOV on M33M33] { Mosaic of the 10 Fabry-Perot pointings in the direction of M33, shown with the sky field to the NW. The FOV of each field is $\sim 14' \times 14'$, with an overlap of $\sim 5$\arcsec. The background image is a R band archive image from KPNO \citep{Hoopes2001}.}
\label{fig:Couverture}      
\end{figure*}

The wide field-of-view of $\sim$14\arcmin\ $\times$14\arcmin\  allows to map the bright inner disk regions of Messier 33 with only ten pointings. The different pointings overlap by a few arcseconds to allow the derivation of one final large mosaic with no gaps  (\S\ref{sec:mosaic}). Figure~\ref{fig:Couverture} shows those pointings, whose centre coordinates are listed in Table~\ref{tab:obs}. 
Notice that the central field has been observed more than once to yield very deep \ha\ data for the innermost regions of Messier 33. A region of the sky, free from apparent  \hII\ regions from Messier 33, has been observed before and after each galaxy pointing whose duration was larger than 30 minutes. Those acquisitions were done with the same interference filter as the one used for the galaxy, in order to perform the subtraction of the night-sky  emission lines.
\begin{table}
 \centering
 \caption{ Characteristics of the observations}
    \begin{tabular}{lccLllll}
    \hline
   Fields ID 		& RA		& DEC  			&DD		& T		&S 	&  am 	&$\rm V_{h}$ \\
   (1)				&(2)		&(3)				&(4)		&(4)	 	&(5)		&(6)		&(7)		\\
   \hline
    M33$_-$01a 	& 1:33:59 & +30:40:48.54		&  10	 	& 2.0		&2.7	& 1.1		&19.6	\\
    M33$_-$01b		& 1:33:59 & +30:40:48.54 	&  16	 	& 1.4		&2.2	&  1.3	&17.5	\\
    M33$_-$01c		& 1:33:59 & +30:40:48.54 	&  16 	& 2.0		&2.2	& 1.1		&17.5	\\
    M33$_-$02a  	& 1:33:22 & +30:29:22.27 	&  11 	& 1.4		&2.8	& 1.1		&17.0	\\
    M33$_-$02b  	& 1:33:22 & +30:29:22.27		&  16 	& 2.7		&2.5	&  1.3	&17.0	\\
    M33$_-$03  	 	& 1:34:29 & +30:51:52.41		&  11 	& 2.4		&2.4	& 1.3		&19.3	\\
    M33$_-$04    	& 1:33:06 & +30:40:16.96 	&  12 	& 2.5		&2.4	& 1.1		&19.2	\\
    M33$_-$05    	& 1:34:50 & +30:40:30.72 	&  12 	& 2.5		&2.3	& 1.4		&18.9	\\
    M33$_-$06    	& 1:34:15 & +30:28:52.98 	&  13 	& 3.0		&2.3	& 1.1		&19.0	\\
    M33$_-$07    	& 1:33:35 & +30:52:04.25 	&  13 	& 2.6		&2.1	& 1.3		&18.5	\\
    M33$_-$08    	& 1:32:33 & +30:28:16.19 	&  14 	& 2.7 	&2.1	& 1.1		&18.5	\\
    M33$_-$09    	& 1:33:26 & +30:18:26.23 	&  17 	& 2.4		&2.2	& 1.1		&17.4	\\
    M33$_-$10  	 	& 1:34:18 & +31:02:01.11 	&  18 	& 2.6		&2.5	& 1.1      	&17.2	\\
    \hline
    & & & & & \\
    Sky  		 	& 1:32:14& +31:00:09.48 		& 		& 0.2		&	& 		&17.5	\\
    Dark			& 		&  				& 		& 0.1		&	&      &		\\
    Noise 	 		&		& 				& 	 	& 0.1   		&	&     		&		\\
   Gain 			&  		&				&  	 	& 0.1		&	&  		&     		\\
    Flat 			& Dome	&  Dome 			& 		& 0.1   		&	&   		&     		\\
    \hline \\
  \end{tabular} \\
  {
  Columns notes:  (1) Identification of the field. (2) and (3) Coordinates of the field center (J2000). Each field-of-view is $14 \arcmin \times 14\arcmin$.  Observations presented  in this paper have been taken  in September 2012. (4) Exposure time in hours,  (5) Seeing FWHM during observations.  (6) Air mass correction factor. (7) $\rm V_{h}$ is the heliocentric velocity correction (\kms).
  }
   \label{tab:obs}%
   \end{table}

The data were obtained by operating the camera at two seconds exposure per frame. 
A gap of 0.4s between two consecutive channels was necessary to move the reflective plates of the FP etalon in order to avoid  overlapping frames during the fast transfer.   

Dark current, gain, flat field and noise calibration observations have been acquired at the beginning and the end of each night. The ``Dark'' observation consisted in a series of at least fifty images during which the detector is exposed for 
two seconds without light, and the gain is set to its largest amplitude, like during the observations of every sky/galaxy fields-of-view.  
The ``Gain'' observations consisted of about 200 frames, again acquired with the largest gain value.     
The ``Noise''  observations consisted in an integration of zero second, with the lowest gain value. 
A minimum of fifty frames were collected in order to calculate the read out noise of the CCDs. The noise, gain and dark were observed in off-light  mode in the dome. 

The total exposure time performed for each field is listed in Table~\ref{tab:obs}.  In total, the time spent to integrate on the Messier 33 fields was $>$30 hours. The seeing of the observations was $\lesssim 3\arcsec$. To perform the wavelength calibration, the system (filter+lenses+FP+camera) has been illuminated by a Neon (Ne) lamp.  A narrow band filter centered on 6598\AA\ and of FWHM=16.3\AA\  was used to isolate the Ne line at $\lambda$=6598.95\AA.

%% file: reductionV2.tex
\section{Data Reduction}
\label{sec:reduction}
 
\subsection{Wavelength calibration, spectral smoothing}

The raw frames of the observations contain interferograms giving the information on the number of  photons  per frame per channel and per cycle.   
 We used Interactive Data Language (IDL) routines to integrate raw 2D files into a 3D datacube. 
 A phase correction consisting in shifting every pixels such that they are at the same wavelength across the field  has been 
applied to the raw datacubes to yield wavelength-calibrated datacubes. For that purpose, a phase map has been derived from a Ne line calibration observation. 
 Datacubes are then wavelength-sorted and corrected for guiding shifts and cosmic rays. 
The ''Noise'', ''Dark'' and ''Gain'' observations (\S\ref{sec:data}) were then used to correct and calibrate the detector response. 
 
A Hanning  filter with a width of three channels has then been applied to the spectra to increase the sensitivity. 
 This reduction step is the same as the one applied to the   Virgo Cluster, GHASP or SINGS samples \citep{Chemin2006,Epinat2008b,Epinat2008,Daigle2006b, Dicaire2008a}.
 The same process is used for the sky observations, whose datacubes have been subtracted of all the other fields  to produce M33-only \ha\ emission line datacubes. 
 For pointings having more than one observation, all night-sky corrected and wavelength calibrated datacubes have been combined into a single datacube, 
 to increase the sensitvity. 
 
\subsection{Airmass and ghosts corrections}

 Because the center of the interference rings of the observations is near the center of the detector,
 there are some reflections about the optical axis, called ghosts.  The size and the intensity of the ghost depend on the shape of the region which produces the effect. %
 We have used the method described in \cite{Epinat2009} to reduce the number of pixels dominated by those ghosts. The typical residuals after ghost removal are about 10\% of the initial ghosts flux.
 
Then, an airmass correction has to be applied. Indeed, differences in airmass during the multiple sessions of observations can affect the number of counts received by the detector,
  which implies field-to-field sensitivity variations. In Table~\ref{tab:obs},  
  we give the airmass correction factor (AM) which  has to be applied to each field in order to get the counts 
  equivalent to an airmass of 1.0.  Starting from   the observed counts $\rm C_{obs}$    at a given airmass, the true value of the counts $\rm C_{true}$ is given by:  
\begin{equation}
\rm C_{true}=10^{-0.4 A X} \times C_{obs}
\label{eq:count}
\end{equation} 
 The parameter A represents the  \ha\ extinction coefficient. 
 Our Fabry-Perot system has already been used by \cite{Larrondo2009} 
 to determine the value of that coefficient. We used their value of 1.03.  
 The X quantity  is given by $\rm X =  AM  - 1.0$, with the airmass correction factors AM listed in Tab.~\ref{tab:obs}. 

\subsection{Mosaicing and binning the datacubes}
\label{sec:mosaic}

Due to the large angular size of Messier 33, there are two possibilities to generate the final set of moment maps (\ha\ integrated emission, line-of-sight velocity and velocity dispersion maps, continuum map). 
A first one is to compute for each of the ten pointings its own   moment maps, 
and combine all the maps to produce a single field-of-view moment map. The second approach 
is first to combine all datacubes into a mosaic datacube, and then generate moment maps. The first approach is easier to do, but it  introduces more errors in radial velocities in regions of overlap. 
 Instead, we think it is better to combine and correct the \ha\ spectra directly, than radial velocities or velocity dispersions.  We thus chose the second approach, which is a more powerful process, though less straightforward to implement.  Combining the data cubes increases the signal to noise per pixel, which results in increasing the numbers of pixels for which we can measure a line. To minimize  the errors in the overlap regions, astrometry information have been added in each single cube headers. The different steps  in the process are :

\begin{enumerate} 
\item White light image (sum of all channels)
\item Accurate astrometry on the white light image
\item Field positioning from astrometry
\item Generation of an exposure map with the sum of exposures
\item Accurate spectral calibration, taking into account heliocentric correction
\item Generation for each channel of a mosaic
\item Weighting of each channel by the exposure map
\item Combination of all channels to generate the mosaic cube

 \end{enumerate}
 
 In order to process  projections and make an accurate mosaic, some parts of the Montage packages  from IPAC (Infrared Processing and Analysis Center) have been used  to develop our  tools.
 
Once the  mosaic  datacube has been obtained, an adaptive spatial binning has been performed.
That adaptive spatial binning  is based on Voronoi~3D binning \citep[e.g.][]{Daigle2006b, Chemin2006, Dicaire2008b}.  
The Voronoi technique consists in combining pixels to larger bins up to a given value of S/N (signal-to-noise ratio) or more. 
The pixels/bins with a S/N lower than the threshold S/N are combined with the neighbouring pixels until 
the threshold S/N value is reached.  A S/N of 7 has been targeted for this study. This particularly allows to preserve the angular resolution where the S/N is high.
 
\begin{figure*}
\centering
	\includegraphics[width=1.6\columnwidth]{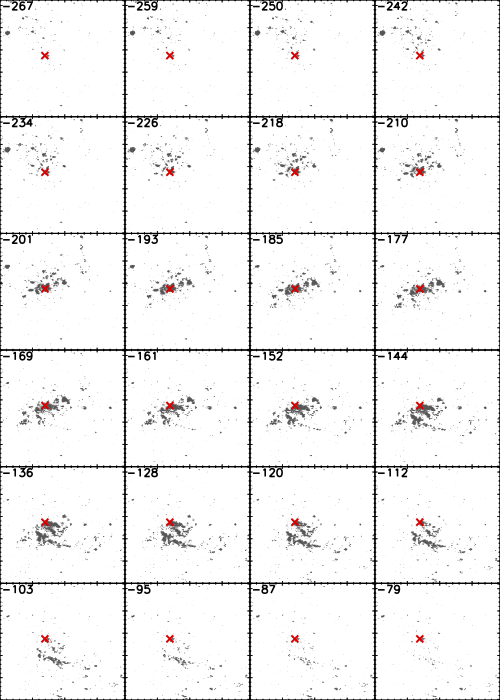}
	\caption[Chanel map] {\ha\ channel maps. The radial velocity (\kms) is indicated in the upper left corner of each channel  and the cross indicate the kinematical center.}
\label{fig:chanelmap}
\end{figure*}

Figure~\ref{fig:chanelmap} gives the channel maps of the data cube used to determine the kinematical parameters of M33. The detections plotted in  Figure~\ref{fig:chanelmap} are those above 3 sigmas.  It represents the wavelength variation from 6556.5\AA\ to 6560.6\AA\ (from channel 11 to channel 35 in our data) where the  channels width is 0.18 \AA.  The \hII\ regions appear progressively from the North East side to the South-West side.
\subsection{Moment maps derivation}
 \label{sec:momentmap}
The integrated emission field is the 0-th moment map derived from  the spectra and the velocity field the first moment map. Radial velocities are given in the heliocentric rest frame. Making the assumption that the PSF has a gaussian profile, the velocity dispersion field is the second moment map:

\begin{equation}
\sigma_{corr}= \rm \sqrt{{\sum_{i} \lambda_i^2 F(\lambda_i) \over \sum_{i} F(\lambda_i)} - \Lambda^2 - \sigma_{PSF}^2}
\end{equation}
 where $\rm F(\lambda_i)$ is the flux corrected for the continuum  level at wavelength $\lambda_i$,  $\Lambda$ is the  barycenter of the  emission line, and $\rm \sigma_{PSF}$  is the instrumental dispersion.  
Velocity dispersions have not been corrected for thermal broadening of the medium, nor for the natural width of the \ha\ line.  While the natural width of the \ha\ line remains negligible ($\sim $3 \kms), the reason for not correcting for the thermal dispersion of the gas was that the   temperature of  the ionized interstellar medium of Messier 33 is not known accurately. A typical broadening of  $\sim 10$ \kms\ is often given in the literature, for a temperature of $\sim 10^4$ K for the ionized gas. However, we have observed many bins of  dispersion (after instrumental broadening correction) lower than this usual value of $\sim 10$ \kms, implying that a typical temperature of $\sim 10^4$ K cannot be valid everywhere.

Finally, the moment maps have been cleaned to get rid of all possible unrealistic patterns, like those  
from reflection, noise and background emission residuals.  For that purpose, we have masked all pixels with data values lower than twenty counts. 
We are left with maps having 546941 independent bins.

\subsection[]{Flux calibration}
\label{sec:fluxcalib}
\begin{figure}
\centering
	\includegraphics[width=\columnwidth]{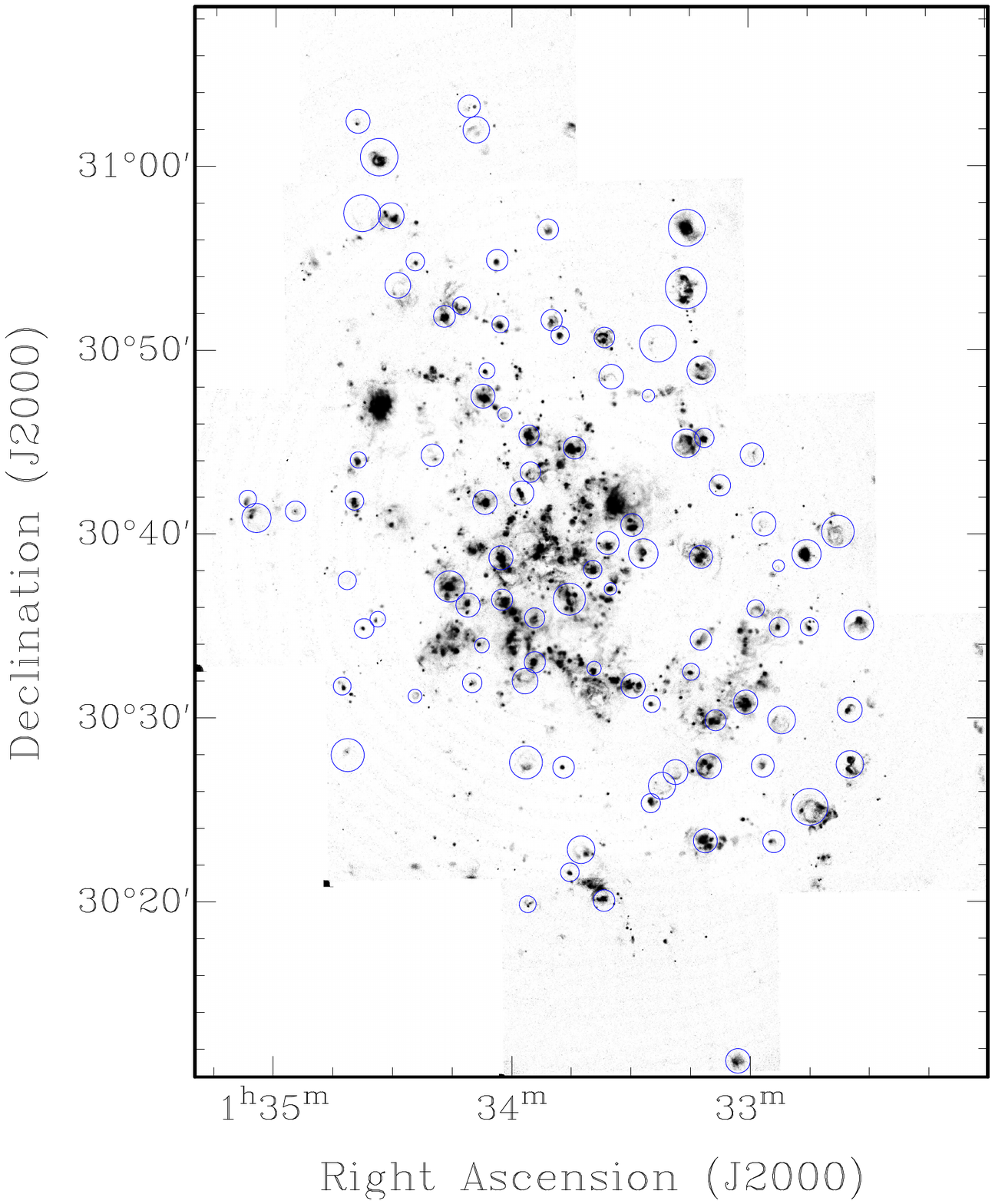}
	\caption[Calibration Sources]{ Selected \hII\ regions for the \ha\ flux calibration over-plotted on the \ha\ line flux map obtained from our set of observations. The aperture size around each region is the same as in \protect\cite{Relano2013}. }    
\label{fig:calibfov} 
\end{figure} 

The flux calibration that turns Fabry-Perot data counts into surface brightness (in $ \rm erg/s/cm^{2}/arcsec^2$) is given by:
\begin{equation}
\rm SB =   \rm Cst \times  SB_{fp} 
\label{eq:flux}
\end{equation}
Here, Cst is a calibration  constant, SB is the corresponding surface brightness value.  and  SB$_{\rm fp}$ is the calibrated value of FP count/pixel/s. The FP fluxes calibration using a \ha\ map is a linear relation. We have used the catalog of \hII\ regions of \cite{Relano2013} as reference fluxes to calibrate our interferometric data.  Figure~\ref{fig:calibfov} shows the selected regions and their associated aperture size used to integrate the \ha\ counts. For each region the total integrated flux is computed using the counts and  the mean exposure time. 

\begin{figure}
\centering
	\includegraphics[width=\columnwidth]{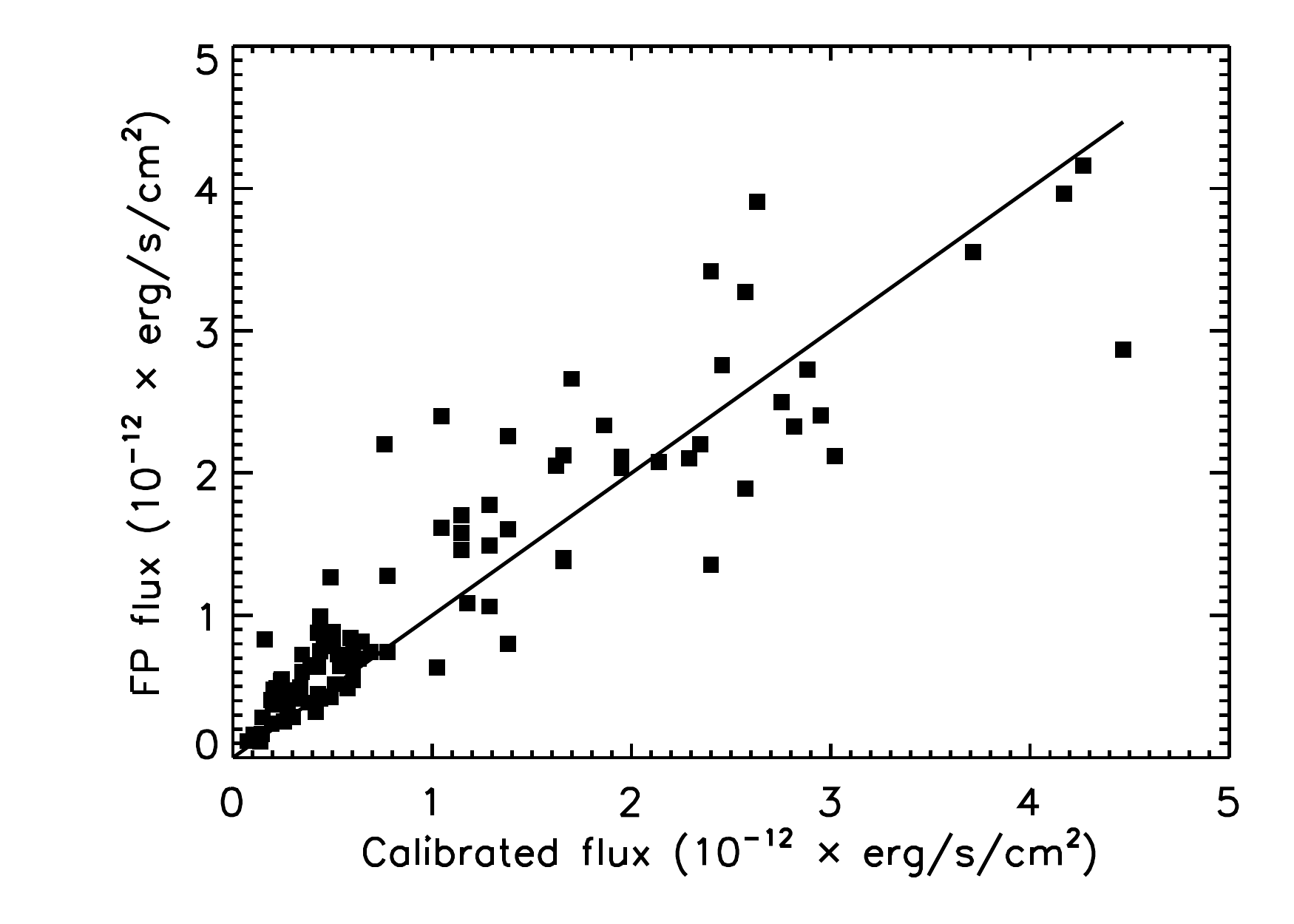}
	\caption[Calibration diag] {Comparison between the derived Fabry-Perot \ha\ fluxes and the reference \ha\ fluxes of \cite{Relano2013} used for the FP flux calibration. Square symbols represent data from individual \hii\ regions (Fig.~\ref{fig:calibfov}) and the solid line is the linear fit to the data.}
\label{fig:calibsources} 
\end{figure}

The calibration constant for the M33 \ha\ FP image is given by 1 $\rm count/pixel/s=2.45 \pm 0.03 \times10^{-17} erg/s/cm^{2}/arcsec^2$. The  typical error  that can be noticed on the flux determination at an aperture used, is  $\sim$$\rm2.4\times 10^{-16} erg/s/cm^{2}/arcsec^2$. It takes into account the errors on the \ha\ flux provided in \cite{Relano2013} and the dispersion of the fit.
 Figure~\ref{fig:calibsources} shows the comparison bewteen our Fabry-Perot calibrated data, and the \cite{Relano2013} data  in units of flux, $erg/s/cm^{2}$. 
%___________________________________________________________________________________ 

%% file: analyseV2.tex
\subsection{\ha\ maps}
\begin{figure}
\centering
\includegraphics[height =3.3in, width=\columnwidth]{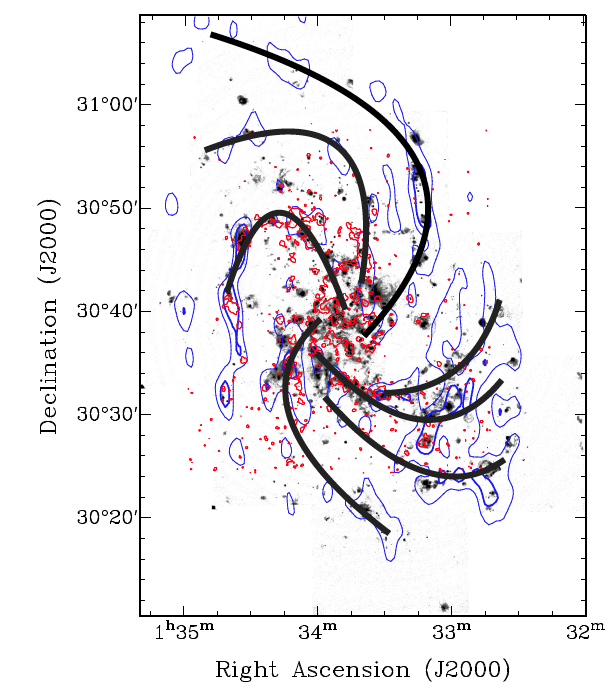} 
\caption[Contours] {\ha\  map of M33 derived from our Fabry-Perot data. The  CO peak contours are in red  (120 $\rm K~km/s$) from \cite{Tosaki2011} and the HI peak contours are in blue (9500 and 1200 $\rm K~km/s$) from \cite{Chemin2012}. The black sketchs are drawn to follow the \hII\ regions and trace the multi arms profile of M33. These arms are close to those shown in \cite{Li2011} or \cite{Boulesteix1974}. } 
\label{fig:subcubes} 
\end{figure}

\begin{figure*}
\centering 
 \includegraphics[width=\columnwidth]{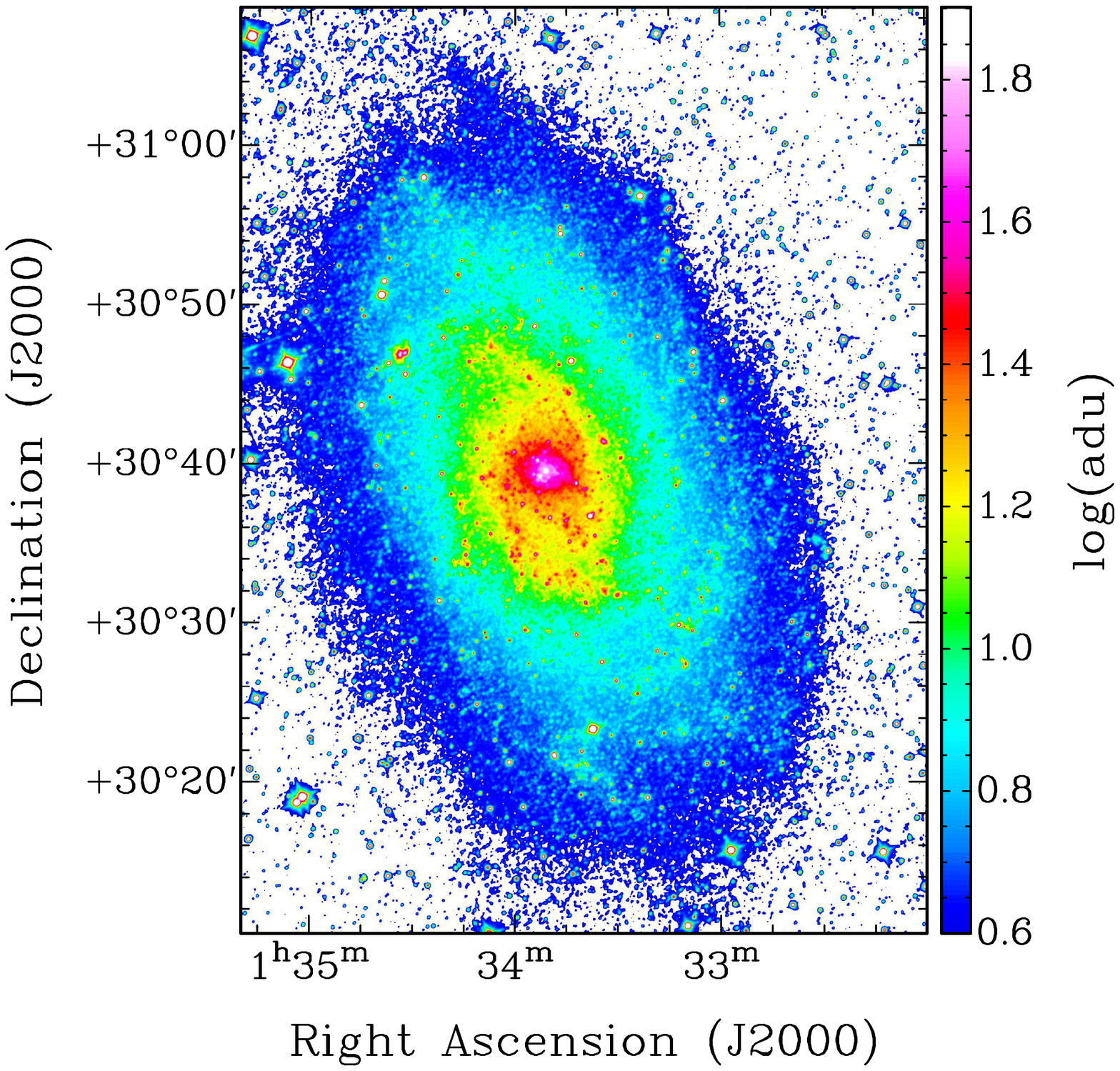}
 \includegraphics[width=\columnwidth]{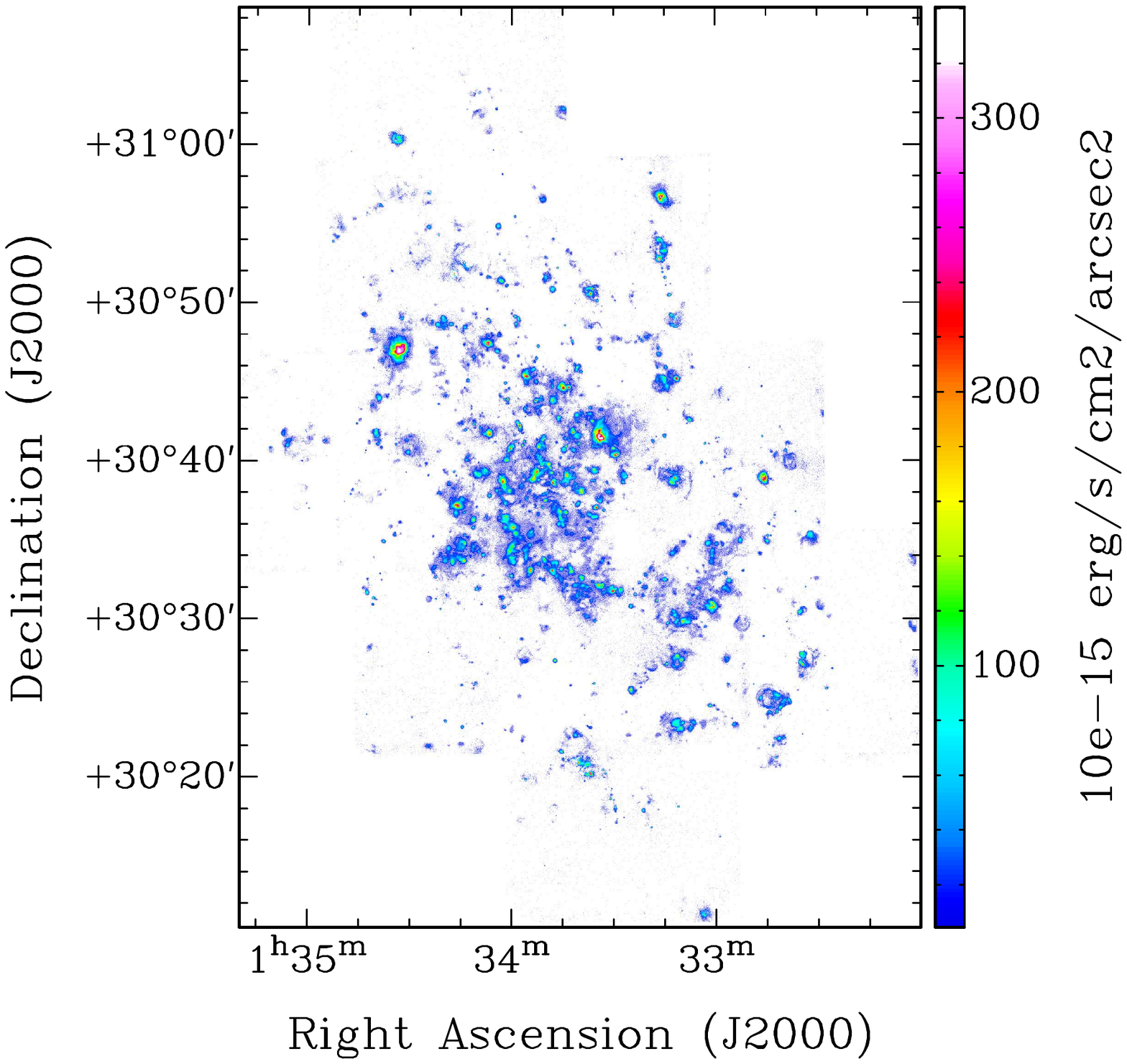}
 \includegraphics[width=\columnwidth]{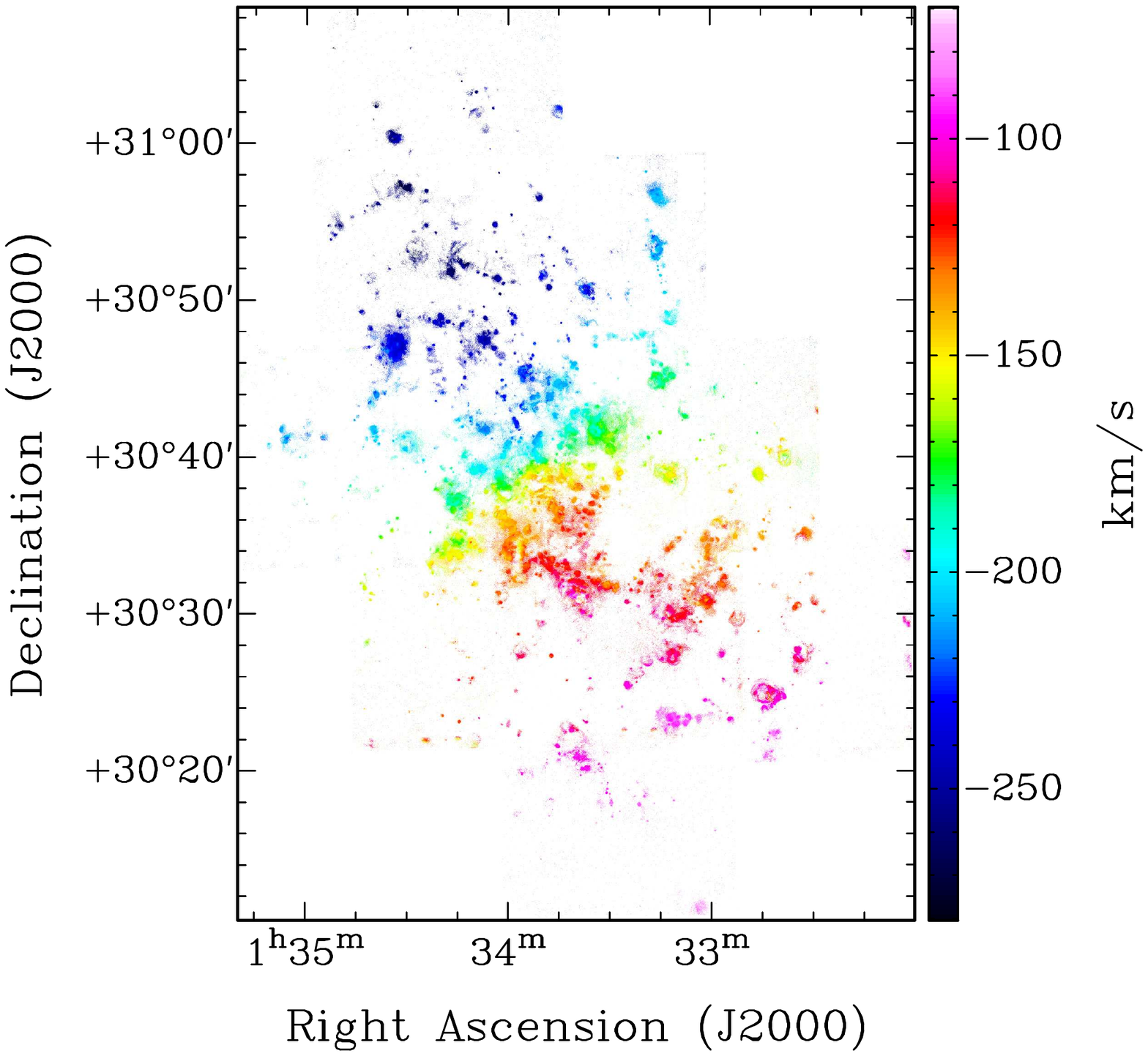}
 \includegraphics[width=\columnwidth]{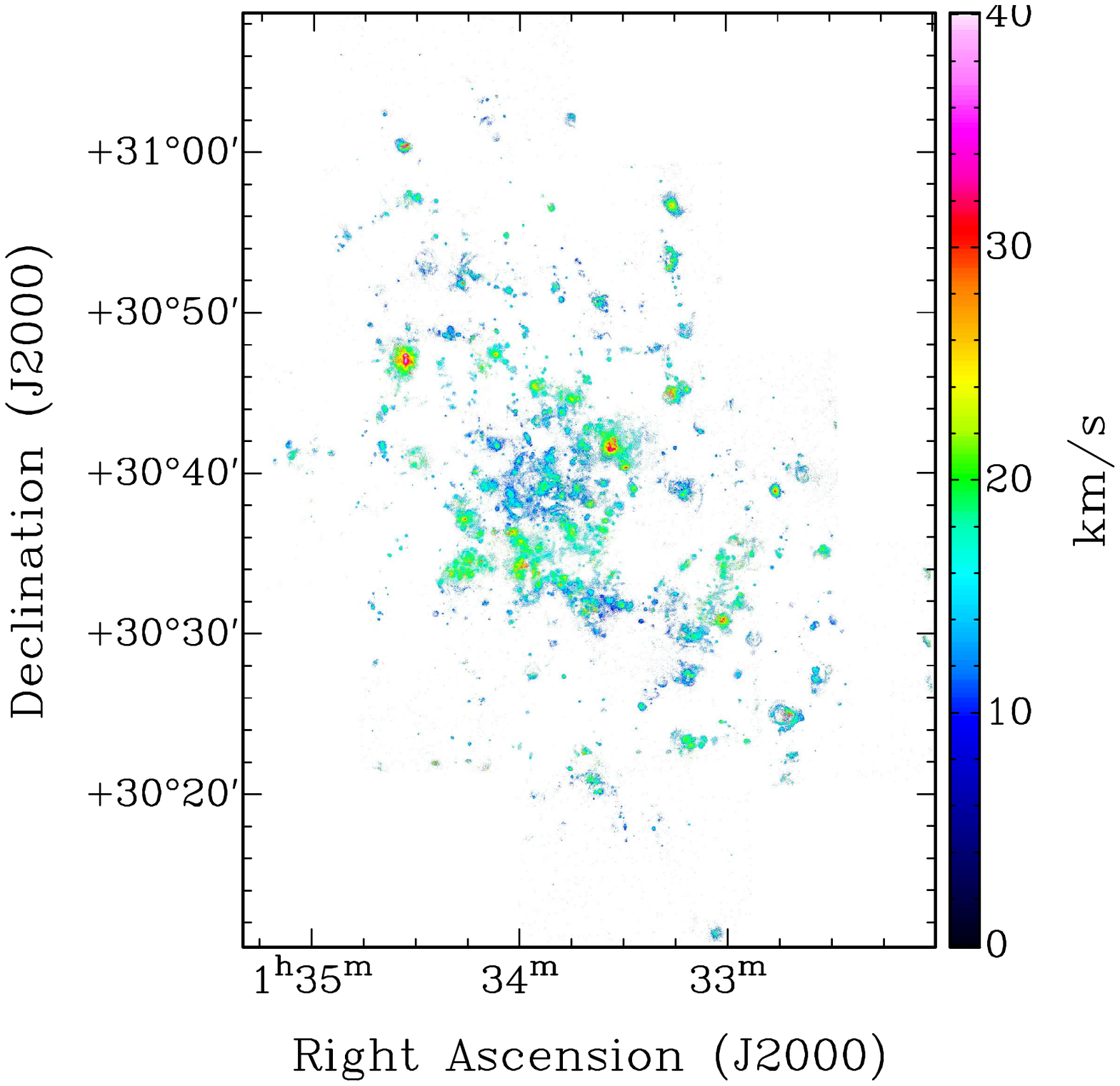}
\caption[ rv] {3.4 $\mu$m and narrow-band \ha\ maps of Messier 33. The NIR image from WISE I (top-left) is shown with a logarithmic stretch.  The Fabry-Perot interferometry images are the \ha\ integrated emission map, velocity field and velocity dispersion field (top-right, bottom-left and bottom-right, respectively). }
\label{fig:rvmap}
\end{figure*}

The observations obtained at the OMM give 13 cubes on ten fields. Those cubes were used to build a mosaic cube as described in section~\ref{sec:mosaic}.   Figure~\ref{fig:subcubes}  shows the \ha\ emission of M33 in the $\sim$42\arcmin\ $\times$ 56\arcmin\ field. The contours are from \hi\ and CO emission. 
We can see that while the \ha\ emission follows roughly the  \hI\ emission, it is even more so for the CO emission. The \hI\ and CO structures  seem to follow the arms described by the \hII\ regions. 
The final cube was produced with the same FSR and same spectral resolution as the small cubes observed. 
The \ha\ monochromatic image (moment zero map) is presented at the top right of Figure~\ref{fig:rvmap} and can be compared to the WISE I NIR image on the left.   
The discrete  \hII\  regions have different  sizes and shapes, filled and  clear shell regions as described  by \cite{Relano2013}. Two main strong arms are clearly defined along with the multi arms structure as presented by \cite{Boulesteix1974}.
 
The velocity field (first moment map) at the bottom left of Figure~\ref{fig:rvmap} was obtained by using the zeroth moment map as a mask. The bins of the voronoi binning with data values greater  than 20 counts are shown. This criteria was chosen in order to avoid all probable unreal pattern introduced by ghosts of strong \ha\ regions, noise and/or background.   The velocity dispersion map (second moment) is shown in the bottom right of Figure~\ref{fig:rvmap}. 
 
\begin{figure*}
\centering
 \includegraphics[width=\columnwidth]{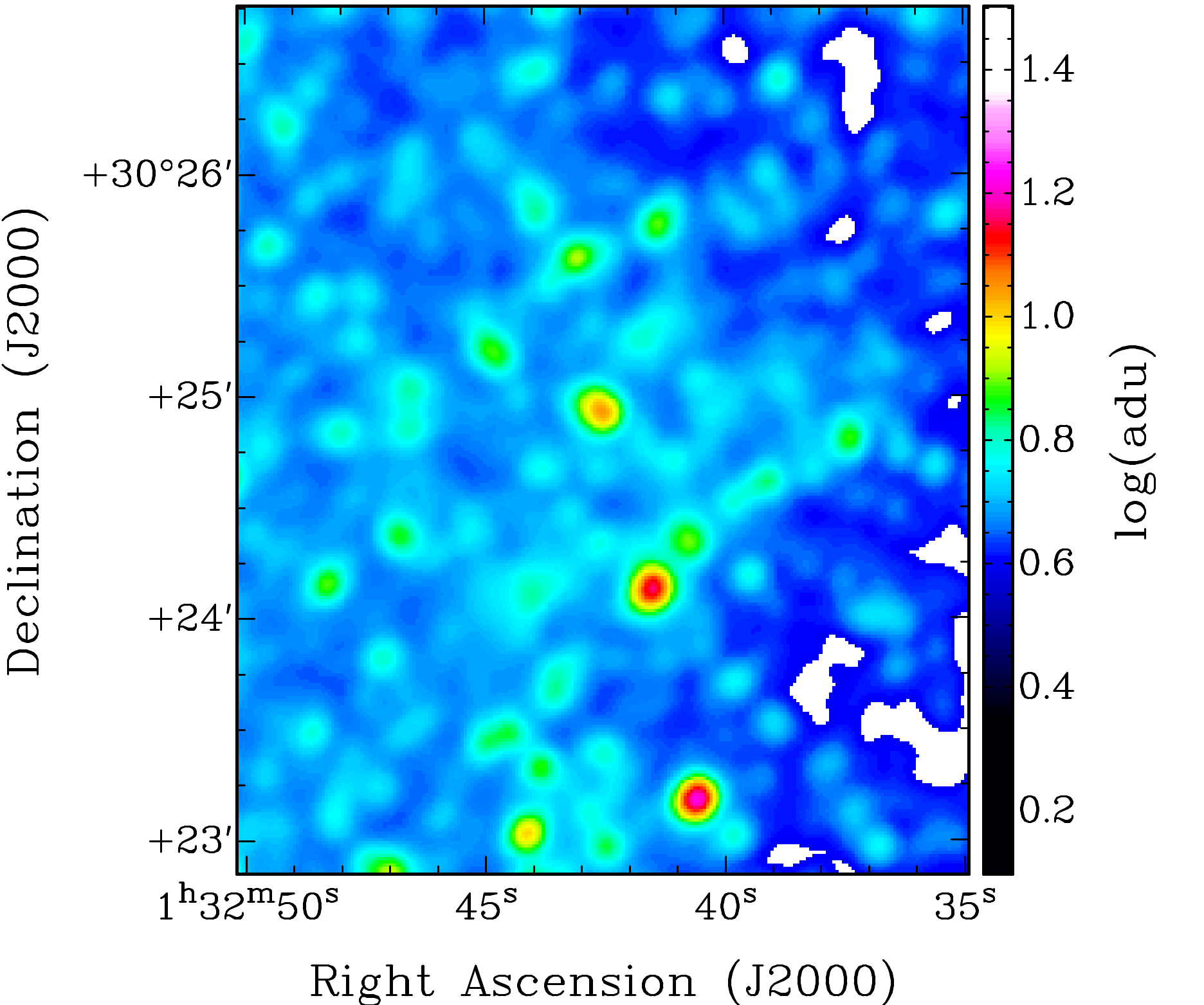}
 \includegraphics[width=\columnwidth]{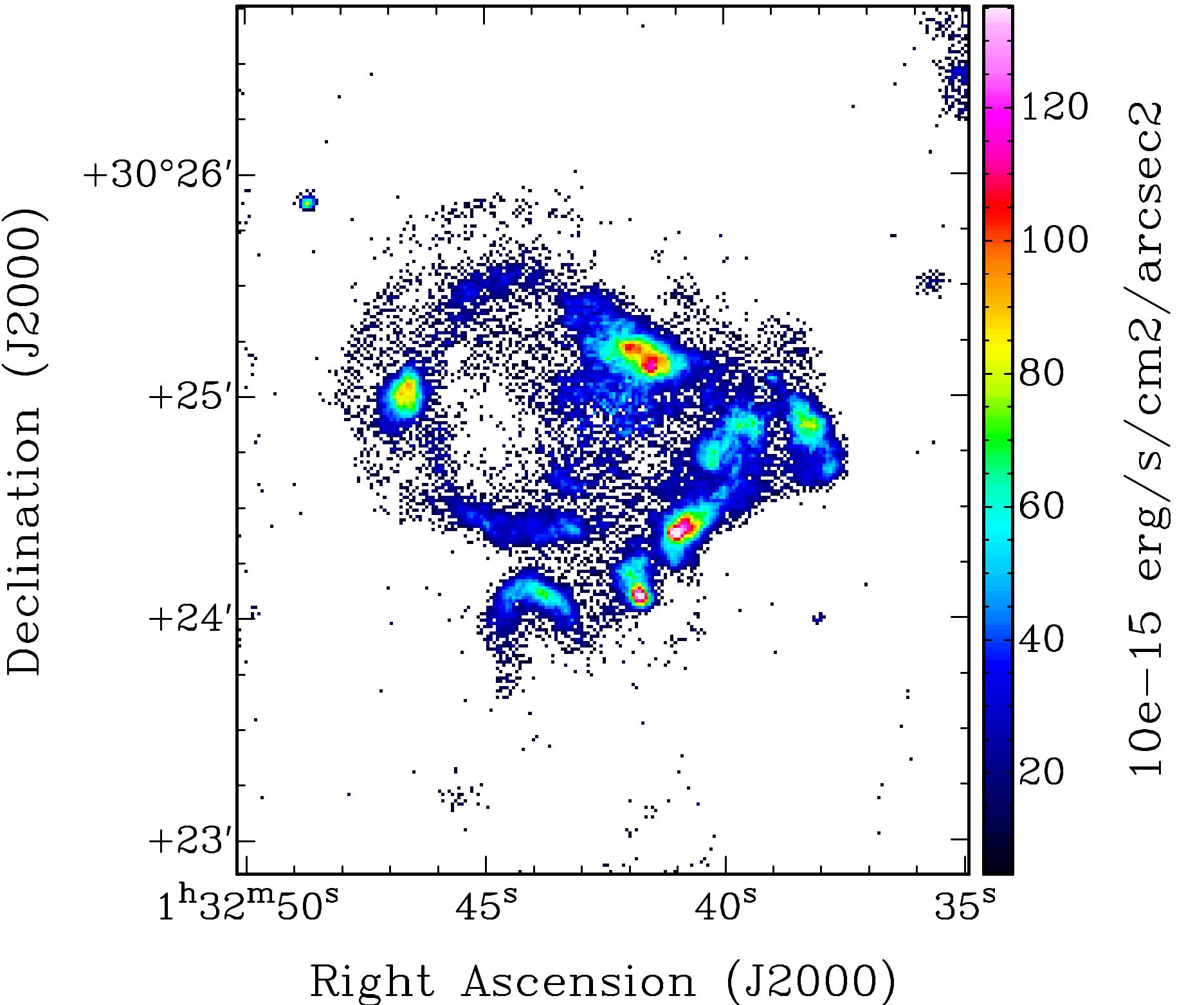}
 \includegraphics[width=\columnwidth]{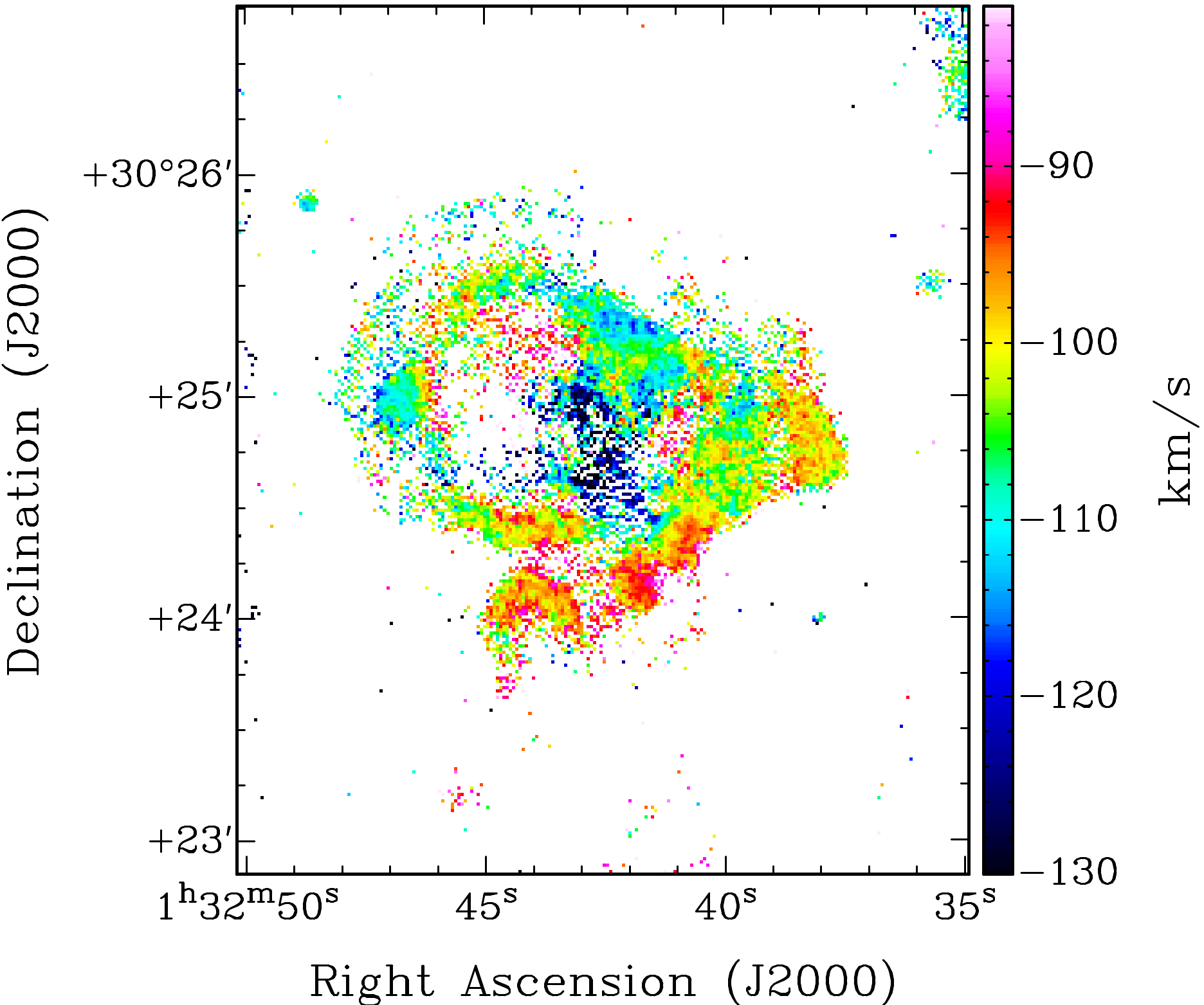}
 \includegraphics[width=\columnwidth]{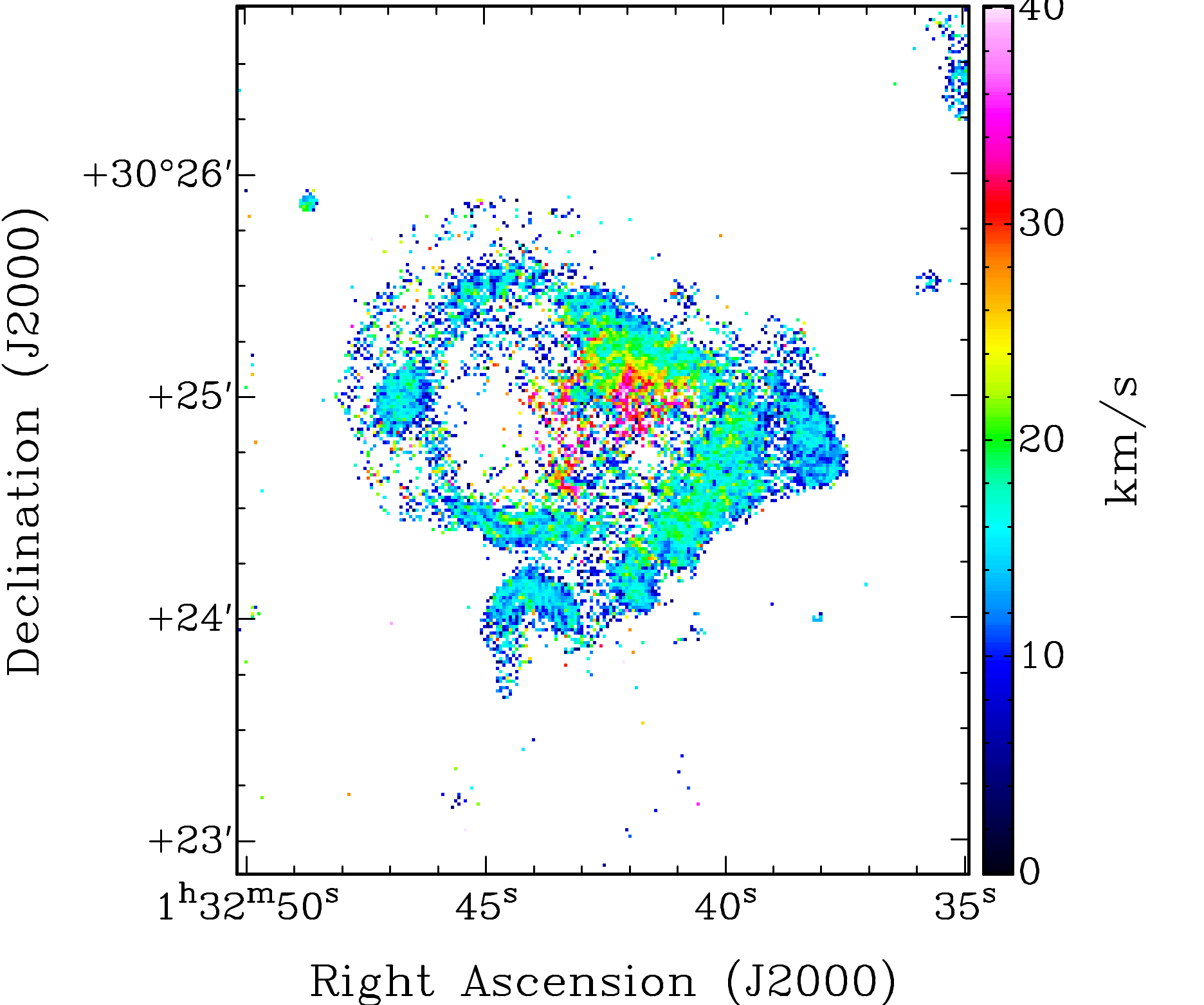}
\caption[NGC604] {Same as in Figure~\ref{fig:rvmap}, zoomed on the star forming region located at $\alpha$=01h32m43s, $\delta$=30\degr24'57''.}  
\label{fig:mapngc}
 \end{figure*}

For this study, the data will be used mainly to derive the overall kinematics and derive the optical rotation curve. Figure~\ref{fig:mapngc} shows that the resolution of this data set ($\sim$ 3\arcsec) could be used to study the detailed kinematics of \hii\ regions, shells, cavities, bubbles, filaments, loops, outflows and ring-like structures. This will be done in another publication. Except within such local features, the large-scale H$\alpha$ velocity field of M33 seems regular, typical of nearby disk galaxies, without significant apparent twist of the major axis, and with mild streamings inherent to spiral arms (see sect. \ref{subsec:rotcur}). 

\begin{figure*}
\centering
 \includegraphics[width=\columnwidth]{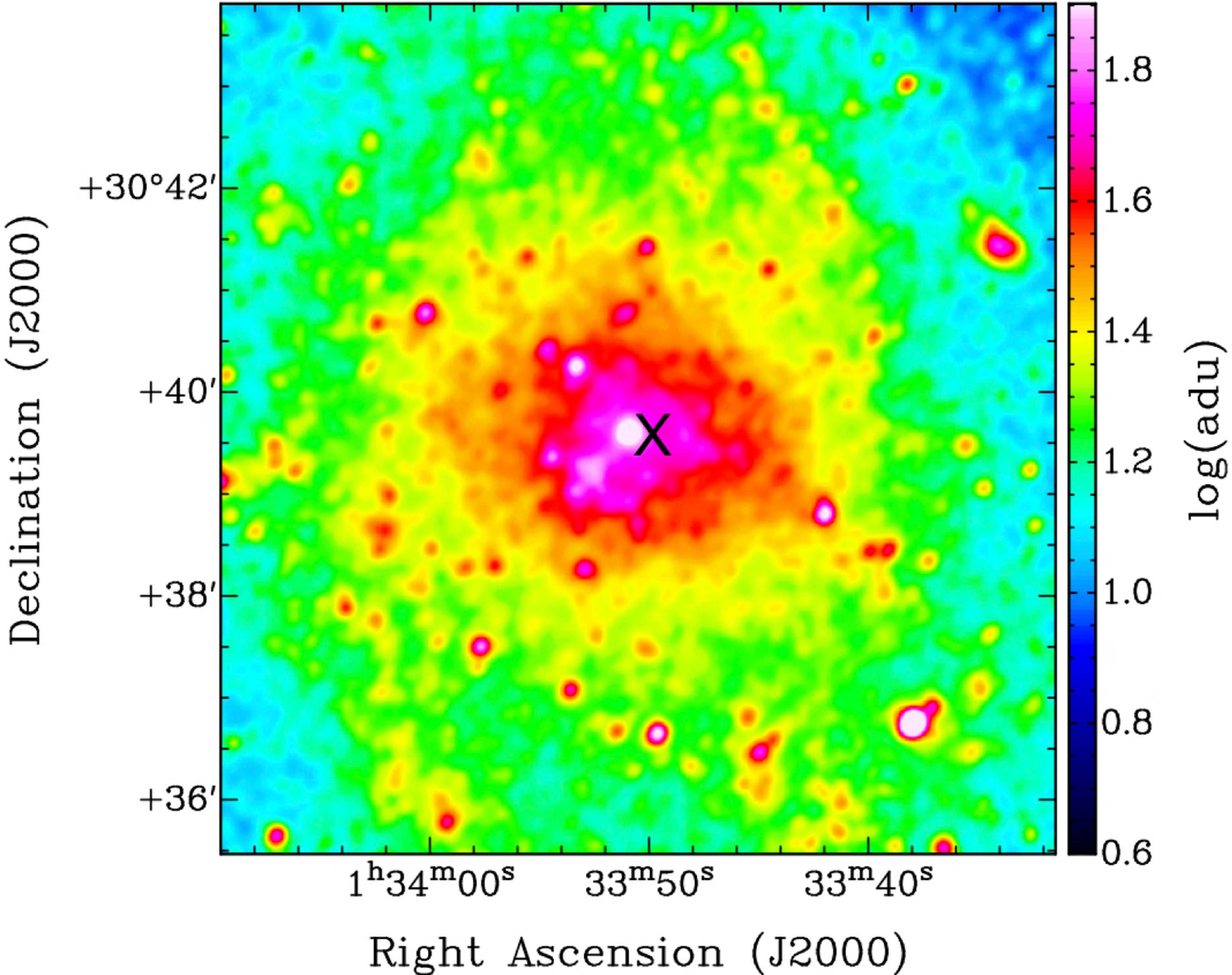}
 \includegraphics[width=\columnwidth]{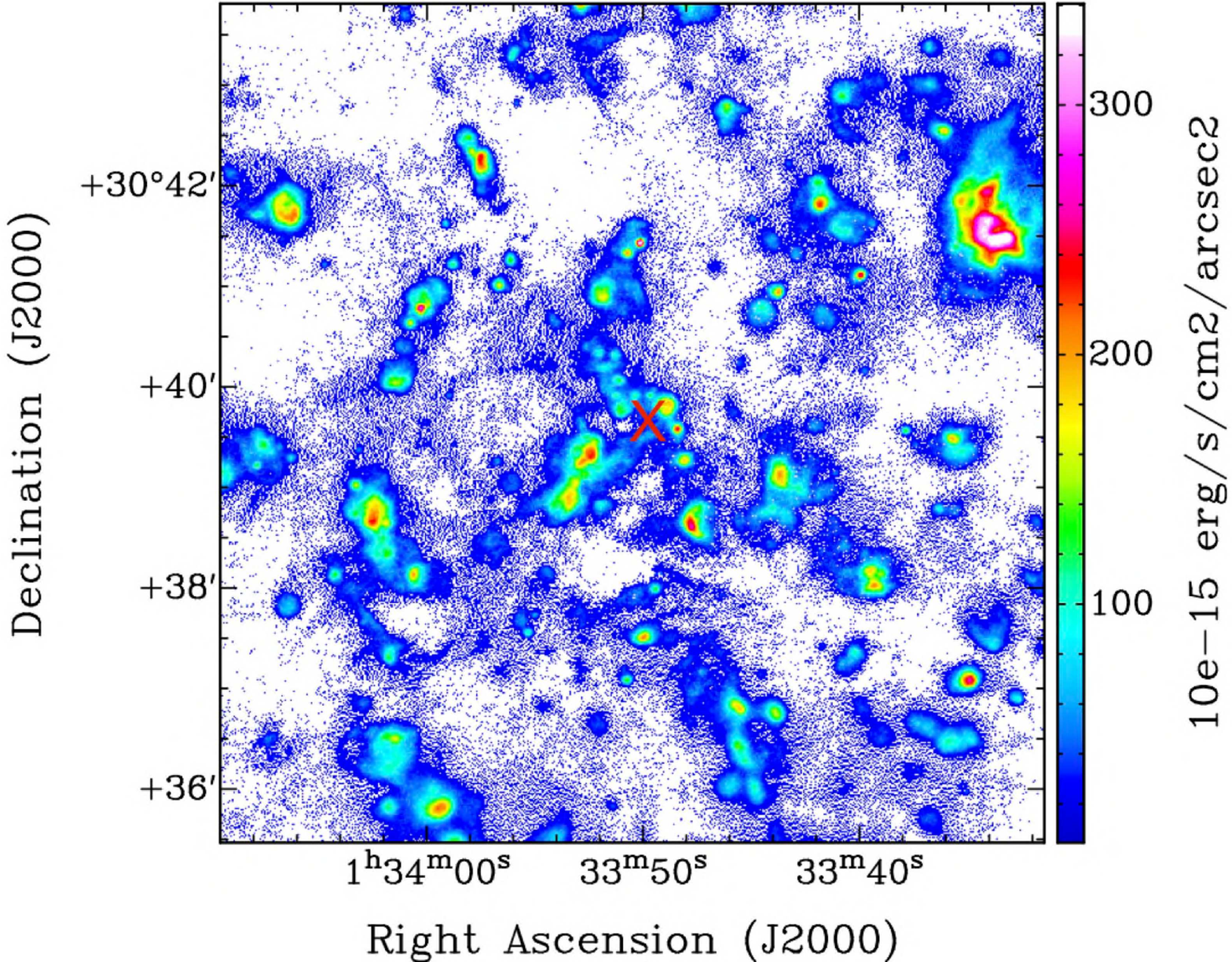}
 \includegraphics[width=\columnwidth]{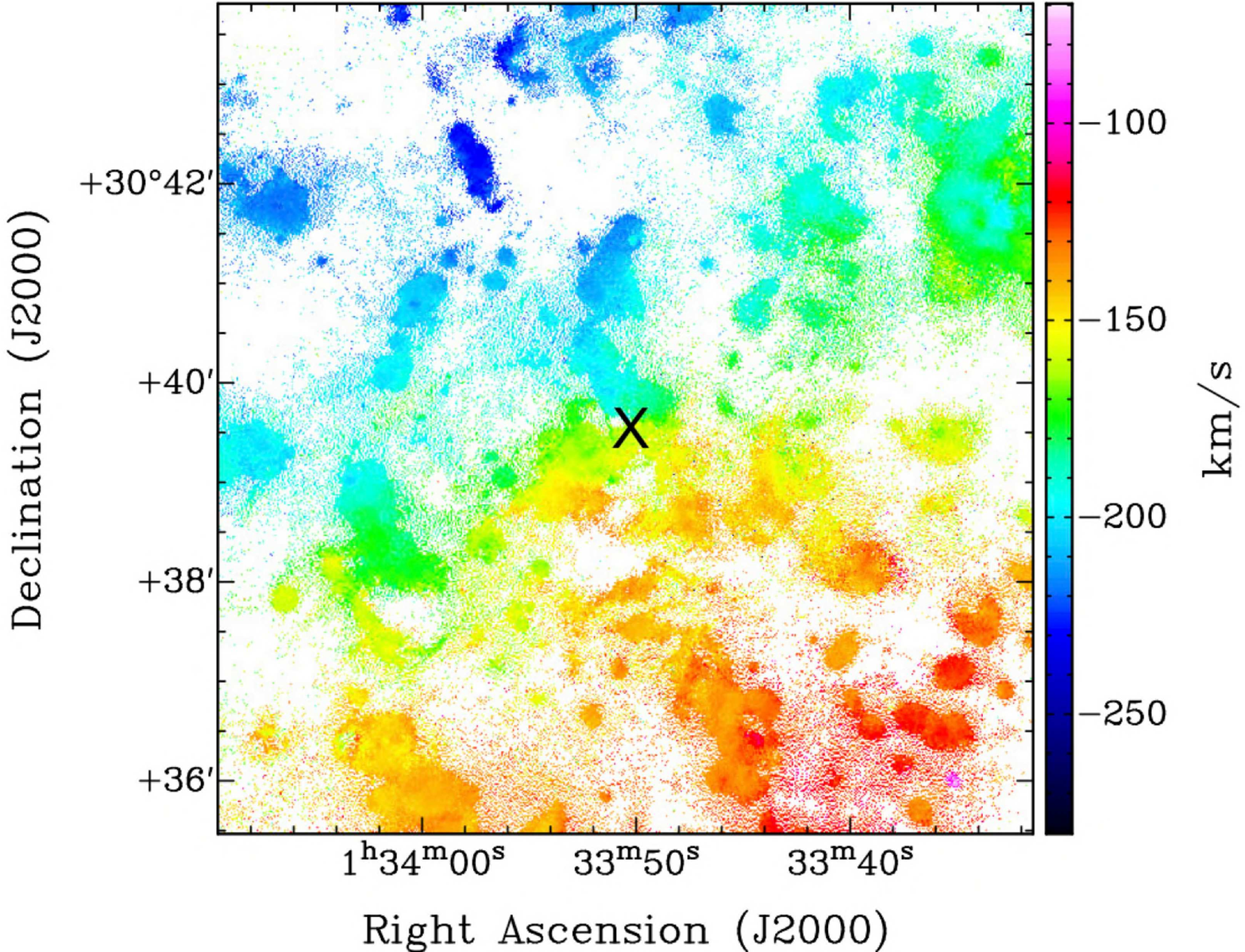}
 \includegraphics[width=\columnwidth]{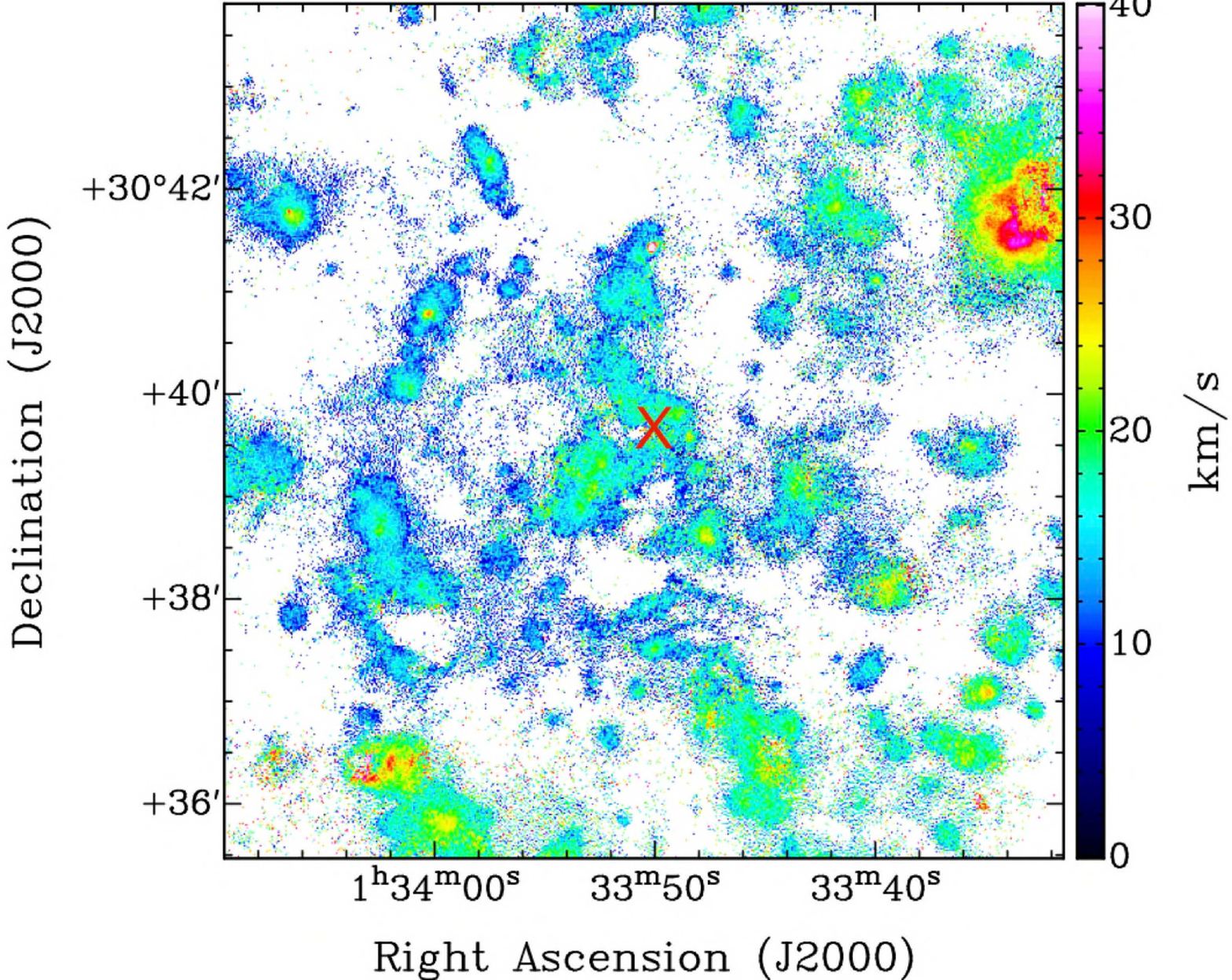}
 \caption[Center] { M33 central $\sim$8\arcmin\ field - same caption as in Figure~\ref{fig:rvmap}. The datapoints presented are the barycenter after an adaptive binning with S/N=7. The cross  indicate the kinematical center.} 
\label{fig:mapcenter}
 \end{figure*}

Figure~\ref{fig:mapcenter} provides a zoom of the $\sim 8\arcmin \times 8\arcmin$ central regions of M33.
When compared to the WISE I NIR image to the lop left,  it can be seen clearly that emission is detected all the way to the very center of the galaxy. 
The velocity dispersion map shows that brighter HII regions exhibit larger dispersion (20-30 \kms), though the largest dispersions (up to 60 km/s) are only seen in bins with the faintest  \ha\ emission. The map also shows that the inner regions exhibit smaller dispersion (see Section 6 for more details).
 \begin{figure*}
\centering
 \includegraphics[width=\columnwidth]{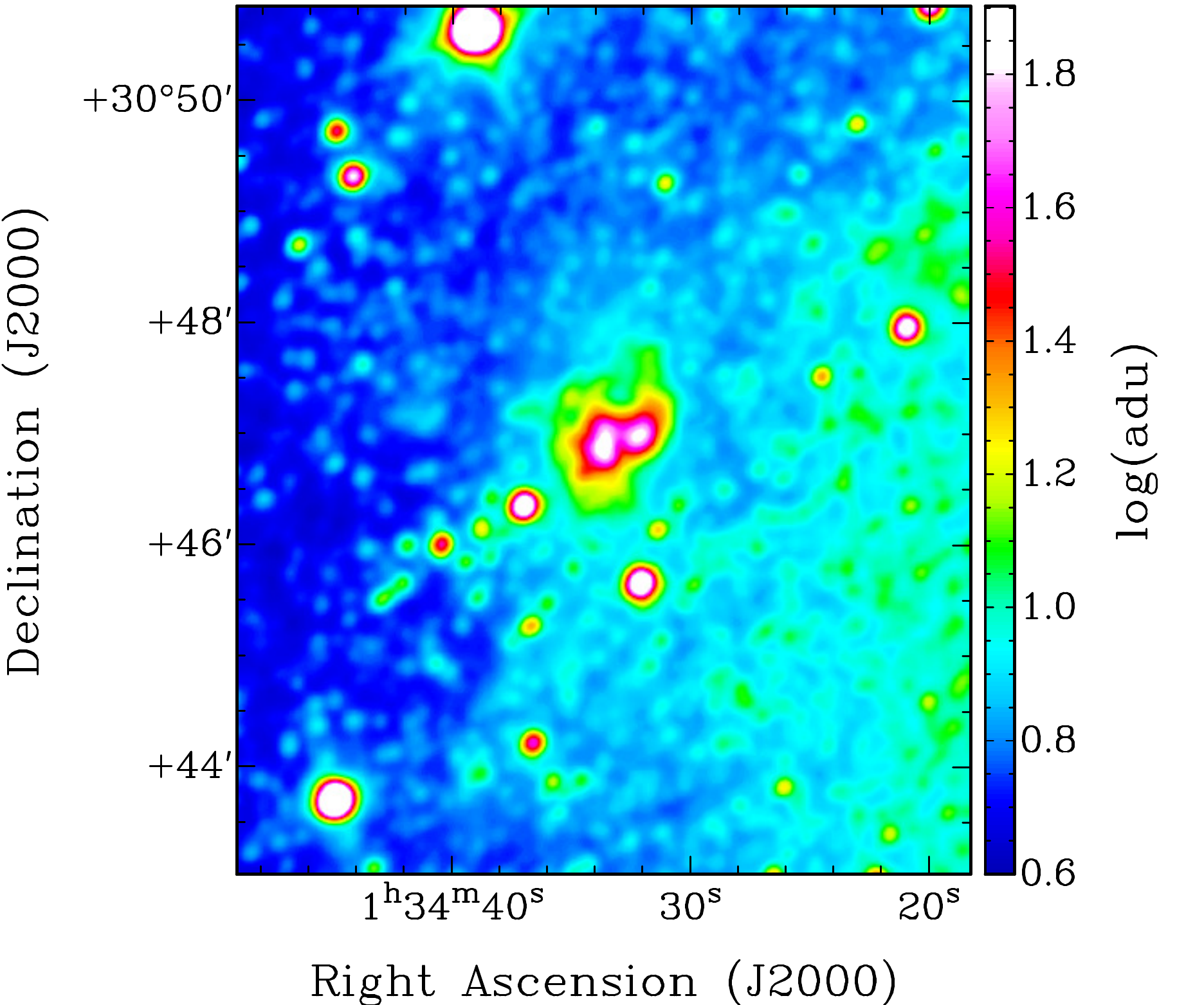}
 \includegraphics[width=\columnwidth]{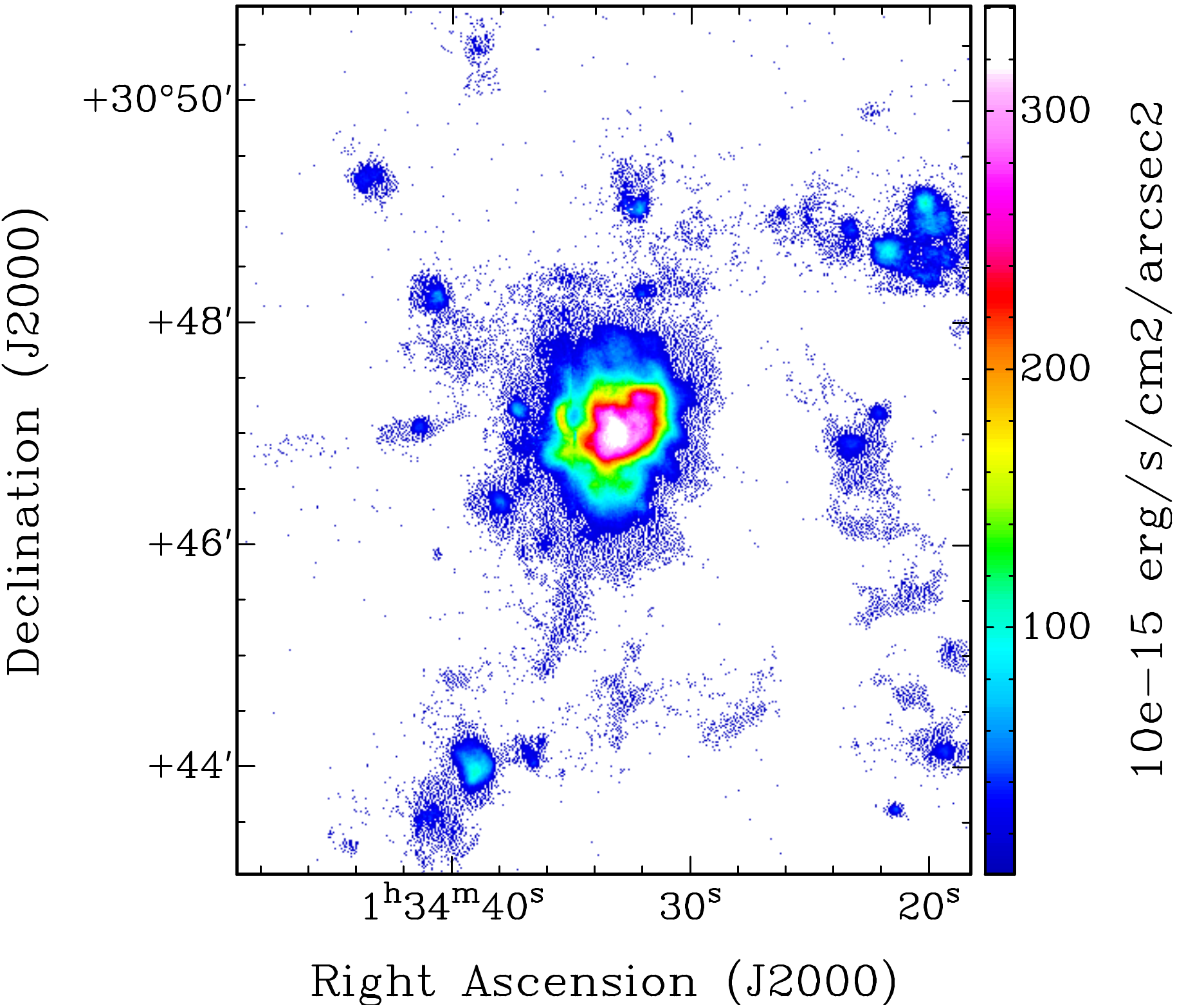}
 \includegraphics[width=\columnwidth]{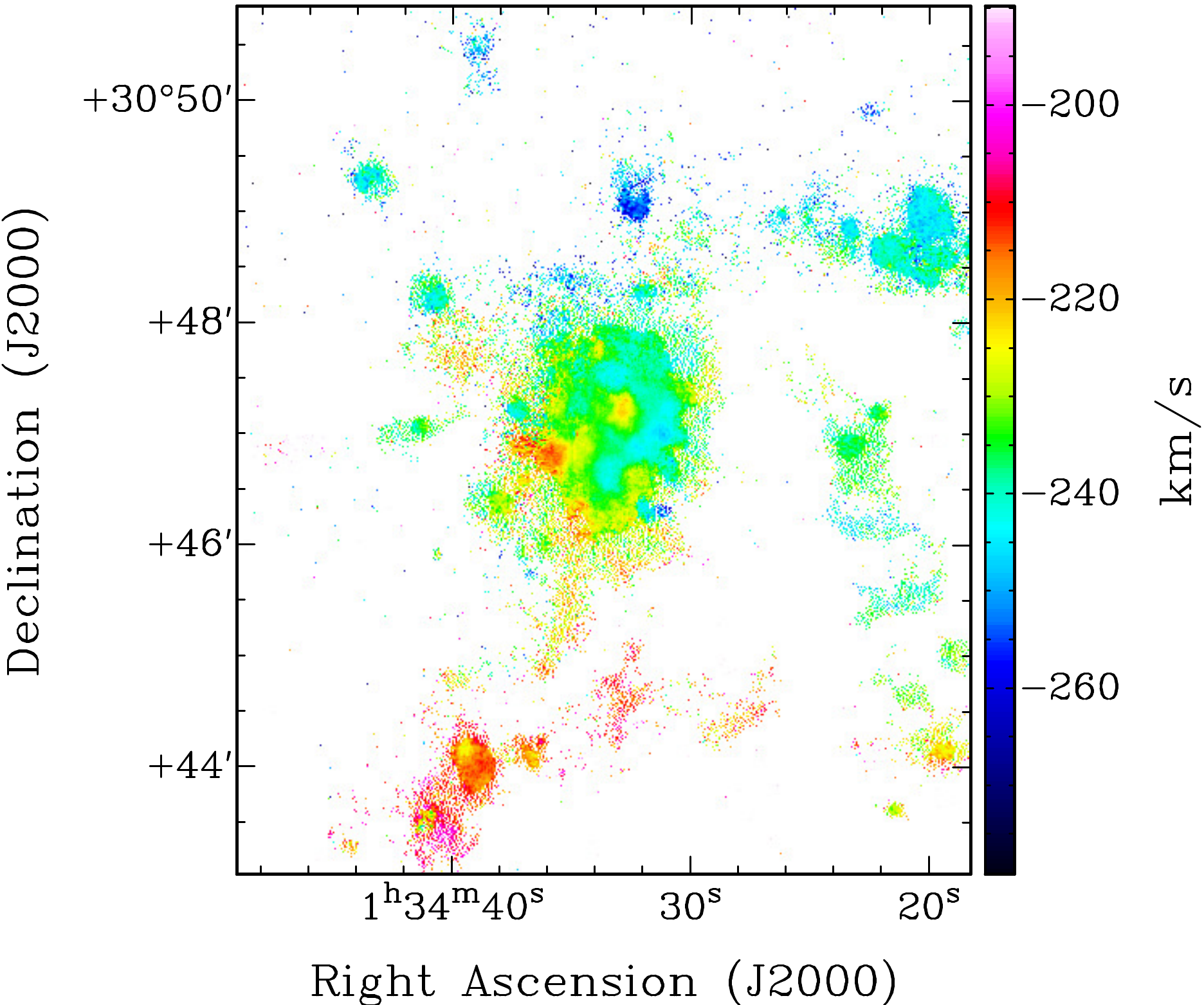}
 \includegraphics[width=\columnwidth]{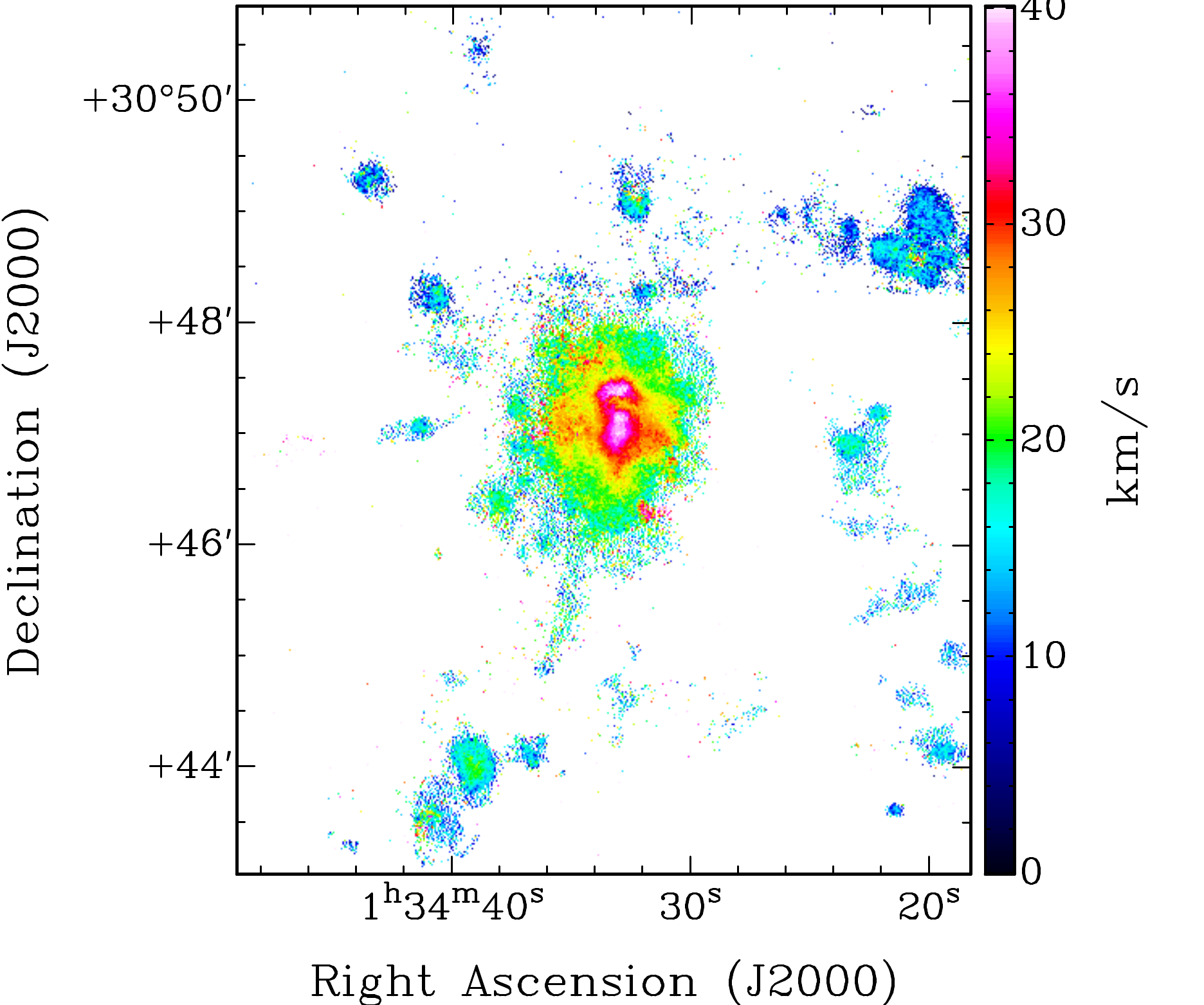}

\caption[NGC604] { Same as in Fig. \ref{fig:mapngc}, zoomed on the star forming region NGC 604} 
\label{fig:mapngc604}
 \end{figure*}
 
\subsection{Kinematics}

We performed a kinematical analysis in order to recover the set of kinematical parameters and the best possible rotation curve with meaningful uncertainties.

\subsubsection{Initial parameters} 
%$\bullet$ \textit{Initial parameters} \\

The ROTCUR algorithm implemented in the reduction package GIPSY (Groningen Image Processing System; \citealt{Vogelaar2001}) has first been used to find the mean initial values of the kinematical parameters, and to verify that no significant warp of the ionized gas disc exists.  ROTCUR is based on the tilted-ring model described in  \cite{Begeman1987}.  Starting from initial   values of $i$ = 51\degr\  and PA = 200\degr\ , $\rm V_{sys}$  and the rotation center $\rm (x_0,y_0)$ are first determined.  A second run allowed us to derive the inclination and PA profiles. From those profiles, we measured an inclination  $i$ = 52\degr\ $\pm\ 2\degr$, and a PA = 202\degr $\pm 2\degr$, both values being very close to the optical morphological values (see Table~\ref{tab:oparam}). In both cases, the errors are determined using the deviation around the mean values. No  trend has been detected in the inclination and position angle profiles, and the small standard deviations are clear indications that no significant warp of the \ha\ disk exists (at least in the regions covered by our observations), as is usually the case for the optical disks (warps are mainly seen in the extended \hi\ disks).

\subsubsection{Rotating disk  model \label{sec:diskmod}}

Since no warp is detected in the \ha\ disc using ROTCUR (tilted ring model), we used a model in which the gas is supposed to lie in a unique plane in order to derive the kinematical parameters and the rotation curve.  Thus we use the whole 2D information to derive the projection parameters and their uncertainties. The model is explained in \cite{Epinat2008} and was used for the  $GHASP$ sample data analysis. When the  vertical motion velocities are not considered,  the observed line-of-sight velocities are expressed as: 
\begin{equation}
\rm V_{obs}=V_{sys} +(V_{rot}(R) \cos(\theta)  + V_{exp}(R)\sin(\theta)) \sin(i); 
\label{eq:vobs_mo}
\end{equation}
where $R$ and $\theta$ are the polar coordinates in the plane of the galaxy, $i$ is the disk inclination, $\rm V_{rot}(R)$ is the azimuthal velocity (i.e. the rotation curve) and  $\rm V_{exp}(R)$ is the radial velocity  in the galaxy plane (often referred to as the expansion velocity). Defining the kinematical position angle $\rm PA$ as the anticlockwise angle  between the North and the direction of the receding side, the azimuthal angle $\theta$ can be deduced at each position (more details in the annexes of \citealp{Epinat2008}).
 
With the hypothesis that the radial velocities of the ionized gas are negligible with respect to the azimuthal rotation, the observed velocities become $\rm V_{obs}=V_{sys} +V_{rot}(R) \cos(\theta) \sin(i)$.  We therefore build a 2D model with a set of projection parameters (center, position angle and inclination) and a set of kinematics parameters describing the rotation curve and the motion of the galaxy with respect to the Earth (systemic velocity). All these parameters have no dependency with radius, contrary to what is done in tilted ring models \citep[e.g.][]{Begeman1987}.

The rotation curve we used in our model is described by the Zhao function \citep{Kravtsov1998}  with reduced parameters:

\begin{equation}
\rm V_{rot}=V_t{(R/r_t)^g \over 1+ (R/r_t)^a};
\label{eq:vrot_opt}
\end{equation}
where  $\rm r_t$  and  $\rm V_t$ define the transition (``turnover'') radius and velocity, $a$ and $g$ describe the sharpness of this transition and the shape before and after the transition.  Therefore, the 2D model is described by a set of 9 parameters (i, PA, $X_{cen}$, $Y_{cen}$, $V_{sys}$, $V_t$, $r_t$, g, a). The optimization starts with the initial parameters i, PA, Vsys and the rotation center previously found using ROTCUR, which are compatible with morphological parameters.  The method uses a $\chi^2$ minimization calling the IDL LMFIT routine based on the Levenberg-Marquardt method in order to find the best fit model. The uncertainties on the parameters are derived using a Monte Carlo method based on the power spectrum of the residual velocity map (see \citealp{Epinat2008} for details).

\subsubsection{Derived kinematical parameters \label{sec:dervkinparam}}

With  thousands of degrees of freedom, the optimized model converges rapidly towards a stable solution. The optimized center is found at  R.A. = 01$\rm^h$ 33$\rm^m$ 54.1$\rm^s$, DEC. = 30\degr 39\arcmin42\arcsec. That location is $\sim 42\arcsec$ (168 pc) from the photometric centre (sky projected distance), towards the NE direction, with an angle of 60\degr\ with respect to  the semi-major axis of the approaching side. This offset corresponds to a deprojected distance of 63\arcsec\ (252 pc) in the galaxy plane of Messier 33.  As a comparison, sky-projected offsets between photometric and kinematical centres of bright spirals in the Virgo cluster of galaxies  have been found between 70 and 800 pc \citep{Chemin2006}. The offset we find for Messier 33 is thus comparable with other spirals, at the low end of the distribution for other galaxies.  

The derived systemic velocity  is $\rm V_{sys} = -178 \pm 3$ \kms, the  position angle  $\rm PA= 202 \pm 4 \degr$ and  the disc inclination  $i = 52 \pm 2\degr$. These parameters are in excellent agreement with those found with ROTCUR, and with literature values.
The values for the other parameters are $g=1.2\pm0.7$  and $a=1.01 \pm 0.6$; those parameters being in good agreement with best fits parameters range of rotating discs described by \cite{Kravtsov1998}. Rotation can be approximated to a linear function of radius in the inner parts where $R<r_t=1.1$\arcmin and velocities $\rm V <  V_t$ = 50 \kms.

\begin{figure*} 
\centering
\includegraphics[width=5.4in]{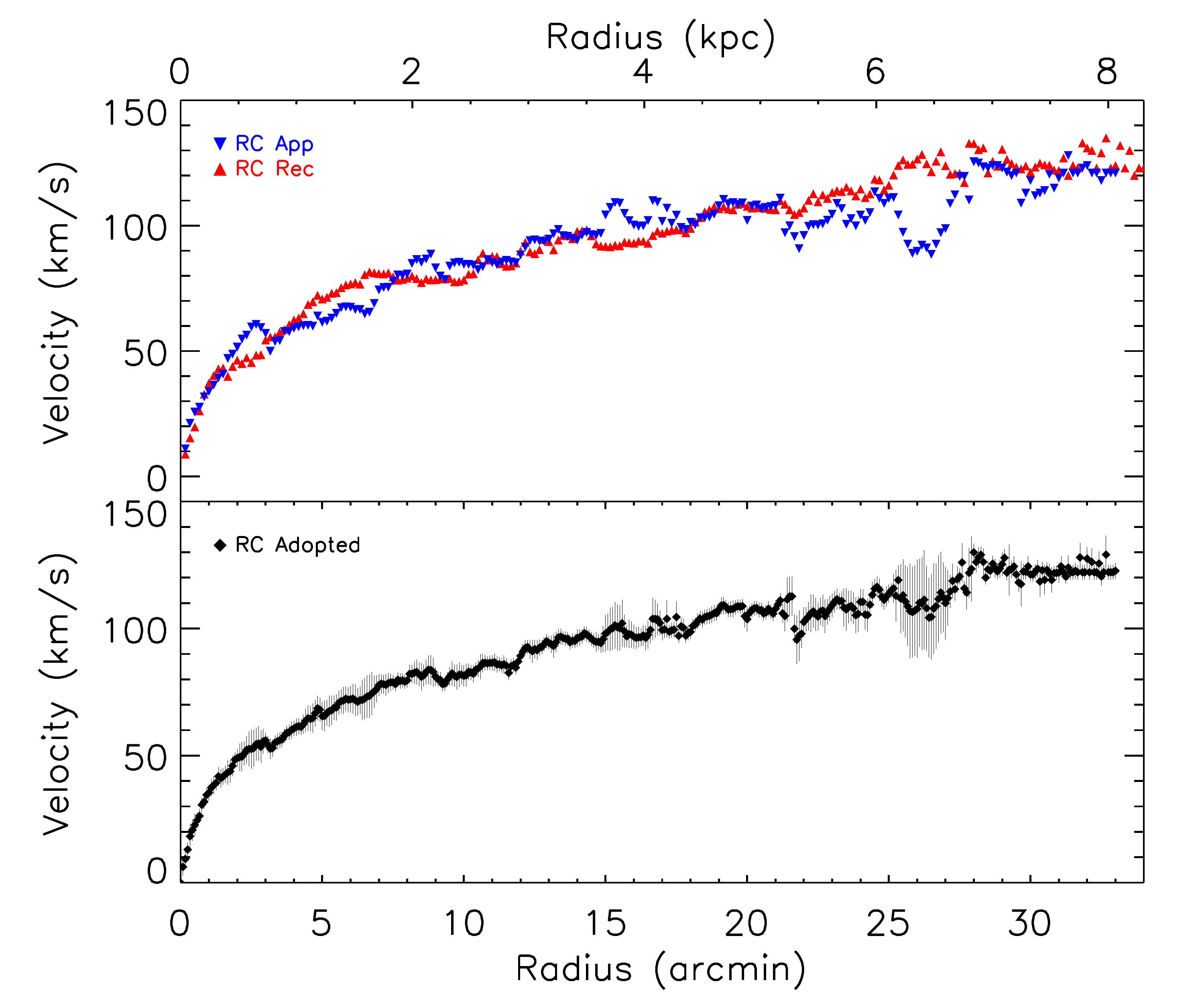}
\caption[RC] {\ha\ rotation curve of Messier 33. The top panel shows the curves for the approaching and receding disk sides (blue downward triangles and red upward triangles, respectively). 
The adopted rotation curve of M33 is shown in the bottom panel.}  
\label{fig:rc} 
%\end{minipage}
\end{figure*}

\subsubsection{Rotation curve \label{subsec:rotcur}}  
%$\bullet$ \textit{Rotation curve}\\
Rotation curves for the entire disk, and for the separate  approaching and receding sides of the disk have been extracted using the projection parameters obtained from the model fitting described above.  As detailed in \cite{Epinat2008},  the rotation curve is extracted in rings whose width is optimized by gathering a minimum of fifteen velocity measurements in the velocity field within each ring. Because the velocity field contains more than 500000 independent pixels, it was very easy to define rings with such a number of pixels.  As a result, the rotation curves were incredibly resolved, with thousands of rings for $0 < R < 35\arcmin$. Therefore, for practical reasons  we decided to rebin the resulting curves at radii regularly spaced by 5\arcsec, and of 5\arcsec\ width.  
\begin{table*}
\caption{ Sample of the \ha\ rotation curve of M33. Full table available online.} 
\begin{tabular}{@{}llllllllllllllll@{}} 
\hline
  Rad & V &$\Delta$V  & $\sigma$&$\Delta \sigma$ &Rad & V  &$\Delta$V  & $\sigma$&$\Delta \sigma$ &  Rad  & V &$\Delta$V & $\sigma$&$\Delta \sigma$ \\ %\hline
   (1)&(2)&(3)&(4)&(5)&(6)&(7)&(8)&(9)&(10)&(11)&(12)&(13)&(14)&(15)\\
    \hline
    0.08 & 6  & 4  & 14 & 10 & 11.08 & 87 & 4  & 16 & 5  & 22.08 & 104 & 5  & 17 & 7 \\
    0.17 & 9  & 4  & 12 & 6  & 11.17 & 86 & 3  & 16 & 5  & 22.17 & 105 & 6  & 17 & 7 \\
    0.25 & 13 & 3  & 12 & 4  & 11.25 & 86 & 3  & 16 & 6  & 22.25 & 106 & 8  & 18 & 7 \\
    0.33 & 18 & 5  & 14 & 5  & 11.33 & 86 & 3  & 16 & 6  & 22.33 & 107 & 7  & 19 & 8 \\
    0.42 & 21 & 4  & 15 & 4  & 11.42 & 86 & 4  & 15 & 6  & 22.42 & 106 & 7  & 18 & 8 \\
    0.50 & 23 & 4  & 16 & 4  & 11.50 & 85 & 3  & 15 & 6  & 22.50 & 105 & 5  & 17 & 7 \\
    0.58 & 25 & 3  & 15 & 4  & 11.58 & 83 & 3  & 15 & 6  & 22.58 & 107 & 5  & 16 & 7 \\
    0.67 & 26 & 3  & 16 & 5  & 11.67 & 85 & 3  & 16 & 6  & 22.67 & 107 & 6  & 17 & 8 \\
    0.75 & 31 & 3  & 16 & 5  & 11.75 & 86 & 3  & 16 & 6  & 22.75 & 105 & 7  & 16 & 7 \\
    0.83 & 32 & 3  & 15 & 5  & 11.83 & 85 & 3  & 16 & 6  & 22.83 & 107 & 5  & 15 & 7 \\
    0.92 & 35 & 3  & 15 & 4  & 11.92 & 87 & 3  & 16 & 6  & 22.92 & 108 & 5  & 16 & 7 \\
    1.00 & 35 & 3  & 15 & 4  & 12.00 & 89 & 3  & 17 & 7  & 23.00 & 109 & 5  & 15 & 6 \\
    2.00 & 49 & 4  & 14 & 4  & 13.00 & 94 & 3  & 18 & 8  & 24.00 & 110 & 5  & 17 & 8 \\
    3.00 & 56 & 3  & 14 & 5  & 14.00 & 96 & 4  & 16 & 6  & 25.00 & 114 & 4  & 16 & 7 \\
    4.00 & 61 & 3  & 15 & 5  & 15.00 & 98 & 7  & 15 & 7  & 26.00 & 108 & 19 & 17 & 7 \\
    5.00 & 65 & 6  & 16 & 5  & 16.00 & 97 & 5  & 16 & 6  & 27.00 & 110 & 14 & 14 & 8 \\
    6.00 & 72 & 5  & 17 & 5  & 17.00 & 99 & 4  & 17 & 6  & 28.00 & 130 & 3  & 19 & 12 \\
    7.00 & 78 & 4  & 17 & 6  & 18.00 & 99 & 3  & 17 & 7  & 29.00 & 125 & 5  & 19 & 9 \\
    8.00 & 80 & 3  & 17 & 6  & 19.00 & 107 & 3  & 17 & 7  & 30.00 & 121 & 3  & 14 & 10 \\
    9.00 & 81 & 4  & 17 & 6  & 20.00 & 104 & 5  & 16 & 6  & 31.00 & 122 & 4  & 15 & 11 \\
    10.0 & 81 & 4  & 15 & 6  & 21.00 & 108 & 3  & 15 & 8  & 32.00 & 128 & 4  & 22 & 21 \\
 \hline   
  \end{tabular}
  
{Notes : Column (1): radius  in arcmin, (2) the rotation velocities in \kms, (3)  errors on V  (4): $\sigma$ ( velocity dispersion) profile and (5) errors on the  velocity dispersion. The following  columns (up to 15) have the same definitions as the previous columns}  
 %\hline
\label{tab:rc}
\end{table*}
The top panel of Figure~\ref{fig:rc} shows the \ha\ rotation curves for the approaching and receding sides, while the bottom panel shows the global rotation curve, as fitted using both sides of the disk simultaneously.  The adopted rotation curve is given in Table~\ref{tab:rc} with the associated velocity errors $\rm \Delta V$.  The adopted rotation velocity uncertainty is given by $\rm \Delta V= \sqrt{\epsilon^2 +|V_a-V_r |^2/4}$.  The term $\epsilon$ is the dispersion around the mean value within each ring (statistical error for the both sides velocity calculation). The second part $\rm |V_a-V_r|/2$ is the systematic uncertainty that expresses the asymmetry between rotation velocities for the approaching ($\rm V_a$) and receding ($\rm V_r$) disk halves. The formal statistical error is usually smaller than the systematic error.  The RC used for the mass models is the adopted RC of Figure~\ref{fig:rc} (bottom panel). Only the points where data are present on both sides are used. As shown in Figure~\ref{fig:rc} (top panel), the RC on the approaching side goes out to $\ge8.5$~kpc and only out to $\sim8.1$~kpc on the approaching side.  So, the adopted RC stops at $\sim8.1$~kpc. The RC was smoothed at a binning  of 5\arcsec to get constant step in the mass modeling. 
The \ha\ rotation curve exhibits a regular rising gradient within the inner $R=8$ kpc, reaching  a maximum velocity of $\sim 125\ $\kms\ at  the last data points. Many wiggles are obviously seen, as probable consequences of the crossing of the  spiral arms of Messier 33 or due to co-rotation effects. The axisymmetry of the rotation is very good. Indeed, the most significant differences ($\sim 35$ \kms) between $\rm V_a$ and $\rm V_r$ are only observed in a narrow range around $R=6.5$ kpc.
\begin{figure*}
  \centering
  \includegraphics[width=\columnwidth]{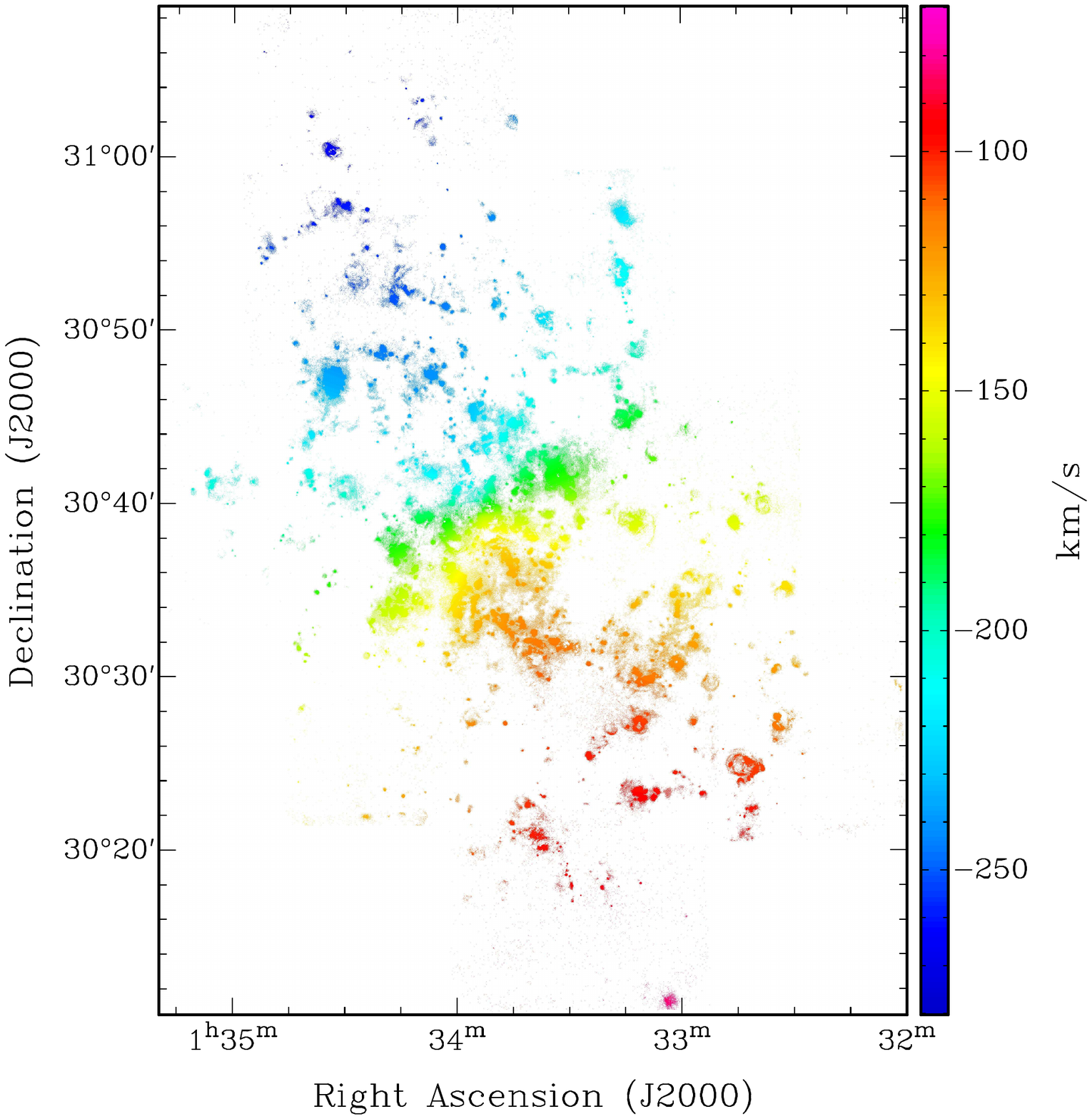}
  \includegraphics[width=\columnwidth]{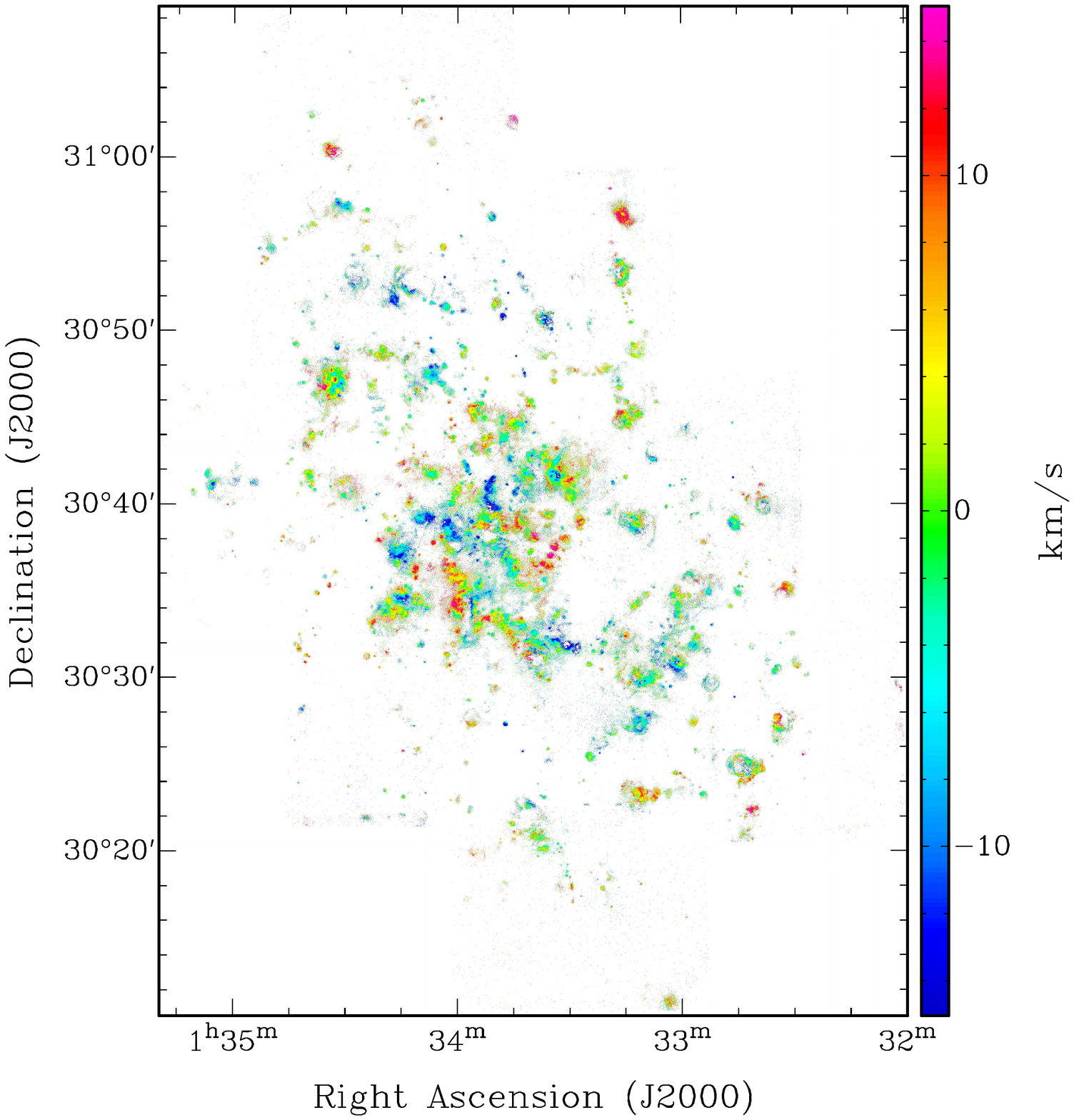}
  \caption[rvmodel] {The  left image gives the axisymmetric model  obtained with the kinematical parameters obtained with the Zhao model (See Sect.~\ref{sec:diskmod}). The map was masked with the \ha\ monochromatic map (top right map in figure~\ref{fig:rvmap}). The  right image gives  the residual  map (observed-model).}
  \label{fig:rcmodel} 
\end{figure*}
The axisymmetric velocity field model, as deduced from the adopted rotation curve with all kinematical parameters kept constant with radius, and its subtraction to the observed velocity field are shown in Fig.~\ref{fig:rcmodel}.
The distribution of residuals is centered on $0$~ \kms, with a standard deviation of 8 \kms, implying a very accurate kinematical model for most of the disk. We note that locally, bright \ha\ emission in   spiral arms can exhibit larger velocity differences.  For instance, the   star forming region at  R.A. = 01$\rm^h$ 33$\rm^m$ 15.4$\rm^s$, DEC. = 30\degr 56\arcmin48\arcsec\  has an average residual of $\sim 10$ \kms. This likely shows the limit of the axisymmetric rotation model at those angular resolution ($ 5\arcsec$, i.e 20 pc).  Our observations are indeed sensitive to very local, non-axisymmetric motions inherent to such   star forming regions (expanding shell, etc.), which cannot be modeled correctly by the large-scale  rotation of the galaxy.  Notice also that larger residuals in spiral arms  indicate the presence of asymmetric or streaming motions. Such motions are often observed in disk galaxies, but cannot be modeled by the axisymmetric rotation. Finally, larger residuals at the outskirts of the H$\alpha$ disk could indicate that some \hII\ regions may not necessarily lie in the main equatorial plane of inclination i = 52$\degr$, so that deprojection of their  velocities has not been performed correctly.
\subsubsection{Velocity dispersions \label{sec:velodisper}}
 \begin{figure}
\centering
	\includegraphics[width=\columnwidth]{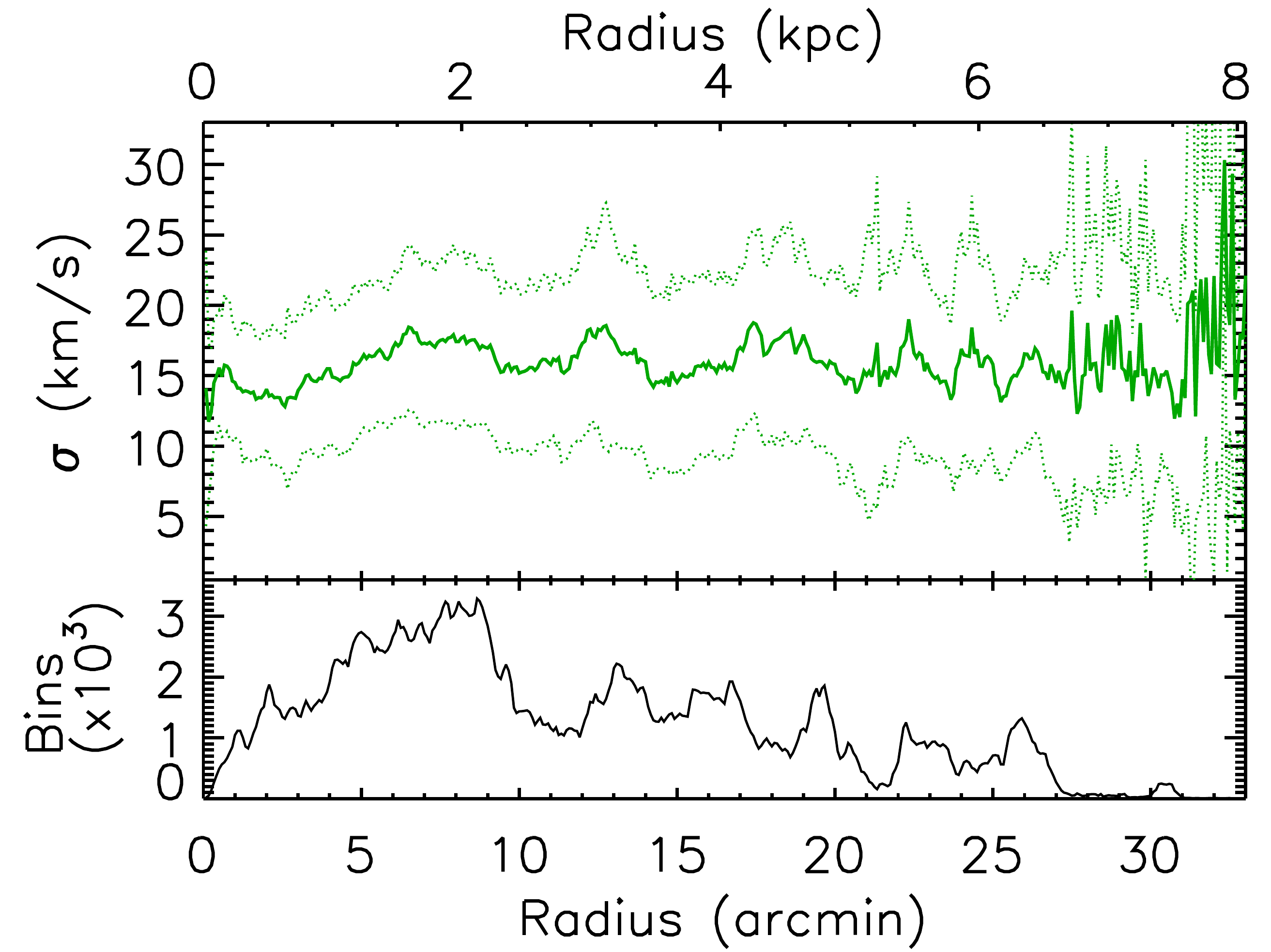} 
	\caption[ Dispersion vs r] { \ha\ velocity dispersion profile of M33. Dotted lines display values at +/- one standard deviation derived in each radial bin. The bottom panel shows the number of independent bins used to calculate the dispersion.}
\label{fig:veldisp} 
\end{figure}

 The velocity dispersion profile has been derived by azimuthally averaging the velocity dispersions $\sigma$ in annuli of 5\arcsec\ width,  in 5\arcsec\ steps (Figure~\ref{fig:veldisp}).  Annuli have been computed using the projection parameters determined from the kinematical model (sect.~\ref{sec:dervkinparam}) The $\sigma$ profile is almost consistent with a flat profile. The mean dispersion is 16 \kms. Like for the rotation curve,  wiggles are detected. They are likely caused by increased dispersion when crossing the spiral arms.  They are nonetheless not significant  to make a noticeable increase of the scatter.  We measure a standard deviation of the disperion profile of $2$ \kms.  The profile marginally decreases from R=1.5 kpc towards the centre.  From $R \sim 7.5$ kpc, the velocity dispersions increase to $\sim 20-25$ \kms. This radius is the location of the beginning of the warp of the \hi\ disk where the twist of the  position angle starts \citep{Corbelli2000}. However, it would be interesting to have \ha\ data with good SNR beyond these radius to better understand the \ha\ dispersions behaviour in regions where the  \hi\ disk warp is more pronounced. 	

%% file: massmodV2.tex
 %The galaxies mass model give 
The  \ha\ rotation curve describes  accurately the velocity gradient in the center of galaxies, usually better than 
any other kinematical tracer. Such optical high-resolution is crucial to test different inner shapes of dark matter haloes, like
cuspy or shallow models. 
This section only focuses on the modelling of the mass distribution of Messier 33 within the inner 8~kpc from our newly derived \ha\ rotation curve. 
However, with a RC only derived out to 8 kpc, we do not expect strong constraints on the halo's parameters.
We postpone to a forthcoming article a more complete modelling of the mass distribution from a more extended rotation curve that will 
merge our inner \ha\ rotation curve with a new  \hi\ rotation curve for the outer regions of Messier 33 (Kam et al., in preparation).

\subsection{\label{sec:gas_stars}Luminous mass components}
\subsubsection{The neutral gaseous disc }
\begin{figure*}
%\begin{minipage}{165mm}
\centering
 \includegraphics[width=\columnwidth]{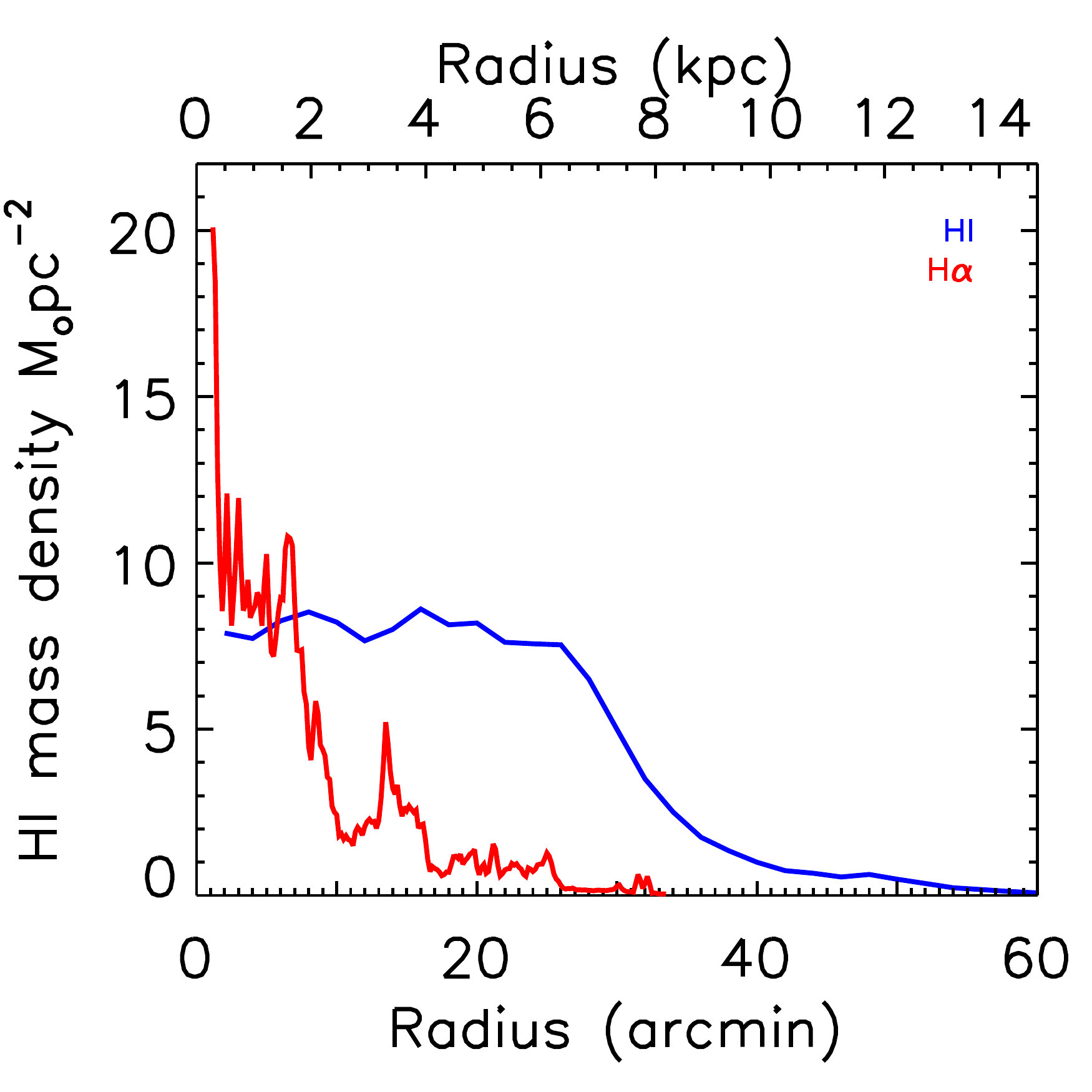}
 \includegraphics[width=\columnwidth]{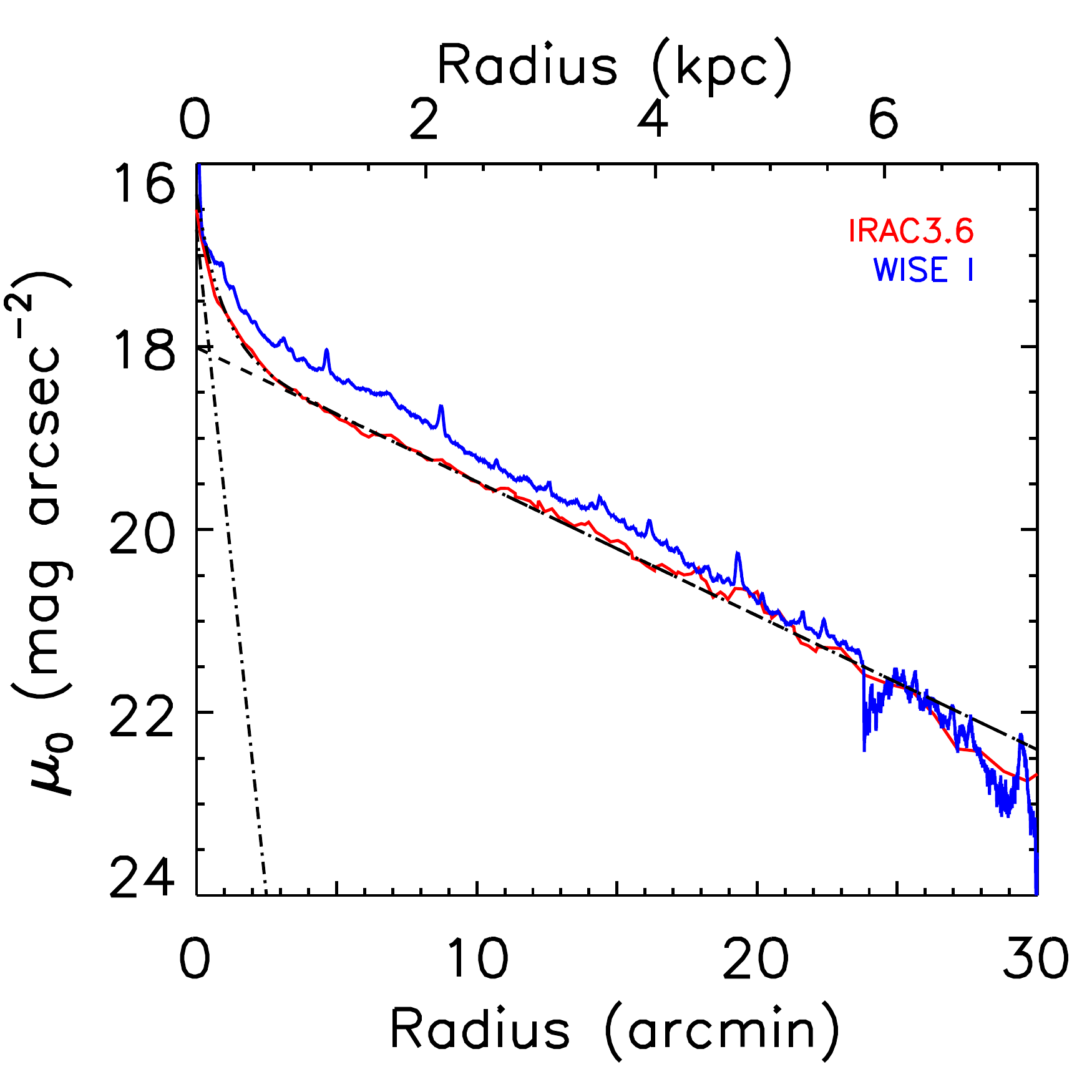}
\caption[ mass density] {Left panel: \hi\ surface density profile (blue line) and \ha\ brightness profile  (red line, in arbitrary units). Right panel: in red the M33 surface brightness profile using the  Spitzer/IRAC $3.6\mu m$ and in blue using the WISE I (Tom Jarrett, private communication) data. The dashed-dotted line is the bulge  component, the dashed line is the disk component,  the other line is the contribution of both components.}  
\label{fig:mdensity} 
%\end{minipage}
\end{figure*}

Provisional data on the \hI\ gas component  have been  presented in \cite{Chemin2012}. Those observations will be fully described in a future Kam et al.'s paper.  Figure~\ref{fig:mdensity} presents (left-hand panel) the \hi\ mass density profile overlaid on the ionized gas brightness profile (in arbitrary units).  Both profiles have been derived with the task {\it ELLINT} in GIPSY. The \hI\ disc mass is $\sim 2 \times 10^9$ \msol. The surface density only slightly varies around 8 \msol\ pc$^{-2}$ within $R=7$ kpc, then decreases at larger radii.

\subsubsection{The bulge-disk decomposition}

For the stellar contribution, the surface brightness profile is derived from the Spitzer/IRAC $3.6\mu m$ data. The archive mosaic file from guaranteed time observations of  Robert Gehrz, Observer Program ID 5 was used \citep{Gehrz2004}. After removing the  stars and the background, the CHANDRA's CIAO\footnote{http://cxc.harvard.edu/ciao/}  tools have been used to derive the profile of Figure~\ref{fig:mdensity}. 
The  Spitzer/IRAC~$3.6~\mu m$ profile is compared to the WISE I ($3.4~\mu m$) data. The profile presented in the right-hand panel of Figure~\ref{fig:mdensity} (red),  reaches a surface brightness of $\mu_{3.6}\sim$ 21.0~$\rm mag.arcsec^{-2}$ at a radius  of 21~arcmin ($\sim$5~kpc), which is similar to the  WISE I profile (blue). The surface brightness $\mu_{cor}$, corrected for the inclination and dust, is obtained by: 
\begin{equation}
\rm \mu_{cor}= \mu_{obs} +A_{\lambda} + 2.5\times b_{\lambda} ~log(cos(i))
\label{eq:sbcor}
\end{equation}
 where  the correction parameters $A_{\lambda}$  and $b_{\lambda}$  are tabulated in \cite{Graham2008}. The mean inclination of 52\degr\ is used for the density correction. In this correction,  the influence of PAH at $3.3~\mu m$ is considered as very weak  and the maximum dust correction used in the IR comes from the J band  \citep{Graham2008}.   In the right panel of  Figure~\ref{fig:mdensity}, the central 3~arcmin  shows a  small spheroidal component in the $3.6~\mu m$ band.  This small central bulge has been discussed by many authors \citep{Regan1994,  Gebhardt2001, Minniti1994, Seigar2011}.  \cite{Kent1987}  found that the nucleus in M33 is similar to a point source and the  rising inner 3\arcmin\ part of the surface brightness profile suggests the presence of a small bulge. \cite{Corbelli2007b} found for M33 a bulge component extending up to 1.7~kpc and  \cite{Seigar2011} only to 0.39~kpc. In view of those different results, it is clear that a disk-bulge decomposition has to be done using the IR profile from this study.

%%%OUR decomposition
The best fit of the bulge-disk decomposition, shown in  Figure~\ref{fig:mdensity}, is obtained using an exponential disk with a Sersic model for the bulge.  The black dot-dot-dot-dash line gives the best fit of the sum of the decomposition. The disk is described by:  

\begin{equation}
\mu (r)=\mu_0 +1.10857~ {R\over R_d} 
\label{eq:bisk_expo}
\end{equation}
where  $\mu_0$ is the central surface brightness and $R_d$ the scale-length of the disk. The disk parameters found are $R_d=1.82\pm0.02$~kpc (slightly smaller than the scale-length in the optical) and $\mu{_0}=18.01\pm0.03$ $\rm mag.arcsec^{-2}$. As seen in Figure~\ref{fig:mdensity}, the Wise I profile has a slightly shorter scale-length.  The disk parameters found for the Wise I profile are $R_d$ = 1.70~kpc and $\mu_0=17.60~ \rm mag.arcsec^{-2}$.

The bulge is described by :  
\begin{equation}
\mu (r)=\mu_e +2.5~b_n \left[ \left(   R\over R_e \right)^{1/n}+1   \right]
\label{eq:bulb_sersic}
\end{equation}
where, $\mu_e$ is the effective surface brightness  at $R_e$,  the effective radius. $R_e$  defines the radius that contains half of the total light. The parameter $ n $ determines the  “curvature”  of the luminosity profile. $b_n$ is defined as $b_n = 1.9992n-0.3271$ for $\displaystyle0.5<n<10$  \citep{Capaccioli1989}. The best fit in Figure~\ref{fig:mdensity} gives $\mu_e=20.8\pm0.3$~$\rm mag.arcsec^{-2}$, $n=1.12\pm0.1$ and an effective radius $R_e=0.35 \pm0.05$~kpc for the IRAC profile and $\mu_e=21.5\pm0.4$~$\rm mag.arcsec^{-2}$, $n=1.20\pm0.05$ and an effective radius $R_e=0.40\pm0.05$~kpc for the Wise profile. 
With these values, our study leads to a bulge-to-disk ratio of B/D=$0.04\pm0.012$ that is in agreement with B/D=0.03 obtained by \cite{Seigar2011}  with IRAC data at 3.6 $\mu m$. The bulge component is subtracted from the total profile in order to get the disk contribution. 

Stars in a disk have a vertical thickness. For  the vertical distribution of the  stellar component, we adopted  a $sech^2(z/z_0$) law \citep{vanderKruit1981}. We used a vertical scale height  of $\sim$ 365~pc, which is $\sim$20\% of the stellar disk scale length. 

\subsubsection{The mass-to-light ratio}
 With such a small bulge-to-disk ratio (and consequently $Re$), M33 can be considered as nearly a pure disk galaxy.  In this paper, both cases are considered; the pure disk case and the disk + small bulge case.  However, with such a small bulge,  they should be fairly similar. The color mass-to-light ratios (M/L) are defined separately for the bulge and the disk and are used to obtain the actual total stellar mass contribution using the method described in \cite{Oh2008}:  
\begin{eqnarray}
\Upsilon_*^{3.6}=B^{3.6}~\times~\Upsilon_*^{K} +A^{3.6}
\label{eq:m1} 
\end{eqnarray}
where $\Upsilon_*^{3.6}$  is the M/L at $3.6\mu m$, $\Upsilon_*^{K}$, the M/L in the K band and  $B^{3.6} $ and $A^{3.6}$ the correction coefficients.
For the K band M/L, we used the relation $\rm log(\Upsilon_*^{K})=1.46(J-K)-1.38$ taken  from \cite{deBlock2008}, obtained by using an extrapolation of \cite{Bell2001}, where the stellar mass synthesis uses a Salpeter initial mass function (IMF).  The $\Upsilon_*^{3.6}$ for nearby galaxies is obtained using $\Upsilon_*^{K}$ \citep{deBlock2008}:
\begin{equation}
\Upsilon_*^{3.6} =0.92~\times~\Upsilon_*^{K} -0.05
\label{eq:ml1} 
\end{equation}

The K band M/L is obtained using the 2MASS (J-K) color computed by \cite{Jarrett2003}. The (J-K)  color has been computed separately for the bulge in the inner part and for the disk in the outer parts. The first data points (J-K) are used for the bulge.   For the pure disk case,  the mean color from 65 to 570 arcsec has been used. The (J-K) decomposition gives  for the bulge and disk 0.94 $\pm 0.03$ and 0.86 $\pm 0.03$ respectively. However, it is quite likely that the bulge color is underestimated, being contaminated by disk light. Those values give the mean  M/L in the ${3.6\mu m}$ band : $\Upsilon_{d}^{3.6}=0.72\pm0.1$ for the disk and $\Upsilon_{b}^{3.6}=0.80\pm0.1$ for the bulge. The effective mass density profile is obtained by \citep{Oh2008}:  
\begin{eqnarray} 
\Sigma [ {\rm M_{\odot}} pc^{-2}] = \Upsilon_{*}^{3.6} \times 10^{-0.4 \times(\mu_{3.6}-24.8).} 
\label{eq:massdensity} 
\end{eqnarray}
 where $\Upsilon_{*}^{3.6}$ is the mass-to-light ratio in the {\it Spitzer}/IRAC 3.6 micron band. The density profile is used in $GIPSY$  to compute the contribution of the  stellar  component.  

\subsection{Dark matter halo density profile}  

The total rotation velocity is given by:
\begin{equation}
\label{eq:vtot}
	\rm V_{rot}^2 =   V_{\star}^2 +  V_{gas}^2 +   V_{DM}^2  
 \end{equation} 
where $\rm V_{\star}^2$ is the contribution of the stars, $ \rm V_{gas}^2$ the contribution of the gas component and  $\rm V_{DM}^2$ the contribution of the dark matter halo.  
The dark matter contribution is required to explain the outermost flat part of the rotation curves in galaxies \citep{Bosma1978, Carignan1985}. %The equation~(\ref{eq:vtot},
The dark matter distribution can be defined by different  types of density profile.  
We will limit this study to the most commonly used halo density profiles, the pseudo-isothermal (ISO) and the Navarro, Frenk and White (NFW) halo distributions, which show the largest differences in the inner parts, where the RC is well defined by the \ha\ data.

\subsubsection{ISO density profile}
The pseudo-isothermal (ISO) dark matter halo is a  core-dominated type of halo. The ISO density profile is given by:% equation~(\ref{eq:ro_iso}): 
\begin{equation}
\rho_{ISO}(R)={\rho_0 \over 1+({R\over R_c})^2}
\label{eq:ro_iso}
\end{equation}
where $\rho_0$ is the central density and $R_c$  the core radius of  the halo. The velocity  contribution of a ISO halo is given by: %by equation~(\ref{eq:v_iso}):
\begin{equation}
\rm V_{ISO}(R)=\sqrt{4\pi~ G~\rho_0~ R_c^2(1-{R_c \over R}~ atan ({R\over R_c}))}
\label{eq:v_iso}
\end{equation}

\subsubsection{NFW density profile}
The NFW model is derived from $\Lambda CDM$  simulations \citep{Navarro1996,Navarro1997}.
This density profile (so-called "universal halo") is known as the cuspy type  and follows an $R^{-1}$ law \citep{deBlok2010} in the innermost regions.  The NFW halo density  profile is described by: 

\begin{equation}
\rho_{NFW}(R)={\rho_i \over {R\over R_S} (1+{R\over R_S})^2}
\label{eq:ro_nfw}
\end{equation}
where $\rho_i \thickapprox 3H_0^2/(8 \pi G)$ is the critical density for closure of
the universe and $R_S$ is a scale radius. The  velocity contribution corresponding to
this halo  is given by: % in the equation~(\ref{eq:v_nfw}). The RC of the NFW halo is given by: 
\begin{equation}
\rm V_{NFW}(R)=V_{200}~\sqrt{{ln(1+cx)-cx/(1+cx) \over x(ln(1+c) -c/(1+c))]}}
\label{eq:v_nfw}
\end{equation}
where $\rm V_{200}$ is  the velocity at the virial radius $\rm R_{200}$, $\rm c = R_{200}/R_S$ gives the concentration parameter of the halo and  x is defined as $\rm R/R_S$.   The relation betwen $\rm V_{200}$ and $\rm R_{200}$ is given by:
\begin{equation}
\rm V_{200}= {R_{200} \times H_0 \over 100}
\label{eq:v200}
\end{equation}
where $\rm H_0$ is the Hubble constant taken as  $\rm H_0~=~72$~\kms$\rm Mpc^{-1}$ \citep{Hinshaw2009}.

\subsubsection{Results}

\begin{figure*}
% \begin{minipage}{160mm}
 \begin{centering}
 \includegraphics[width=6.5in]{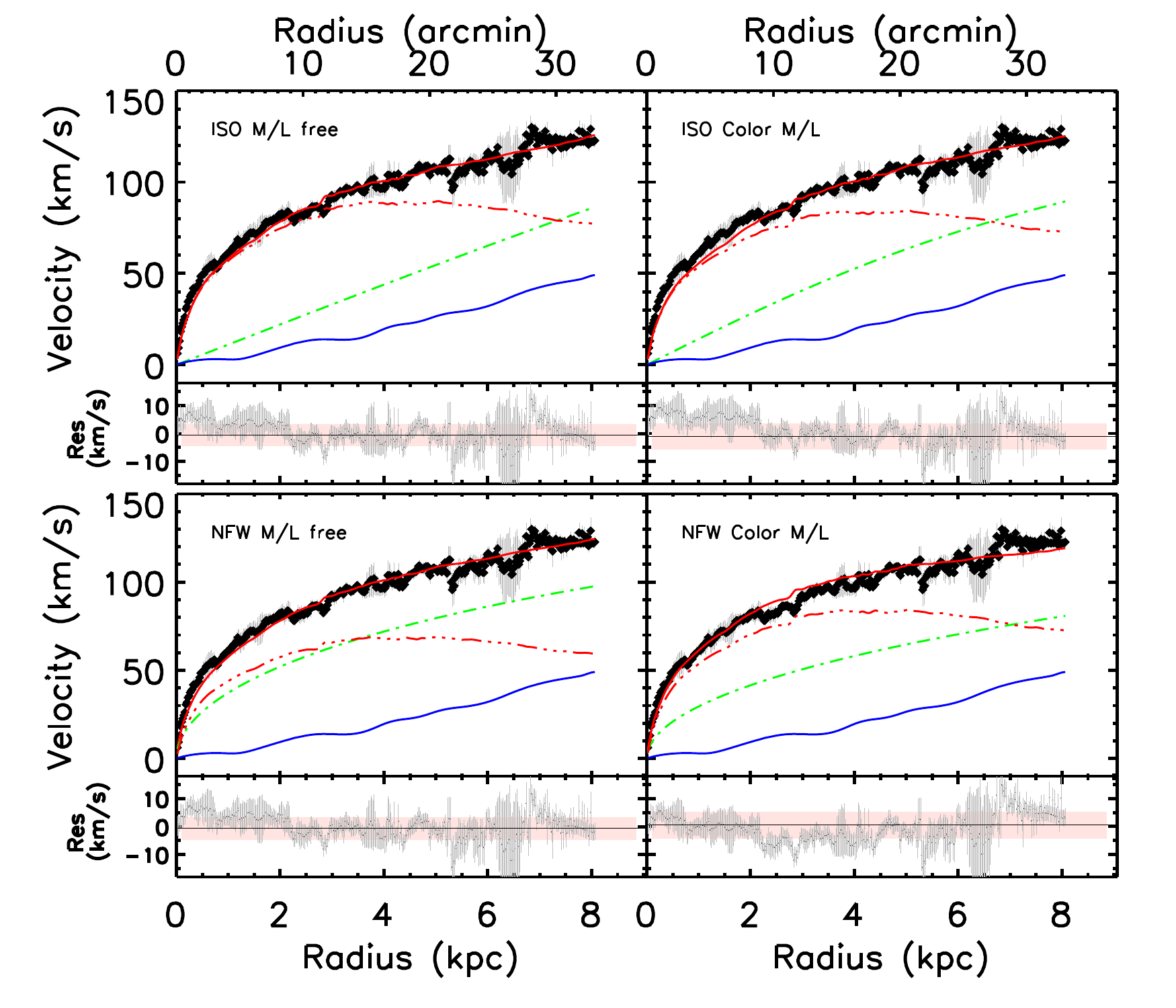}
  \caption{M33 mass models with a DM halo for the pure disk case. Models on the left are best fits while models on the right have a fixed M/L determined by the IR color (see sec.~\ref{sec:gas_stars}). The top panels represent the models with the ISO  and the bottom panels those with the NFW density distributions. The stellar disk contribution is in broken red, the \hI\ contribution in blue, the halo contribution in broken green and the sum of the 3 components is shown as a continuous red line. At the bottom of each model the mean residuals (observation - model) are presented.}
   \label{fig:m33dm_all}
  \end{centering}
%  \end{minipage}
\end{figure*}

\begin{figure*}
 \begin{minipage}{160mm}
 \begin{centering}
 \includegraphics[width=6.5in]{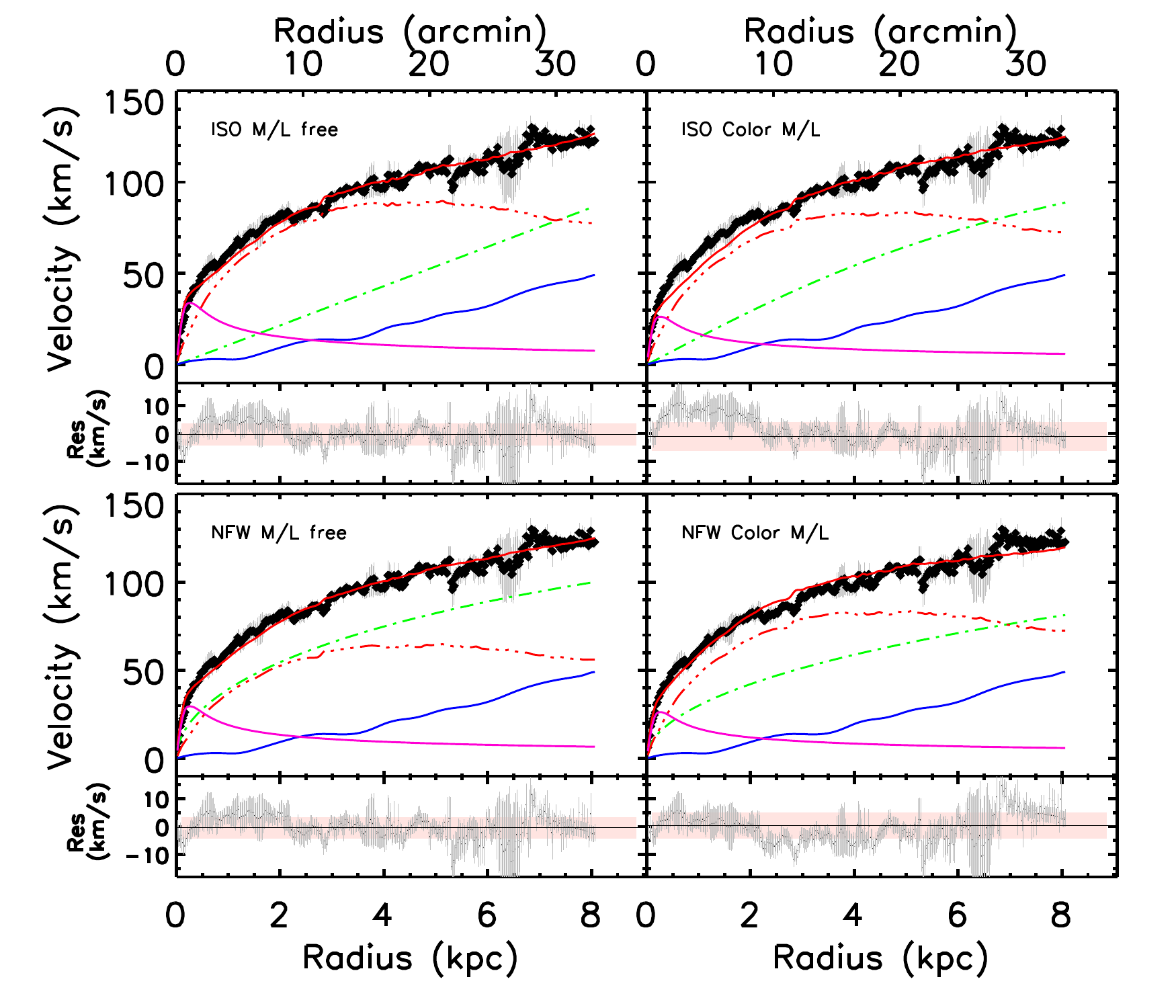}
  \caption{ Same as in Figure~\ref{fig:m33dm_all} but with the bulge contribution (in cyan)}
  \label{fig:m33dm_allbd} 
  \end{centering}
 \end{minipage}
 \end{figure*}

The \ha\ rotation curve of Figure~\ref{fig:rc} (bottom panel) will be used  for the mass modeling.   Figure~\ref{fig:m33dm_all} shows the models using
the ISO (top) and the NFW (bottom) DM distributions for the pure disk case and Figure~\ref{fig:m33dm_allbd} for the bulge-disk decomposition.  
 The left panels of the figures give the best fit models and the right panels, the models with the mass-to-light ratio constrained by the IR color and population synthesis models \citep{Oh2008}.  Both fits use Levenberg-Marquardt least-squares fitting techniques. 
At the bottom of each model, the mean residuals (observation - model) are represented by a black line with the same error bars as the velocity errors of the top panels.  The part colored in pink gives the dispersion of the residuals around the black regression line. 

Table~\ref{resulmassmodelDM} and Table~\ref{resulmassmodelDM2}
give the results of the mass models: column (1) gives the type of halo density profile used;  column (2) gives the parameters of the halo. In the tables,  $\Upsilon$ gives the mass-to-light ratios obtained from the fits and $\chi_r^2$ the goodness of the fit.  Column (3) shows the results using the best fits and column(4) the results when the mass-to-light ratios are kept fixed at the value obtained using the (J-K) color and population synthesis models. In  Table~\ref{resulmassmodelDM2}, the parameter for the bulge mass-to-light ratio is given by $\Upsilon_b$. The results of the mass models will be discussed in the next section.
 
 \begin{table}
\caption{ \ha\ mass  model results with DM halos for the pure disk case.}
\begin{tabular}{llccc} 
\hline
Halo Model&Params&  Best Fit& Color M/L* \\ \hline
(1)&(2)&(3)&(4)&\\ %\hline
\hline
ISO	
	&$\rho_0$			&$6.75\pm0.57$	&$7.46\pm0.43$\\
	& $R_c$ 				&$24.23\pm17.02$ 	&$11.13\pm  0.40$\\
	& $\Upsilon$			& $0.81\pm0.01$	& 0.72 \\
	&  $\chi_r ^2$			&  0.88			&1.16\\
	&\\
NFW 
	&$R_{200}$			&$240.00\pm188.34$	&$300.26\pm106.55$\\
	& c 				 	&$4.41\pm4.15$ 		&$2.37\pm0.91$\\
	& $\Upsilon$			&$0.48\pm 0.11$		&0.72  \\
	& $\chi_r ^2$ 			&0.87				&1.56\\	
%	&\\
\hline
 
\label{resulmassmodelDM}
%\hline
\end{tabular}

*M/L fixed by the color of the disk; \\
$\rho_0$, the central DM density, is given in units of $10^{-3}$ \msol/$pc^3$; \\
$R_c$  and $R_{200}$ are in kpc.

\end{table}%	
\begin{table}
\caption{\ha\ mass  model results with DM halos, using the bulge-disk decomposition.}
\begin{tabular}{llccccc} 
\hline \hline
Halo Model&Params& Best Fit& Color M/L* \\ \hline
(1)&(2)&(3)&(4)&\\ %\hline
\hline
ISO	
	&$\rho_0$			&$6.41\pm0.60$	&$12.52 \pm 0.53$		\\
	& $R_c$ 			 	&$119.50\pm \ga 200$ 	&$6.32\pm0.40$	\\
	& $\Upsilon_d$			& $0.82\pm0.01$	& 0.72 				\\
	& $\Upsilon_b$			&$1.32\pm0.06$	& $0.80$ 		 \\
	&  $\chi_r ^2$			&0.85			&1.34				\\

	&\\

NFW 
	&$R_{200}$			&$211.55\pm0.72$	&$2.78\pm0.78$		\\
	& c 					&$8.67\pm0.12$ 	&$0.76\pm0.45$		\\
	& $\Upsilon_d$			&$0.43\pm 0.01$	&0.72  				\\
	& $\Upsilon_b$			&$1.01 \pm0.11$	& $0.80$ 		\\
	& $\chi_r ^2$ 			&0.75		& 1.43				\\

	&\\
\hline
\label{resulmassmodelDM2}
%\hline
\end{tabular}
*M/L  fixed by the color of the disk  and of the bulge; \\
$\rho_0$, the central DM density is given in units of $10^{-3}$ \msol/$pc^3$; \\
$R_c$ and $R_{200}$ are in kpc. 

\end{table}%	

%% file: discussionV2.tex
\subsection{M33 velocity dispersion}
\begin{figure*}
\centering
  \includegraphics[width=3.2in]{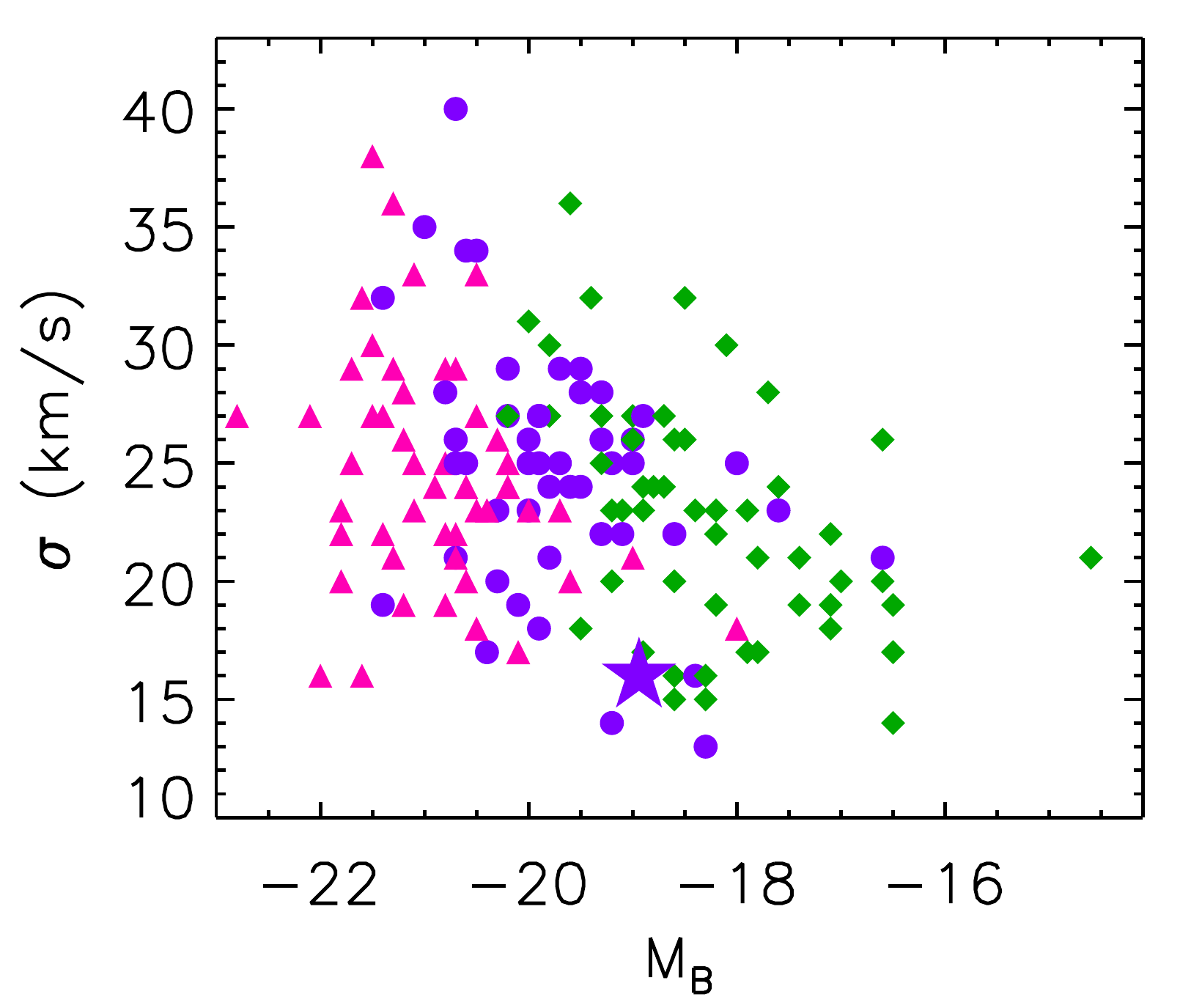}
  \includegraphics[width=3.2in]{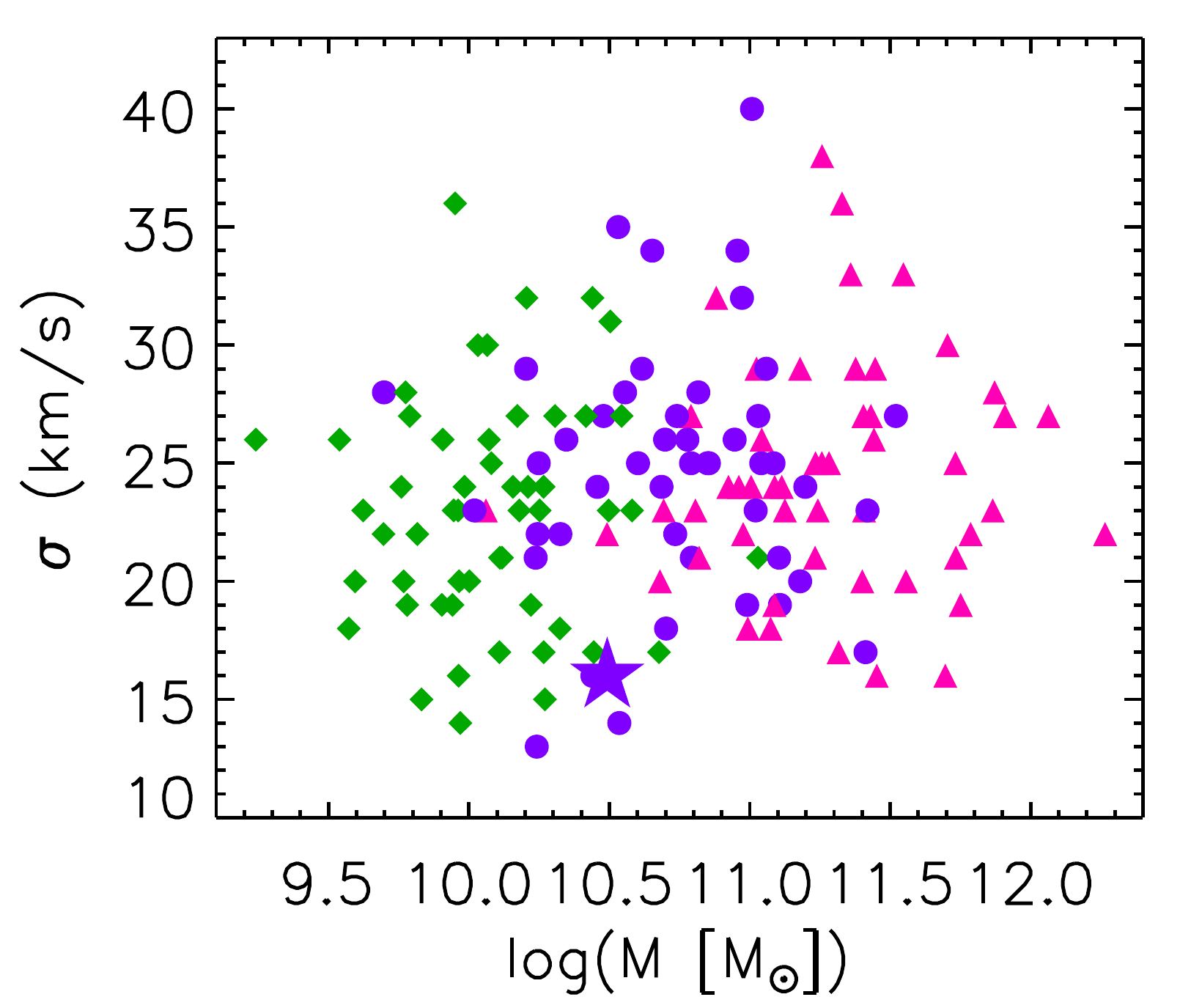}
  \caption{ Mean velocity dispersion of Messier 33 (star) and of other nearby star-forming disc galaxies from \protect\cite{Epinat2010}. On the right, the mass is given by $\rm M=V_{max}^2 R_{25}/G$, where $\rm V_{max}$ is the maximum velocity of each \ha\ rotation curve.}
\label{fig:compghasp}
\end{figure*} 
Our new dataset is ideal to study the internal velocity dispersion of HII regions in M33, and the mean velocity disperion of the galaxy.
In Figure~\ref{fig:compghasp}, we compare the mean velocity dispersion of M33 to a subsample of 151 nearby star-forming galacitc disks from the GHASP survey \citep{Epinat2008} studied in \cite{Epinat2010}
It is obvious how the ionized interstellar medium of Messier 33 appears   dynamically colder (lower velocity dispersion) than in other galaxies. 
We have divided the  sample into three classes of size: small galaxies with ${\rm R}_{25} < 7.5$ kpc, intermediate size galaxies with $7.5 < {\rm R}_{25} < 11.5$ kpc
and large galaxies with $ {\rm R}_{25}>11.5$ kpc. Though being within the class of intermediate galaxy size, Messier 33 curiously behaves like smaller galaxies that have  lower mass and absolute magnitude. 

Giant extragalactic \hII\ regions seen in gas-rich spiral and dIrr galaxies are regions of strong star formation with a size ranging between 0.1 and 1 kpc, a H$\alpha$ luminosity of $10^{39} - 10^{40}$ erg/s, a mean density of $1-10$ cm$^{-3}$ and an ionized mass of $10^4-10^5$ \msol,  which embeds a population of $100-200$ ionizing stars \citep{Kennicutt1984a, Kennicutt1984b}. In addition, the velocity dispersion of the gas presents distinct kinematics with subsonic and supersonic emission-line widths \citep{Smith1970, Smith1971b}. Uniform expansion of \hII\ regions into a medium of constant density results in subsonic expansion velocities \citep{Spitzer1968}.  Supersonic motions are inferred for velocity dispersion typically larger than (kT$/m)^{0.5} \sim10-13$ \kms, where $\rm k$ is the Boltzmann constant, $m$  the typical hydrogen mass weighted by the molecular content and T = $10^4$K, the characteristic temperature of the \hII\  region. Supersonic velocity dispersions ranging from $15$ to $40$ \kms\ are observed and give rise to different interpretations (see discussion in \cite{Bordalo2010}. In the "champagne" model \citep{Tenorio1979}, the expanding \hII\  bubble bursts thought the surface of the surrounding molecular cloud where the sharp gas density discontinuity generates a shock wave that accelerates the gas to supersonic speeds. Density gradient within the HII regions  may generate supersonic motions prior to their acceleration through the surface of the cloud \citep{Mazurek1982}.
%========Disc P. Amram    

Our data confirm that the region NGC 604 (Figure~\ref{fig:mapngc604}), like many other giant \hII\ regions, displays both single Gaussian and complex line profiles \citep{Munoz1996, Yang1996}.  Single Gaussian profiles essentially come from the bright regions and show the broadening mechanics due to self-gravitation. More complex lines come from faint regions and emanate from the wind-driven mechanical energy injection due to massive stars that produce expanding shells, cavities and bubbles, filaments and outflows, loops and ring-like regions.

\begin{figure}
\centering
	\includegraphics[width=\columnwidth]{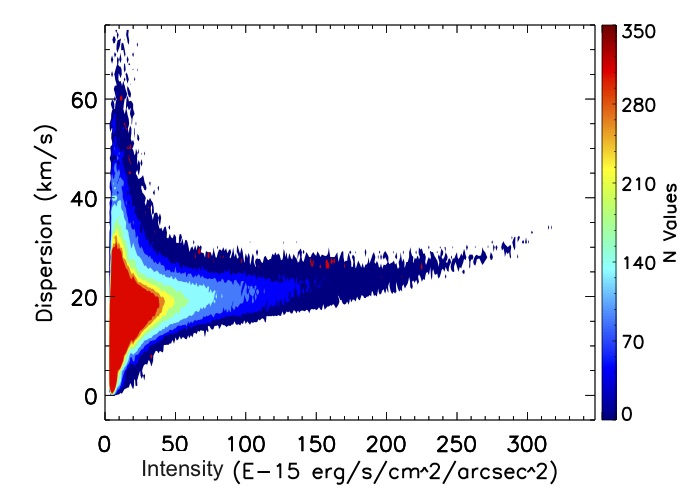}
\caption[ Dispersion vs r] { Corrected velocity dispersion ($\sigma_{cor}^2=\sigma_{obs}^2-\sigma_{PSF}^2$), computed using a single Gaussian, vs integrated line intensity for the whole field-of-view of M33.  The number of pixels having a given intensity and dispersion is given by the color scale.}
\label{fig:dispflux}
\end{figure}

%An intensity-weighted 
A velocity dispersion versus intensity  ($\sigma-I$) diagram provides diagnostics used to separate the main broadening mechanisms detected in the emission lines of giant \hII\ regions, such as differentiating the broadening produced by virial motions resulting from the total gravitational potential and that resulting from the superposition of shells and loops generated by massive stars, which ends up dispersing the parent clouds \citep{Munoz1996, Yang1996, Martinez2007, Moiseev2012}. 
Such a diagram is given for the whole disk of the galaxy in Figure~\ref{fig:dispflux}. We used the total intensity in the line rather than the intensity peak because the total intensity is independent from the spectral resolution (which is not the case of the intensity peak). In the following, we use intensity to refer to the total line intensity. 
Figure~\ref{fig:dispflux} shows the strong relation between the \ha\  velocity dispersion and the \ha\  intensity.  Only low intensity regions show a broad range of velocity dispersions. In other words, the range of velocity dispersions decreases when intensity increases. Regions showing larger velocity dispersions arise from diffuse emitting areas rather than from intense \hII\ regions.  
Subsonic dispersions are only observed in low intensity regions, whereas supersonic motions are both observed in the low surface brightness medium and the brightest star forming regions. These latter have a roughly constant velocity dispersion (20-30 \kms), while the largest supersonic dispersions (40-60 \kms) are only seen among the lowest intensities.

As described in \cite{Munoz1996}, in the frame of the CSM for stellar cluster formation \citep{Tenorio1993}, the gravitational collapse fragments the gas clouds and first forms the low-mass stars.  The bow shocks and wakes caused by their stellar winds suspend the collapse of the cloud and communicate the stellar velocity dispersion to the surrounding gas. The core of the cloud is thus virialized and the supersonic gas velocity dispersion traces its gravitational potential, despite dissipation. Massive stars form later in a cloud at equilibrium in which the velocity dispersion of the gas is constant and does not depend on the intensity of the newly formed giant \hII\ regions. In the $\sigma-I$ diagram,  this corresponds to the horizontal area (with the almost constant velocity dispersion of $20\pm3$~\kms) displaying a broad range of intensities.  

As massive stars evolve, strong mechanical energy sweeps the ISM into shells displaying supersonic velocities higher than the ambient stellar velocity dispersion.  The velocity dispersion of these shells decreases with age while their luminosity increases: young and faint shells produce a large range of velocity dispersions and low intensities while older and bright shells display a narrower velocity dispersion amplitude and higher intensities.  If the shells are embedded in the virialized core, their velocities cannot be lower than the stellar velocity dispersion but, if the shells are outside the core, their velocities can decrease to lower values, down to subsonic velocities.  In the $\sigma-I$ diagram,  these shell phases, ages and locations are found within the broad range of velocity dispersions at low intensities. To fully understand how this last area of the $\sigma-I$ diagram is filled by expanding shells, this zone should be understood as the result of different projection effects. 

On the one hand, emission from a line-of-sight passing across the center of an expanding spherical shell of gas will present the largest velocity difference across the shell due to the large distance separating its two edges and an intermediate intensity.  On the other hand, a line-of-sight going through the inner-edge of the shell will present a lower velocity difference but a very high intensity due to the large quantity of gas integrated along the line-of-sight.  Finally, a line-of-sight running through the outer-edge of the shell will display even a much lower velocity difference and a very low intensity due to the small quantity of gas integrated along the line-of-sight. In addition to these shells that cannot be distinguished individually on this plot, this area of the $\sigma-I$ diagram mainly corresponds to diffuse \ha\ emission.

\subsection{Comparison of the \ha\ kinematics with \hi\ results}
\begin{figure} 
\centering
 \includegraphics[width=\columnwidth]{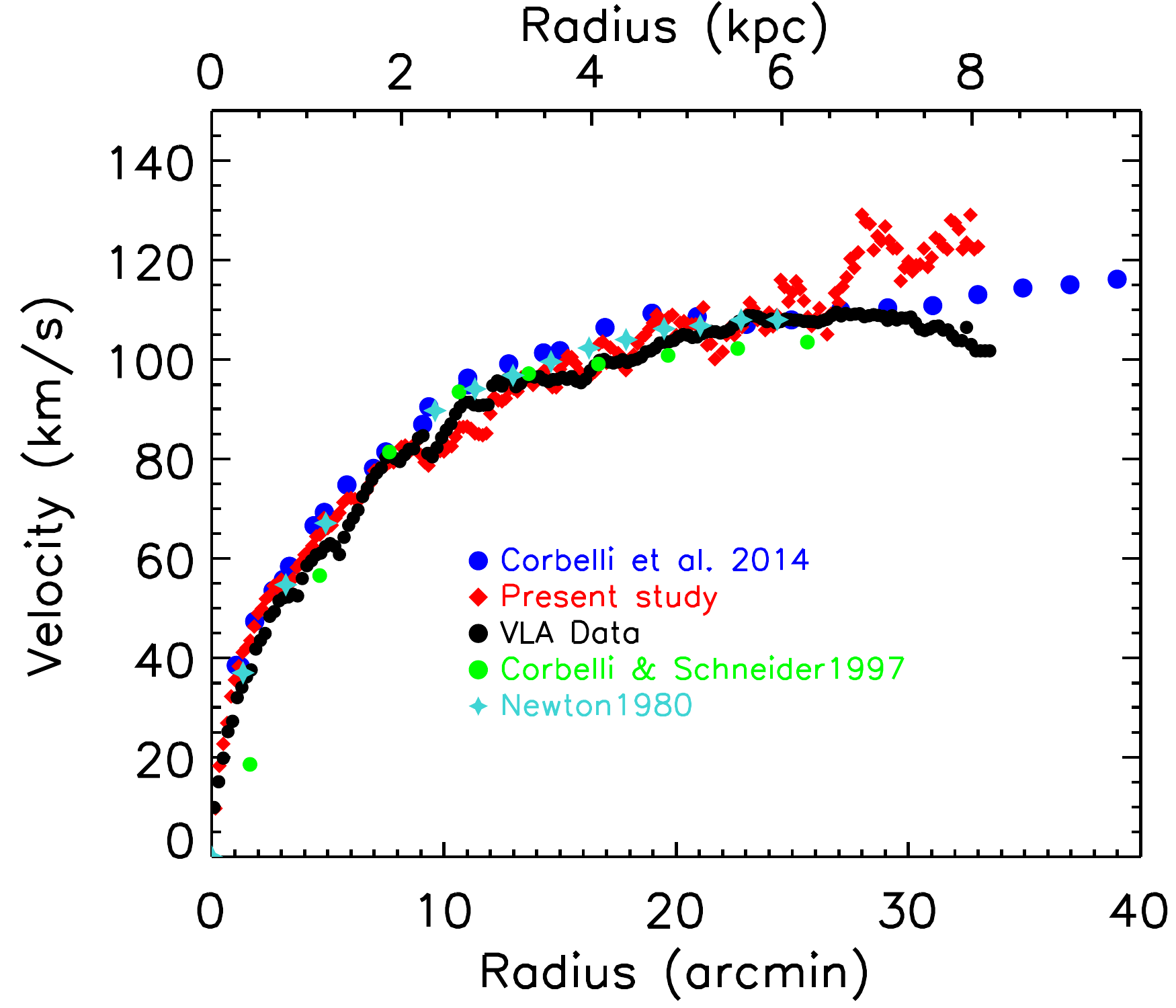}
\caption{Comparison of \ha\  and \hI\ rotation curves. Blue  represents  the \hI\ (squares) RC from \protect\cite{Corbelli2014}; black is for  \protect\cite{Gratier2010} VLA data;  green circles are for \protect\cite{Corbelli1997} data; green crosses for \protect\cite{Newton1980} data and red symbols are for our \ha\ RC.} 
\label{fig:rv_comp}
\end{figure}

 In this section, a comparison of our \ha\ kinematical results is done with previous \hi\ studies.  Normally, our 5" binning \ha\ RC should give the optimal representation of the kinematics in the inner parts of M33, without suffering from e.g. the beam smearing that may affect earlier \hI\ data. 
 Figure~\ref{fig:rv_comp} compares  the \hI\ RC derived by \cite{Corbelli2014}, as well as older \hi\ studies by   \cite{Corbelli1997} and \cite{Newton1980} and recent VLA data at 12\arcsec\ resolution \citep{Gratier2010} with the \ha\ RC obtained in this study. It is not surprising that the Arecibo data of \cite{Corbelli1997} suffer from beam smearing in the inner kpc and underestimate the rotational velocities while the 1.5\arcmin x 3.0\arcmin\ resolution
 Newton (1980) data are quite consistent with the 12\arcsec\ resolution ($\sim$ 48 pc) \hi\ data and our 5\arcsec\ step ($\sim$ 20 pc) \ha\ data, at least out to 6.5 kpc. Comparing our \ha\ data to the \hi\ for R $>$ 6.5 kpc seems to suggest that we may overestimate the RC in the outer parts. 
 Since the difference occurs at radii with the lowest number of independent bins (see bottom panel of Fig.~\ref{fig:veldisp}), it could simply reflect the difficulty to derive accurate \ha\ rotation velocities by our fitting procedure within these regions. However it should be  noted that the 2010 VLA data of \cite{Gratier2010} have lower velocities than \cite{Corbelli2014} in those same outer parts. 

\subsection{Mass modeling results}

 The extent of the spheroidal component shown in the surface brightness profile is very small $ r \le 1.5$ arcmin~($\le 350$ pc); with a bulge-to-disk ratio of only $\sim$0.04, large differences between the mass models with and without a bulge component are not expected. In each case,  a best-fit model and a fixed  M/L model are explored. The reduced chi-square  $\chi^2_r$ is used to determine the goodness of the fit.

In the pure disk case (Figure~\ref{fig:m33dm_all}), the two DM best-fit models (ISO and NFW) give very similar good fits as shown by the values of the $\chi^2_r$  in Table~\ref{resulmassmodelDM}. The redisuals (bottom panels) shows that the only discrepancy is in the very inner parts. When using the fixed M/L value of the disk, it can be seen that ISO gives a better fit ($\chi^2_r$ = 1.16 vs 1.56), despite the fact that it still slightly underestimates the velocities for R $<$ 2 kpc. This may points out  that the method used to derive the M/L may underestimates it slightly (0.72 vs 0.83). This could also point out that an additional mass component (a small bar or nuclear disk) is missing in our models in the innermost parts of the disk.

As expected, in the bulge-disk case, it is clear that the addition of a bulge component does not bring anything more than the disk only model due to its very small Re and the extra uncertainty coming from the M/L of the bulge. Looking at Figure~\ref{fig:m33dm_allbd} and Table~\ref{resulmassmodelDM2},  the conclusions are that both functional forms (ISO and NFW) yields very similar results. Because in the bulge+disk case, we add one free parameter ($\Upsilon_b$ of the bulge) and because the uncertainty on that parameter is large since the IR color used to derive it is polluted by disk light with the result of underestimating the IR color and thus the $\Upsilon_b$ of the bulge, we favor the results of the disk-only models.

We conclude, that the pure-disk ISO models (best-fit and/or fixed M/L) give a slightly better representation of the mass distribution using the \ha\ rotation curve that is derived out to $\sim$8 kpc  with a M/L for the luminous disk  $\sim 0.72 - 0.82$,  a core radius $\sim12 - 24$ kpc and a central density $\rho_0 \sim 0.007 - 0.008$  \msol .$\rm pc^{-3}$  for  the dark isothermal halo. It is clear that when we will combine those \ha\ data to more extended \hi\ kinematics, we should be able to constrain much better the parameters of the mass distribution.

%% file: M33RCadoptedV2.tex
\onecolumn
\begin{longtable}{lllllllllllllll}
%First page
  \caption{ The \ha\ rotation curve of M33 }\\ \hline \hline
Rad & V &$\Delta$V  & $\sigma$&$\Delta \sigma$ &Rad & V  &$\Delta$V  & $\sigma$&$\Delta \sigma$ & Rad  & V &$\Delta$V & $\sigma$&$\Delta \sigma$ \\ \hline
\endfirsthead
%All pages
 \multicolumn{9}{c}{{\centering{Table~\ref{tab:rctotal} Continued}}}\\ \hline
Rad & V  &$\Delta$V   & $\sigma$&$\Delta \sigma$ &Rad & V  &$\Delta$V & $\sigma$&$\Delta \sigma$ & Rad  & V &$\Delta$V & $\sigma$&$\Delta \sigma$ \\ \hline
(1)&(2)&(3)&(4)&(5)&(6)&(7)&(8)&(9)&(10)&(11)&(12)&(13)&(14)&(15)\\ \hline
\endhead
\hline \multicolumn{8}{c}{{\centering{Continued on next page}}}\\ 
\endfoot
\endlastfoot %  \hline
%(') &(km/s) & (km/s)& (') &(km/s) & (km/s)  & (') &(km/s) & (kms)\\ 
(1)&(2)&(3)&(4)&(5)&(6)&(7)&(8)&(9)&(10)&(11)&(12)&(13)&(14)&(15)\\
%    \hline
    \hline
    0.08 & 6  & 4  & 14 & 10 & 11.08 & 87 & 4  & 16 & 5  & 22.08 & 104 & 5  & 17 & 7 \\
    0.17 & 9  & 4  & 12 & 6  & 11.17 & 86 & 3  & 16 & 5  & 22.17 & 105 & 6  & 17 & 7 \\
    0.25 & 13 & 3  & 12 & 4  & 11.25 & 86 & 3  & 16 & 6  & 22.25 & 106 & 8  & 18 & 7 \\
    0.33 & 18 & 5  & 14 & 5  & 11.33 & 86 & 3  & 16 & 6  & 22.33 & 107 & 7  & 19 & 8 \\
    0.42 & 21 & 4  & 15 & 4  & 11.42 & 86 & 4  & 15 & 6  & 22.42 & 106 & 7  & 18 & 8 \\
    0.50 & 23 & 4  & 16 & 4  & 11.50 & 85 & 3  & 15 & 6  & 22.50 & 105 & 5  & 17 & 7 \\
    0.58 & 25 & 3  & 15 & 4  & 11.58 & 83 & 3  & 15 & 6  & 22.58 & 107 & 5  & 16 & 7 \\
    0.67 & 26 & 3  & 16 & 5  & 11.67 & 85 & 3  & 16 & 6  & 22.67 & 107 & 6  & 17 & 8 \\
    0.75 & 31 & 3  & 16 & 5  & 11.75 & 86 & 3  & 16 & 6  & 22.75 & 105 & 7  & 16 & 7 \\
    0.83 & 32 & 3  & 15 & 5  & 11.83 & 85 & 3  & 16 & 6  & 22.83 & 107 & 5  & 15 & 7 \\
    0.92 & 35 & 3  & 15 & 4  & 11.92 & 87 & 3  & 16 & 6  & 22.92 & 108 & 5  & 16 & 7 \\
    1.00 & 35 & 3  & 15 & 4  & 12.00 & 89 & 3  & 17 & 7  & 23.00 & 109 & 5  & 15 & 6 \\
    1.08 & 37 & 3  & 14 & 4  & 12.08 & 91 & 3  & 17 & 7  & 23.08 & 111 & 4  & 15 & 6 \\
    1.17 & 38 & 4  & 14 & 4  & 12.17 & 92 & 3  & 17 & 7  & 23.17 & 112 & 4  & 15 & 6 \\
    1.25 & 39 & 3  & 14 & 5  & 12.25 & 93 & 3  & 18 & 7  & 23.25 & 111 & 3  & 15 & 6 \\
    1.33 & 42 & 4  & 14 & 5  & 12.33 & 92 & 4  & 18 & 7  & 23.33 & 111 & 4  & 15 & 5 \\
    1.42 & 41 & 4  & 14 & 5  & 12.42 & 91 & 4  & 18 & 6  & 23.42 & 108 & 7  & 15 & 5 \\
    1.50 & 42 & 3  & 14 & 5  & 12.50 & 92 & 4  & 18 & 7  & 23.50 & 108 & 8  & 15 & 6 \\
    1.58 & 43 & 3  & 14 & 5  & 12.58 & 91 & 4  & 18 & 7  & 23.58 & 108 & 6  & 14 & 5 \\
    1.67 & 43 & 5  & 13 & 4  & 12.67 & 92 & 3  & 18 & 8  & 23.67 & 109 & 6  & 13 & 5 \\
    1.75 & 44 & 5  & 13 & 4  & 12.75 & 93 & 3  & 18 & 9  & 23.75 & 108 & 6  & 14 & 7 \\
    1.83 & 46 & 4  & 13 & 4  & 12.83 & 94 & 3  & 19 & 9  & 23.83 & 105 & 7  & 15 & 7 \\
    1.92 & 48 & 3  & 14 & 5  & 12.92 & 95 & 3  & 18 & 8  & 23.92 & 106 & 7  & 16 & 8 \\
    2.00 & 49 & 4  & 14 & 4  & 13.00 & 94 & 3  & 18 & 8  & 24.00 & 110 & 5  & 17 & 8 \\
    2.08 & 49 & 5  & 14 & 4  & 13.08 & 94 & 3  & 18 & 8  & 24.08 & 111 & 5  & 16 & 7 \\
    2.17 & 50 & 6  & 14 & 5  & 13.17 & 93 & 5  & 17 & 7  & 24.17 & 105 & 6  & 16 & 7 \\
    2.25 & 51 & 5  & 13 & 5  & 13.25 & 95 & 4  & 17 & 7  & 24.25 & 105 & 6  & 16 & 8 \\
    2.33 & 52 & 5  & 13 & 5  & 13.33 & 96 & 3  & 17 & 7  & 24.33 & 109 & 3  & 18 & 9 \\
    2.42 & 53 & 7  & 13 & 5  & 13.42 & 97 & 3  & 17 & 7  & 24.42 & 113 & 5  & 17 & 8 \\
    2.50 & 53 & 8  & 13 & 5  & 13.50 & 96 & 3  & 17 & 7  & 24.50 & 116 & 4  & 17 & 9 \\
    2.58 & 53 & 7  & 13 & 5  & 13.58 & 96 & 3  & 17 & 7  & 24.58 & 116 & 4  & 16 & 7 \\
    2.67 & 54 & 7  & 13 & 6  & 13.67 & 96 & 3  & 17 & 7  & 24.67 & 115 & 4  & 16 & 7 \\
    2.75 & 55 & 7  & 13 & 6  & 13.75 & 95 & 3  & 17 & 8  & 24.75 & 113 & 4  & 16 & 7 \\
    2.83 & 53 & 6  & 13 & 5  & 13.83 & 95 & 3  & 16 & 6  & 24.83 & 111 & 4  & 16 & 7 \\
    2.92 & 56 & 5  & 13 & 5  & 13.92 & 95 & 3  & 16 & 6  & 24.92 & 113 & 4  & 16 & 6 \\
    3.00 & 56 & 3  & 14 & 5  & 14.00 & 96 & 4  & 16 & 6  & 25.00 & 114 & 4  & 16 & 7 \\
    3.08 & 54 & 3  & 14 & 5  & 14.08 & 97 & 3  & 16 & 7  & 25.08 & 115 & 3  & 14 & 6 \\
    3.17 & 53 & 4  & 15 & 5  & 14.17 & 97 & 3  & 15 & 6  & 25.17 & 116 & 6  & 14 & 6 \\
    3.25 & 53 & 3  & 15 & 5  & 14.25 & 98 & 3  & 14 & 7  & 25.25 & 113 & 10 & 13 & 6 \\
    3.33 & 55 & 3  & 15 & 6  & 14.33 & 98 & 3  & 14 & 6  & 25.33 & 119 & 6  & 14 & 6 \\
    3.42 & 56 & 3  & 15 & 5  & 14.42 & 97 & 3  & 15 & 6  & 25.42 & 112 & 14 & 14 & 6 \\
    3.50 & 56 & 3  & 15 & 5  & 14.50 & 96 & 3  & 15 & 6  & 25.50 & 113 & 14 & 14 & 6 \\
    3.58 & 56 & 3  & 15 & 5  & 14.58 & 95 & 3  & 15 & 6  & 25.58 & 111 & 14 & 15 & 6 \\
    3.67 & 58 & 3  & 15 & 6  & 14.67 & 95 & 4  & 14 & 6  & 25.67 & 109 & 16 & 15 & 6 \\
    3.75 & 59 & 3  & 15 & 6  & 14.75 & 95 & 3  & 15 & 6  & 25.75 & 107 & 19 & 15 & 5 \\
    3.83 & 59 & 3  & 15 & 5  & 14.83 & 94 & 4  & 14 & 6  & 25.83 & 106 & 18 & 15 & 6 \\
    3.92 & 60 & 3  & 15 & 6  & 14.92 & 96 & 5  & 15 & 7  & 25.92 & 107 & 18 & 16 & 6 \\
    4.00 & 61 & 3  & 15 & 5  & 15.00 & 98 & 7  & 15 & 7  & 26.00 & 108 & 19 & 17 & 7 \\
    4.08 & 61 & 3  & 15 & 5  & 15.08 & 99 & 7  & 15 & 6  & 26.08 & 110 & 17 & 17 & 6 \\
    4.17 & 62 & 3  & 15 & 5  & 15.17 & 100 & 9  & 15 & 6  & 26.17 & 110 & 18 & 16 & 6 \\
    4.25 & 61 & 3  & 15 & 5  & 15.25 & 100 & 9  & 15 & 7  & 26.25 & 111 & 20 & 16 & 6 \\
    4.33 & 62 & 4  & 15 & 5  & 15.33 & 101 & 9  & 15 & 7  & 26.33 & 108 & 17 & 17 & 6 \\
    4.42 & 64 & 5  & 15 & 5  & 15.42 & 100 & 9  & 15 & 7  & 26.42 & 104 & 16 & 17 & 6 \\
    4.50 & 65 & 5  & 15 & 5  & 15.50 & 100 & 9  & 15 & 7  & 26.50 & 105 & 17 & 16 & 6 \\
    4.58 & 64 & 5  & 15 & 5  & 15.58 & 102 & 9  & 15 & 7  & 26.58 & 108 & 18 & 15 & 6 \\
    4.67 & 65 & 6  & 15 & 5  & 15.67 & 99 & 6  & 16 & 7  & 26.67 & 109 & 17 & 15 & 7 \\
    4.75 & 67 & 5  & 16 & 6  & 15.75 & 97 & 4  & 16 & 6  & 26.75 & 112 & 17 & 15 & 7 \\
    4.83 & 69 & 5  & 16 & 5  & 15.83 & 98 & 6  & 16 & 6  & 26.83 & 114 & 16 & 16 & 8 \\
    4.92 & 68 & 5  & 16 & 5  & 15.92 & 98 & 5  & 16 & 6  & 26.92 & 112 & 14 & 15 & 8 \\
    5.00 & 65 & 6  & 16 & 5  & 16.00 & 97 & 5  & 16 & 6  & 27.00 & 110 & 14 & 14 & 8 \\
    5.08 & 66 & 6  & 16 & 5  & 16.08 & 96 & 4  & 16 & 7  & 27.08 & 112 & 11 & 15 & 8 \\
    5.17 & 67 & 5  & 16 & 5  & 16.17 & 97 & 4  & 16 & 6  & 27.17 & 115 & 7  & 15 & 9 \\
    5.25 & 67 & 6  & 16 & 5  & 16.25 & 97 & 4  & 16 & 6  & 27.25 & 119 & 7  & 14 & 9 \\
    5.33 & 68 & 6  & 16 & 5  & 16.33 & 97 & 4  & 15 & 6  & 27.33 & 116 & 4  & 15 & 9 \\
    5.42 & 69 & 5  & 16 & 5  & 16.42 & 96 & 5  & 16 & 7  & 27.42 & 119 & 5  & 15 & 12 \\
    5.50 & 69 & 5  & 17 & 5  & 16.50 & 98 & 5  & 16 & 6  & 27.50 & 121 & 3  & 20 & 14 \\
    5.58 & 71 & 5  & 17 & 5  & 16.58 & 100 & 5  & 16 & 7  & 27.58 & 126 & 8  & 15 & 7 \\
    5.67 & 71 & 5  & 17 & 5  & 16.67 & 104 & 8  & 15 & 6  & 27.67 & 116 & 3  & 12 & 8 \\
    5.75 & 72 & 5  & 16 & 5  & 16.75 & 104 & 8  & 16 & 7  & 27.75 & 114 & 7  & 13 & 7 \\
    5.83 & 72 & 5  & 16 & 5  & 16.83 & 104 & 6  & 15 & 6  & 27.83 & 122 & 10 & 15 & 8 \\
    5.92 & 72 & 5  & 16 & 5  & 16.92 & 102 & 5  & 16 & 5  & 27.92 & 124 & 13 & 15 & 8 \\
    6.00 & 72 & 5  & 17 & 5  & 17.00 & 99 & 4  & 17 & 6  & 28.00 & 130 & 3  & 19 & 12 \\
    6.08 & 72 & 6  & 17 & 5  & 17.08 & 100 & 3  & 17 & 6  & 28.08 & 126 & 4  & 15 & 9 \\
    6.17 & 72 & 6  & 17 & 5  & 17.17 & 104 & 7  & 17 & 6  & 28.17 & 128 & 4  & 16 & 9 \\
    6.25 & 71 & 5  & 17 & 5  & 17.25 & 99 & 3  & 17 & 6  & 28.25 & 129 & 3  & 16 & 11 \\
    6.33 & 72 & 5  & 18 & 6  & 17.33 & 99 & 3  & 18 & 7  & 28.33 & 126 & 5  & 14 & 6 \\
    6.42 & 72 & 8  & 18 & 6  & 17.42 & 100 & 3  & 19 & 7  & 28.42 & 120 & 3  & 14 & 7 \\
    6.50 & 72 & 8  & 18 & 6  & 17.50 & 105 & 6  & 19 & 7  & 28.50 & 123 & 3  & 17 & 10 \\
    6.58 & 73 & 8  & 18 & 6  & 17.58 & 97 & 3  & 19 & 6  & 28.58 & 123 & 3  & 19 & 13 \\
    6.67 & 73 & 9  & 18 & 6  & 17.67 & 101 & 3  & 18 & 7  & 28.67 & 126 & 3  & 16 & 9 \\
    6.75 & 74 & 8  & 18 & 6  & 17.75 & 100 & 4  & 18 & 7  & 28.75 & 124 & 4  & 19 & 9 \\
    6.83 & 75 & 6  & 17 & 6  & 17.83 & 97 & 3  & 16 & 6  & 28.83 & 123 & 3  & 15 & 9 \\
    6.92 & 76 & 5  & 17 & 6  & 17.92 & 98 & 3  & 17 & 6  & 28.92 & 124 & 4  & 19 & 10 \\
    7.00 & 78 & 4  & 17 & 6  & 18.00 & 99 & 3  & 17 & 7  & 29.00 & 125 & 5  & 19 & 9 \\
    7.08 & 78 & 4  & 17 & 6  & 18.08 & 101 & 3  & 17 & 7  & 29.08 & 128 & 3  & 16 & 7 \\
    7.17 & 78 & 4  & 17 & 6  & 18.17 & 102 & 3  & 17 & 7  & 29.17 & 122 & 3  & 15 & 8 \\
    7.25 & 78 & 4  & 17 & 6  & 18.25 & 103 & 3  & 18 & 7  & 29.25 & 123 & 10 & 15 & 8 \\
    7.33 & 79 & 4  & 17 & 5  & 18.33 & 104 & 3  & 18 & 7  & 29.33 & 123 & 3  & 17 & 10 \\
    7.42 & 79 & 3  & 17 & 6  & 18.42 & 104 & 3  & 18 & 8  & 29.42 & 124 & 4  & 13 & 5 \\
    7.50 & 79 & 3  & 18 & 6  & 18.50 & 104 & 3  & 18 & 7  & 29.50 & 121 & 3  & 15 & 8 \\
    7.58 & 78 & 3  & 18 & 6  & 18.58 & 105 & 3  & 18 & 7  & 29.58 & 118 & 4  & 15 & 5 \\
    7.67 & 80 & 3  & 17 & 6  & 18.67 & 105 & 3  & 18 & 8  & 29.67 & 118 & 9  & 19 & 10 \\
    7.75 & 79 & 3  & 18 & 6  & 18.75 & 105 & 3  & 17 & 7  & 29.75 & 122 & 3  & 15 & 8 \\
    7.83 & 80 & 3  & 18 & 6  & 18.83 & 106 & 3  & 17 & 6  & 29.83 & 122 & 3  & 15 & 15 \\
    7.92 & 79 & 3  & 18 & 6  & 18.92 & 106 & 3  & 17 & 7  & 29.92 & 124 & 4  & 14 & 7 \\
    8.00 & 80 & 3  & 17 & 6  & 19.00 & 107 & 3  & 17 & 7  & 30.00 & 121 & 3  & 14 & 10 \\
    8.08 & 82 & 4  & 17 & 6  & 19.08 & 109 & 3  & 18 & 7  & 30.08 & 121 & 3  & 16 & 10 \\
    8.17 & 82 & 3  & 18 & 6  & 19.17 & 109 & 3  & 17 & 6  & 30.17 & 122 & 3  & 15 & 7 \\
    8.25 & 83 & 4  & 18 & 6  & 19.25 & 108 & 3  & 17 & 6  & 30.25 & 124 & 3  & 14 & 7 \\
    8.33 & 83 & 5  & 17 & 6  & 19.33 & 107 & 3  & 16 & 6  & 30.33 & 119 & 6  & 15 & 6 \\
    8.42 & 82 & 5  & 17 & 6  & 19.42 & 108 & 3  & 16 & 6  & 30.42 & 123 & 3  & 15 & 7 \\
    8.50 & 81 & 5  & 18 & 6  & 19.50 & 108 & 3  & 16 & 6  & 30.50 & 119 & 6  & 15 & 7 \\
    8.58 & 82 & 4  & 18 & 6  & 19.58 & 108 & 3  & 16 & 6  & 30.58 & 123 & 3  & 15 & 7 \\
    8.67 & 83 & 5  & 17 & 6  & 19.67 & 109 & 3  & 16 & 6  & 30.67 & 122 & 4  & 14 & 6 \\
    8.75 & 84 & 6  & 17 & 5  & 19.75 & 109 & 3  & 16 & 6  & 30.75 & 119 & 5  & 12 & 8 \\
    8.83 & 84 & 6  & 17 & 5  & 19.83 & 109 & 3  & 16 & 6  & 30.83 & 122 & 3  & 13 & 6 \\
    8.92 & 83 & 6  & 17 & 5  & 19.92 & 105 & 4  & 16 & 7  & 30.92 & 122 & 3  & 12 & 7 \\
    9.00 & 81 & 4  & 17 & 6  & 20.00 & 104 & 5  & 16 & 6  & 31.00 & 122 & 4  & 15 & 11 \\
    9.08 & 80 & 3  & 17 & 6  & 20.08 & 106 & 3  & 16 & 7  & 31.08 & 124 & 4  & 13 & 8 \\
    9.17 & 80 & 3  & 17 & 6  & 20.17 & 107 & 3  & 15 & 6  & 31.17 & 122 & 3  & 20 & 14 \\
    9.25 & 78 & 3  & 16 & 6  & 20.25 & 108 & 3  & 15 & 8  & 31.25 & 120 & 3  & 20 & 19 \\
    9.33 & 78 & 3  & 16 & 6  & 20.33 & 108 & 3  & 15 & 8  & 31.33 & 124 & 5  & 21 & 23 \\
    9.42 & 80 & 3  & 15 & 6  & 20.42 & 107 & 3  & 14 & 7  & 31.42 & 122 & 3  & 12 & 7 \\
    9.50 & 81 & 4  & 15 & 6  & 20.50 & 107 & 3  & 15 & 7  & 31.50 & 122 & 3  & 19 & 14 \\
    9.58 & 82 & 5  & 16 & 6  & 20.58 & 106 & 3  & 15 & 6  & 31.58 & 123 & 3  & 22 & 16 \\
    9.67 & 81 & 4  & 16 & 6  & 20.67 & 106 & 3  & 15 & 7  & 31.67 & 122 & 3  & 17 & 11 \\
    9.75 & 81 & 5  & 15 & 6  & 20.75 & 108 & 3  & 14 & 7  & 31.75 & 128 & 6  & 22 & 11 \\
    9.83 & 82 & 5  & 16 & 6  & 20.83 & 107 & 3  & 14 & 8  & 31.83 & 122 & 3  & 17 & 12 \\
    9.92 & 82 & 4  & 16 & 6  & 20.92 & 106 & 3  & 15 & 8  & 31.92 & 122 & 3  & 15 & 13 \\
    10.00 & 81 & 4  & 15 & 6  & 21.00 & 108 & 3  & 15 & 8  & 32.00 & 128 & 4  & 22 & 21 \\
    10.08 & 82 & 5  & 15 & 5  & 21.08 & 109 & 3  & 15 & 10 & 32.08 & 122 & 3  & 16 & 7 \\
    10.17 & 83 & 3  & 15 & 6  & 21.17 & 111 & 3  & 15 & 11 & 32.17 & 126 & 5  & 16 & 9 \\
    10.25 & 83 & 3  & 15 & 6  & 21.25 & 106 & 4  & 15 & 9  & 32.25 & 122 & 3  & 25 & 17 \\
    10.33 & 82 & 3  & 16 & 6  & 21.33 & 105 & 6  & 16 & 10 & 32.33 & 122 & 3  & 30 & 3 \\
    10.42 & 84 & 3  & 16 & 6  & 21.42 & 111 & 9  & 17 & 12 & 32.42 & 126 & 5  & 20 & 19 \\
    10.50 & 84 & 3  & 15 & 6  & 21.50 & 113 & 8  & 14 & 7  & 32.50 & 121 & 4  & 19 & 7 \\
    10.58 & 85 & 5  & 16 & 5  & 21.58 & 113 & 8  & 15 & 8  & 32.58 & 122 & 3  & 29 & 20 \\
    10.67 & 86 & 4  & 16 & 6  & 21.67 & 100 & 6  & 15 & 7  & 32.67 & 129 & 8  & 13 & 9 \\
    10.75 & 87 & 3  & 16 & 6  & 21.75 & 96 & 10 & 15 & 8  & 32.75 & 122 & 3  & 15 & 4 \\
    10.83 & 86 & 3  & 16 & 6  & 21.83 & 97 & 10 & 16 & 8  & 32.83 & 122 & 3  & 18 & 8 \\
    10.92 & 87 & 3  & 16 & 6  & 21.92 & 98 & 6  & 15 & 8  & 32.92 & 122 & 3  & 18 & 13 \\
    11.00 & 87 & 3  & 16 & 5  & 22.00 & 103 & 6  & 15 & 7  & 33.00 & 123 & 3  & 22 & 3 \\
  \label{tab:rctotal}
\end{longtable}